\documentclass[12pt,english]{article}
\usepackage{jheppub_modified}

\usepackage[latin9]{inputenc}
\usepackage[USenglish]{babel}
\usepackage{esint}
\usepackage{dsfont}

\usepackage[usenames,dvipsnames]{color}

\usepackage{multirow}

\usepackage{suffix}

\usepackage[normalem]{ulem}

\newcommand{\Dcal}{\mathcal{D}}
\newcommand{\Lcal}{\mathcal{L}}

\newcommand{\Ncal}{\mathcal{N}}
\newcommand{\Rcal}{\mathcal{R}}

\newcommand{\phitt}{\tilde{\phi}}
\newcommand{\psitt}{\tilde{\psi}}
\newcommand{\Ftt}{\tilde{F}}
\newcommand{\Btt}{\tilde{B}}
\newcommand{\lambdatt}{\tilde{\lambda}}
\newcommand{\Dtt}{\tilde{D}}
\newcommand{\xitt}{\tilde{\xi}}

\newcommand{\abb}{\bar{a}}
\newcommand{\bbb}{\bar{b}}
\newcommand{\Wbb}{\bar{W}}

\newcommand{\lp}{\left (}
\newcommand{\rp}{\right )}
\newcommand{\lb}{\left [}
\newcommand{\rb}{\right ]}
\newcommand{\lc}{\left \{}
\newcommand{\rc}{\right \}}
\newcommand{\lr}{\left .}
\newcommand{\rr}{\right .}

\newcommand{\Tr}{\text{Tr}}
\newcommand{\diag}{\text{diag}}
\newcommand{\pderiv}[2]{\frac{\partial #1}{\partial #2}}

\newcommand{\duind}[3]{{#1}_{#2}^{\phantom{#2}#3}}
\newcommand{\udind}[3]{{#1}^{#2}_{\phantom{#2}#3}}
\newcommand{\fduind}[3]{(#1)_{#2}^{\phantom{#2}#3}}
\newcommand{\ket}[1]{|#1\rangle}
\newcommand{\var}[1]{{\color{blue} #1}}

\newcommand{\beq}{\begin{equation}}
\newcommand{\eeq}{\end{equation}}
\newcommand{\bes}{\begin{equation*}}
\newcommand{\ees}{\end{equation*}}
\newcommand{\bea}{\begin{align}}
\newcommand{\ena}{\end{align}}
\newcommand{\eea}{\end{align}}
\newcommand{\beas}{\begin{align*}}
\newcommand{\eeas}{\end{align*}}
\makeatletter
\newenvironment{calc}{\allowdisplaybreaks\start@align\@ne\st@rredtrue\m@ne}
{\addtocounter{equation}{1}\tag{\theequation}\endalign}
\makeatother

\newcommand{\cc}{g_{\text{YM}}}
\newcommand{\rst}{\mathbb{R}\times S^3}
\newcommand{\rsp}{R_\text{s}}

\definecolor{purple}{rgb}{0.5,0,0.5}

\newcommand{\rmv}[1]{}

\DeclareFontShape{OT1}{cmr}{mx}{n}{<->cmr10}{}
\fontseries{mx}\selectfont

\newcommand{\tgorange}[1]{%
{\setlength\fboxrule{1pt}%
\fcolorbox{YellowOrange}{white}{$\displaystyle #1$}}}
\WithSuffix\newcommand\tgorange*[1]{%
{\setlength\fboxrule{1pt}%
\fcolorbox{YellowOrange}{white}{$\displaystyle{\color{red}\cancel{\color{black} #1}}$}}}
\newcommand{\ttorange}[1]{%
{\color{YellowOrange}#1}}

\newcommand{\tggreen}[1]{%
{\setlength\fboxrule{1pt}%
\fcolorbox{Green}{white}{$\displaystyle #1$}}}
\WithSuffix\newcommand\tggreen*[1]{%
{\setlength\fboxrule{1pt}%
\fcolorbox{Green}{white}{$\displaystyle{\color{red}\cancel{\color{black} #1}}$}}}

\newcommand{\tgblue}[1]{%
{\setlength\fboxrule{1pt}%
\fcolorbox{Blue}{white}{$\displaystyle #1$}}}
\WithSuffix\newcommand\tgblue*[1]{%
{\setlength\fboxrule{1pt}%
\fcolorbox{Blue}{white}{$\displaystyle{\color{red}\cancel{\color{black} #1}}$}}}

\newcommand{\tgcyan}[1]{%
{\setlength\fboxrule{1pt}%
\fcolorbox{Cyan}{white}{$\displaystyle #1$}}}
\WithSuffix\newcommand\tgcyan*[1]{%
{\setlength\fboxrule{1pt}%
\fcolorbox{Cyan}{white}{$\displaystyle{\color{red}\cancel{\color{black} #1}}$}}}
\newcommand{\ttcyan}[1]{%
{\color{Cyan}#1}}

\newcommand{\tglime}[1]{%
{\setlength\fboxrule{1pt}%
\fcolorbox{LimeGreen}{white}{$\displaystyle #1$}}}
\WithSuffix\newcommand\tglime*[1]{%
{\setlength\fboxrule{1pt}%
\fcolorbox{LimeGreen}{white}{$\displaystyle{\color{red}\cancel{\color{black} #1}}$}}}
\newcommand{\ttlime}[1]{%
{\color{LimeGreen}#1}}

\newcommand{\tgmagenta}[1]{%
{\setlength\fboxrule{1pt}%
\fcolorbox{Magenta}{white}{$\displaystyle #1$}}}
\WithSuffix\newcommand\tgmagenta*[1]{%
{\setlength\fboxrule{1pt}%
\fcolorbox{Magenta}{white}{$\displaystyle{\color{red}\cancel{\color{black} #1}}$}}}
\newcommand{\ttmagenta}[1]{%
{\color{Magenta}#1}}

\newcommand{\tgpurple}[1]{%
{\setlength\fboxrule{1pt}%
\fcolorbox{Fuchsia}{white}{$\displaystyle #1$}}}
\WithSuffix\newcommand\tgpurple*[1]{%
{\setlength\fboxrule{1pt}%
\fcolorbox{Fuchsia}{white}{$\displaystyle{\color{red}\cancel{\color{black} #1}}$}}}
\newcommand{\ttpurple}[1]{%
{\color{Fuchsia}#1}}

\newcommand{\tgbrown}[1]{%
{\setlength\fboxrule{1pt}%
\fcolorbox{Maroon}{white}{$\displaystyle #1$}}}
\WithSuffix\newcommand\tgbrown*[1]{%
{\setlength\fboxrule{1pt}%
\fcolorbox{Maroon}{white}{$\displaystyle{\color{red}\cancel{\color{black} #1}}$}}}

\newcommand{\tgred}[1]{%
{\setlength\fboxrule{1pt}%
\fcolorbox{Red}{white}{$\displaystyle #1$}}}
\WithSuffix\newcommand\tgred*[1]{%
{\setlength\fboxrule{1pt}%
\fcolorbox{Red}{white}{$\displaystyle{\color{red}\cancel{\color{black} #1}}$}}}
\newcommand{\ttred}[1]{%
{\color{Red}#1}}

\allowdisplaybreaks

\title{Massive quiver matrix models for massive charged particles in AdS}

\author[a]{Curtis T. Asplund,}
\author[a]{Frederik Denef}
\author[b]{and Eric Dzienkowski}

\affiliation[a]{Department of Physics, Columbia University, 538 West 120th Street, New York, New York 10027}
\affiliation[b]{Department of Physics, Broida Hall, University of California Santa Barbara, Santa Barbara, California 93106}

\emailAdd{ca2621@columbia.edu}
\emailAdd{fmd7@columbia.edu}
\emailAdd{eric.m.dzienkowski@gmail.com}

\date{\today}

\abstract{We present a new class of ${\cal N}=4$ supersymmetric quiver matrix models and argue that it describes the stringy low-energy dynamics of internally wrapped D-branes in four-dimensional anti-de Sitter (AdS) flux compactifications. The Lagrangians of these models differ from previously studied quiver matrix models by the presence of mass terms, associated with the AdS gravitational potential, as well as additional terms dictated by supersymmetry. These give rise to dynamical phenomena typically associated with the presence of fluxes, such as fuzzy membranes, internal cyclotron motion and the appearance of confining strings. We also show how these models can be obtained by dimensional reduction of four-dimensional supersymmetric quiver gauge theories on a three-sphere.}
\toccontinuoustrue

\begin{document}
\maketitle


\section{Introduction and summary}
\label{sec:intro}

Type II string theory compactified on a six-dimensional manifold $X_6$ gives rise to a four-dimensional spacetime $M_4$. A D$p$-brane wrapped on a $p$-dimensional cycle in $X_6$ appears as a charged particle in $M_4$.
The wrapped branes interact with each other via strings that end on them. At small separations, the lightest open string modes dominate the interactions. The low-energy dynamics of these modes is captured by quiver matrix mechanics \cite{Douglas:1996sw,Douglas:1996yp,Douglas:2000qw,Denef:2002ru}. A quiver is an oriented graph of which the vertices are called nodes and the edges arrows. In quiver matrix mechanics, the nodes correspond to four-dimensional spacetime degrees of freedom and label wrapped branes; the arrows correspond to internal-space degrees of freedom and label open string modes. If $X_6$ is a Calabi-Yau manifold without fluxes then $M_4$ is flat Minkowski space and the bulk superalgebra has eight supercharges, of which the branes preserve four. In this case, the corresponding one-dimensional ${\cal N}=4$ quiver matrix mechanics Lagrangian may be obtained by dimensional reduction of a four-dimensional ${\cal N}=1$ quiver gauge theory.  
On the other hand if $X_6$ is an Einstein manifold carrying magnetic fluxes, compactifications with eight or more supersymmetries to $M_4 = {\rm AdS}_4$ are possible. A standard example is the type IIA ${\rm AdS}_4 \times {\mathbb CP}^3$ compactification \cite{Watamura:1983hj, Nilsson:1984bj} holographically dual to ABJM theory \cite{Aharony:2008ug}. Thus, a natural question is what the analogous quiver matrix mechanics description is for D-particles in ${\rm AdS}_4$. In this paper we answer this question.

We will argue that the low energy, short distance dynamics of particles in AdS$_4$, obtained as internally wrapped branes preserving at least four supercharges, is captured by a tightly constrained ${\cal N}=4$ supersymmetric {\it massive} quiver matrix mechanics. By ``massive'' we mean the brane position degrees of freedom are trapped near the origin by a harmonic potential, interpreted here as the AdS gravitational potential well. These ${\cal N}=4$ massive quiver matrix models generalize the ${\cal N}=16$ BMN matrix model \cite{Berenstein:2002jq}, which is a mass deformation of the BFSS matrix model \cite{Banks:1996vh}. Although the standard interpretation of the BMN model is quite different from the interpretation we consider here, its Lagrangian can nevertheless be viewed a special case of our general class of models, after a suitable 
field redefinition. As was pointed out in \cite{Kim:2003rza}, the BMN model can be obtained by dimensionally reducing ${\cal N}=4$ super-Yang-Mills theory on ${\mathbb R} \times S^3$. Similarly, we will see that the ${\cal N}=4$ massive quiver matrix models we present can be obtained by dimensional reduction of ${\cal N}=1$ quiver gauge theories on ${\mathbb R} \times S^3$. The details of this reduction are given in section \ref{sec:reduction}.

\begin{figure}
\centering
\includegraphics[width=140mm]{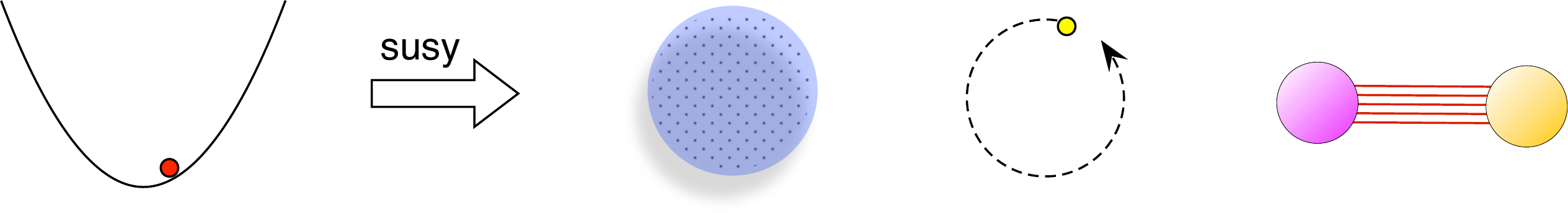}
\caption{Our results in pictures: mass deformation with four supercharges implies flux phenomena such as fuzzy membranes, internal space cyclotron motion, and branes confined by strings. 
We will elaborate on these in the introduction and especially in sections~\ref{sec:interpads}, \ref{sec:NAMyers}, \ref{sec:internalmagnetic}, and \ref{sec:gaussconfines}, respectively. \label{fig:resultpict}}
\end{figure}

The core result of the paper is the general Lagrangian of these ${\cal N}=4$ massive quiver matrix models, presented in section \ref{sec:genlagandsusy}. Besides the parameters already present in the flat-space quiver mechanics of \cite{Denef:2002ru} (particle masses $m_v$, Fayet-Ilopoulos parameters $\theta_v$ and superpotential data), they depend on just one additional mass deformation parameter $\Omega$, appearing in harmonic potentials for the particle positions $\vec x_v$,
\begin{equation}
 V(x) = \sum_v \tfrac{1}{2} m_v \, \Omega^2 \, \vec x_v^2 \, ,
\end{equation} 
as well as in a number of other terms related by supersymmetry. The parameter $\Omega$ has the dimension of frequency. In the context of our AdS interpretation, it equals the global time oscillation frequency of a particle in the AdS gravitational well: 
\begin{equation}
  \Omega = \frac{c}{\ell_{\rm AdS}} \, ,
\end{equation}
where $c$ is the speed of light and $\ell_\mathrm{AdS}$ is the AdS radius.
Under some simplifying assumptions stated in section \ref{sec:genlagandsusy}, we conjecture that this captures, in fact, the most general case consistent with the symmetries imposed.\footnote{More precisely, we conjecture that for connected quivers, and modulo ``$R$-frame'' field redefinitions discussed in \ref{sec:R-symm}, the most general massive quiver matrix mechanics preserving $\mathrm{SO}(3)$ rotation symmetry of the vector multiplets and ${\cal N}=4$ supersymmetry, assuming a flat target-space metric for both vector and chiral multiplets, is given by the Lagrangian \eqref{eq:totalLag}.}
In the context of the AdS interpretation, the isotropic harmonic potential is due to the AdS gravitational potential well. The fact that the deformation introduces just one new parameter, uniform across all connected nodes, can be physically understood as the equality of gravitational and inertial mass, i.e., the equivalence principle. 
In view, however, of the very different (short distance) regime of validity of the quiver picture and the (long distance) bulk supergravity picture, it is by no means a priori obvious that the quiver should retain this feature of gravity. It does so as a consequence of the structure of the interactions and the constraints of supersymmetry. 

Further remarkable consequences are highlighted in figure \ref{fig:resultpict}. Turning on the mass deformation for the position degrees of freedom and requiring ${\cal N}=4$ supersymmetry automatically implies all of the peculiar dynamical phenomena typically  featured by branes in {\it flux} backgrounds, including noncommutative fuzzy membranes, magnetic cyclotron motion in the internal space, and confinement of particles by fundamental strings. In section \ref{sec:examples} we discuss examples explicitly exhibiting these phenomena in simple quiver models. The supergravity counterpart of this is, essentially, that supersymmetric compactifications to AdS require flux \cite{Duff:1986hr,Grana:2005jc,Douglas:2006es}.  

We devote particular attention to the emergence of confining strings, as this is perhaps the most dramatic difference with the flat-space quiver models of \cite{Denef:2002ru}, and one of the main motivations for this work, prompted by problems raised in \cite{Anninos:2013mfa}. The goal of \cite{Anninos:2013mfa} was to demonstrate the existence of multicentered black hole bound states in AdS$_4$ flux compactifications and to investigate their potential use as holographic models of structural glasses. 
A simple four-dimensional gauged supergravity model was considered, with the appropriate ingredients needed to lift previously known, asymptotically flat bound states of black holes carrying wrapped D-brane charges \cite{Denef:2000nb,Denef:2002ru,Denef:2007vg,Anninos:2011vn} to AdS$_4$ with minimal modifications. 
However, as was pointed out already in \cite{Anninos:2013mfa}, this model actually misses an important universal feature of flux compactifications of string theory; the fact that particles obtained by wrapping branes on certain cycles are confined by fundamental strings. 

In the example of AdS$_4 \times {\mathbb CP}^3$, dual to ABJM theory, it was explained in \cite{Aharony:2008ug} how this can be understood from a four-dimensional effective field theory point of view; it is because these particles have a nonzero magnetic charge with respect to a Higgsed $\mathrm{U}(1)$. The Higgs condensate forces the magnetic flux lines into flux tubes, which act as confining strings. Alternatively, their inevitability can be inferred directly from the D-brane action. In the presence of background flux, the Gauss's law constraint for the brane worldvolume gauge field gets a contribution equal to the quantized flux threading the brane, which must be canceled by an equal amount of endpoint charge of fundamental strings attached to the brane. This shows that the confining strings are fundamental strings, and a rather universal feature of flux compactifications. If the brane is considered in isolation, the attached strings extend out from it all the way to the boundary of AdS. For this reason, such branes are often called baryonic vertices \cite{Witten:1998xy}. Note, however, that suitable pairs of charges may allow the strings emanating from one brane to terminate on the other, thus producing a finite-energy configuration. 

In section \ref{sec:two_nodes} we show that all of this is elegantly reproduced by massive quiver matrix mechanics.
Gauss's law for the quiver gauge fields forces  
charged fields to have a nonzero minimal excitation energy that grows linearly with particle separation. The tension of this string is a multiple of the fundamental string tension. 
More precisely, the number of fundamental strings $N_{F,v}$ terminating on a brane corresponding to a quiver node $v$, with Fayet-Iliopoulos (FI) parameter $\theta_v$, is given by the universal formula 
\begin{equation} \label{eq:NF1}
  N_{F,v} = \Omega \, \theta_v \, .
\end{equation}
Quantum consistency requires $N_{F,v}$ to be an integer and hence $\theta_v$ to be quantized, in contrast to the case of flat-space quivers, where the FI parameters are related to continuously tunable bulk moduli. This is consistent with the fact that bulk moduli are typically stabilized in flux compactifications. 
As in the flat space case, the FI parameters also control supersymmetric bound state formation. In particular, for a two-node quiver with all arrows oriented in one direction, a supersymmetric bound state exists for one sign of the FI parameter but not the other. An interesting immediate consequence is that the boundary of the region in constituent charge space where supersymmetric bound states cease to exist is the same as the codimension-one slice through charge space where confining strings between the constituents are absent. In section \ref{sec:ABJM} we interpret these findings in some detail for internally wrapped branes in AdS$_4 \times {\mathbb CP}^3$.

Most of the analysis in this paper is classical, but we provide the complete quantum Hamiltonian and supersymmetry algebra in appendix \ref{app:algebra}. The supersymmetry algebra is $\mathfrak{su}(2|1)$. If the Lagrangian has a $\mathrm{U}(1)_R$ symmetry, the algebra is extended to the semidirect product $\mathfrak{su}(2|1) \rtimes \mathfrak{u}(1)_R$. This algebra arises naturally on the worldline of a superparticle in an ${\cal N}=2$ AdS$_4$ background, as shown in appendix \ref{app:osp2bar4}. This confirms our AdS interpretation and provides the appropriate identifications of the global AdS energy with a particular linear combination of the Hamiltonian and the $R$-charge generator, namely the global AdS $R$-frame identified in section \ref{sec:energydef}.

We note that the chiral multiplet part of the massive quiver matrix mechanics Lagrangian in equation \eqref{eq:totalLag} has been given before, as part of a systematic construction of supersymmetric quantum mechanics models with $\mathfrak{su}(2|1)$ supersymmetry \cite{Ivanov:2014ima,Ivanov:2013ova}. 
This part can also be obtained by dimension reduction of the general four-dimensional ${\cal N}=1$ chiral multiplet Lagrangian of \cite{Festuccia:2011ws} on ${\mathbb R} \times S^3$, and it has been obtained this way in \cite{Assel:2015nca}, for the purpose of computing Casimir energies in conformal field theories on curved spaces.
The dimensional reduction of the four-dimensional vector multiplet is also known from \cite{Kim:2003rza}, but
we explain how to create a general gauged $\mathfrak{su}(2|1)$ quantum mechanics with coupled vector and chiral multiplets and an arbitrary superpotential.
The massive quiver Lagrangian in equation \eqref{eq:totalLag} is a special case of these models as is the BMN matrix model (see \ref{sec:BMN}). 
We explain how to perform the dimensional reduction in section \ref{sec:reduction} and give additional details in appendix \ref{app:reductiondetails}.
In section \ref{sec:reduccomparison}, we give a more detailed comparison of our models and those given in the works \cite{Ivanov:2014ima,Ivanov:2013ova}.


\section{General Lagrangian and supersymmetry}

\label{sec:genlagandsusy}


In this section we give the core results of the paper, the general massive quiver matrix mechanics Lagrangian and its supersymmetries. It represents a general deformation of the quiver models of \cite{Denef:2002ru} (see especially appendix C) preserving $\mathrm{SO}(3)$ rotation symmetry and ${\cal N}=4$ supersymmetry. For simplicity, we also restrict to a flat target-space metric for both the vector and chiral multiplets. We conjecture that this is the most general Lagrangian having these properties.

The field content remains the same as in \cite{Denef:2002ru}. It is encoded in a quiver, with nodes $v\in V$, directed edges (arrows) $a\in A$, and dimension vector $N=(N_v)_{v\in V}$.
To each node is assigned a vector, or linear, multiplet $(A_v,~X_v^i,~{\lambda_v}_\alpha,~D_v)$ with $i=1,~2,~3$.
The field $A_v$ is the gauge field for the group $\mathrm{U}(N_v)$.
The fields $X_v^i$, ${\lambda_v}_\alpha$, $D_v$ transform in the adjoint of $\mathrm{U}(N_v)$. To each edge $a:v\rightarrow w$ is assigned a chiral multiplet $(\phi^a,~\psi^a_\alpha,~F^a)$ transforming in the bifundmental $(N_w, \bar{N}_v)$ of $\mathrm{U}(N_w)\times \mathrm{U}(N_v)$. In a string theory context the nodes $v$ can be thought of as labeling different ``parton'' D-branes wrapped on internal cycles with multiplicity $N_v$, and arrows $a:v\to w$ as labeling light open string modes polarized in the internal dimensions, connecting the parton branes. 

The Lagrangian depends on a number of parameters that are already present in the flat-space quiver models. For each node $v$, there is an inertial mass parameter $m_v$ determining the kinetic terms for the vector multiplet fields, and a Fayet-Iliopoulos (FI) parameter $\theta_v$ setting the D-term potential for the scalars $\phi^a$ connected to the node $v$. The quiver model may also have a superpotential, given by an arbitrary gauge-invariant holomorphic function $W(\phi^a)$ of the $\phi^a$. 

Before imposing any supersymmetry, a general SO(3)-symmetric and gauge invariant mass deformation of the vector multiplets consists of adding harmonic potential terms of the form $\frac{\mu_v}{2} {\rm Tr} (X^i_v)^2$ to the Lagrangian, and similarly for the fermions. Requiring ${\cal N}=4$ supersymmetry to be preserved dictates the inclusion of additional terms for the vector and chiral multiplets, and reduces the a priori arbitrary deformation parameters $\mu_v$ to functions of a single deformation parameter $\Omega$, namely $\mu_v = m_v \, \Omega^2$. In the AdS interpretation discussed in section \ref{sec:examples}, we identify $\Omega = 1/\ell_{\rm AdS}$. 
In appendix \ref{app:susy}, we provide more details on how supersymmetry fixes the form of the mass deformation. In section \ref{sec:reduction}, we explain how it can be obtained from dimensional reduction of ${\cal N}=1$ quiver gauge theories on ${\mathbb R} \times S^3$.

\subsection{Lagrangian} 
\label{sec:Lag}

The Lagrangian of massive quiver matrix mechanics with deformation parameter $\Omega$ is given by
\begin{equation}
\label{eq:totalLag}
\Lcal_\Omega = \Lcal^0 + \Lcal'_\Omega \ ,
\end{equation}
where $\Lcal^0$ is the original, undeformed, flat-space quiver Lagrangian,
identical to the Lagrangian in appendix C of \cite{Denef:2002ru}, and $\Lcal'_\Omega$ is the mass deformation. 
We give our conventions and definitions for this Lagrangian in detail in 
appendix \ref{app:Def}.
\begin{align}
\label{eq:Lag1}
\Lcal^0 &= \Lcal_V^0 + \Lcal_{FI}^0 + \Lcal_C^0 + \Lcal_I^0 + \Lcal_W^0 \\[12pt]
\Lcal_V^0 &= \sum_v  m_v \Tr\lp \tfrac{1}{2} (\Dcal_tX_v^i)^2 + \tfrac{1}{2} D_v^2 + \tfrac{1}{4}[X^i_v, X^j_v]^2  + \tfrac{i}{2}(\lambda^\dag_v \Dcal_t  \lambda_v - (\Dcal_t  \lambda^\dag_v) \lambda_v ) - \lambda^\dag_v\sigma^i[X^i_v,\lambda_v]\rp \nonumber \\
\Lcal_{FI}^0 &= - \sum_v \theta_v \Tr D_v  \nonumber \\
\Lcal_C^0 &= \sum_a \Tr\lp |\Dcal_t \phi^a|^2 + |F^a|^2 + \tfrac{i}{2} (\psi^{a\dag} \Dcal_t \psi^a  - (\Dcal_t \psi^{a\dag}) \psi^a  ) \rp \nonumber \\
\Lcal_W^0 &= \sum_a \Tr\lp \frac{\partial W}{\partial \phi^a} F^a + \text{h.c.}\rp + \sum_{ab}\Tr\lp \frac{1}{2} \frac{\partial^2 W}{\partial\phi^a \partial\phi^b} \psi^b\epsilon\psi^a + \text{h.c.}\rp \nonumber \\
\Lcal_I^0 &= -\sum_{a:v\rightarrow w} \Tr \bigl( (\phi^{a\dag}X_w^i - X_v^i\phi^{a\dag})(X_w^i\phi^a - \phi^aX_v^i) - \phi^{a\dag}(D_w\phi^a - \phi^aD_v)  \nonumber \\
&\qquad  + \psi^{a\dag}\sigma^i(X^i_w\psi^a - \psi^aX^i_v) %
+ i\sqrt{2}\lp (\phi^{a\dag}\lambda_w - \lambda_v\phi^{a\dag})\epsilon\psi^a %
- \psi^{a\dag}\epsilon(\lambda_w^\dag\phi^a - \phi^a\lambda^\dag_v)\rp \bigr) \nonumber \\[12pt]
\label{eq:Lprm}
\Lcal'_\Omega  &= \Lcal_V' + \Lcal_{FI}' + \Lcal_C' \\[12pt]
\Lcal_V' &= -\sum_v m_v  \Tr\lp \tfrac{1}{2} \, \Omega^2  (X_v^i)^2 + \tfrac{3}{2} \, \Omega \, \lambda^\dag_v\lambda_v + i \, \Omega \, \epsilon_{ijk}X_v^iX_v^jX^k_v \rp \nonumber \\
\Lcal_{FI}' &= \sum_v \theta_v \, \Omega \, \Tr A_v  \nonumber \\
\Lcal_C' &= \sum_a \Tr\lp \tfrac{i}{2} \, \Omega \bigl( (\Dcal_t \phi^{a \dagger})  \phi^a - \phi^{a \dagger} \Dcal_t \phi^a \bigr)  - \tfrac{1}{2} \, \Omega\,\psi^{a\dag}\psi^a \rp \, . \nonumber
\end{align}
The covariant derivatives are given by 
\begin{align*}
\Dcal_t X_v^i &= \partial_t X_v^i - i[A_v, X_v^i], \\
\Dcal_t\lambda_v &= \partial\lambda_v - i[A_v, \lambda_v], \quad \Dcal_t\lambda_v^\dag = \partial\lambda_v^\dag - i[\lambda_v^\dag, A_v], \\
\Dcal_t \phi^a &= \partial_t \phi^a - iA_w \phi^a + i\phi^a A_v, \quad \Dcal_t \phi^{a\dag} = \partial_t \phi^{a\dag} + i\phi^{a\dag}A_w - iA_v\phi^{a\dag} \\
\Dcal_t\psi^a &= \partial_t \psi^a - iA_w\psi^a + i\psi^a A_v, %
\quad \Dcal_t\psi^{a\dag} = \partial_t \psi^{a\dag} + i\psi^{a\dag} A_w - iA_v \psi^{a\dag},
\end{align*}
with the arrow $a:v\to w$.


\subsection{Supersymmetry transformations}
\label{sec:susytransf}

The action is supersymmetric with respect to the transformations
\begin{calc}
\delta A_v &= i\lambda^\dag_v\xi - i\xi^\dag\lambda_v \\
\delta X_v^i &= i\lambda^\dag_v\sigma^i\xi - i\xi^\dag\sigma^i\lambda_v \\
\delta \lambda_v &= (\Dcal_tX^i_v)\sigma^i\xi  +\tfrac{1}{2}\epsilon^{ijk}[X^i_v,X^j_v]\sigma^k\xi + iD_v\xi - i \, \Omega \, X^i_v\sigma^i\xi \\
\delta D_v &= -(\Dcal_t\lambda_v^\dag)\xi  - i[X_v^i,\lambda^\dag_v]\sigma^i\xi - \xi^\dag(\Dcal_t\lambda_v)  - i\xi^\dag\sigma^i[X^i_v,\lambda_v] \,
+ \tfrac{3i}{2} \, \Omega \, \lambda_v^\dagger \xi 
- \tfrac{3i}{2} \, \Omega \, \xi^\dagger \lambda_v 
\\
\delta \phi^a &= -\sqrt{2} \,\xi\epsilon\psi^a \\
\delta \psi^a &= -i\sqrt{2} \,\xi^\dag\epsilon(\Dcal_t\phi^a) - \sqrt{2} \,\sigma^i(\xi^\dag\epsilon)(X^i_w\phi^a - \phi^aX_v^i) + \sqrt{2} \,\xi F^a  \\
\delta F^a &= -i\sqrt{2} \,\xi^\dag(\Dcal_t\psi^a) + \sqrt{2} \,\xi^\dag\sigma^i(X^i_w\psi^a - \psi^aX_v^i) - 2i\xi^\dag\epsilon(\lambda^\dag_w\phi^a - \phi^a\lambda^\dag_v) + \tfrac{\sqrt{2}}{2} \,  \Omega \, \xi^\dag\psi^a \\
\label{eq:susyxi}
\xi(t) & = e^{- \tfrac{i}{2} \Omega \, t} \, \xi_0 \, .
\end{calc}


\subsection{$R$-symmetry and $R$-frames}
\label{sec:R-symm}

A notable difference with \cite{Denef:2002ru} is that the supersymmetry parameter $\xi$ in equation \eqref{eq:susyxi} is time dependent. 
A given massive quiver matrix model Lagrangian may or may not possess $R$-symmetry.
If its $R$-symmetry group contains a $\mathrm{U}(1)$ subgroup then $\xi$ can be made time-independent by a field redefinition, which we give below.
Without loss of generality, we can take the subgroup $\mathrm{U}(1)_R$ to act on the fields $Y$ as
\begin{equation} 
\label{eq:Rsym}
Y \to e^{i Q_Y \vartheta} \, Y \, , \qquad Y^\dagger \to e^{-i Q_Y \vartheta} \, Y^\dagger \, ,
\end{equation} 
with the charges $Q_Y$ given in table \ref{tab:rcharges}.
\begin{table}[ht]
\centering
\begin{tabular}{|c|cccccccc|}
\hline 
 $Y$ & $\xi$ & $A_v$ & $X_v$ & $D_v$ & $\lambda_v$ & $\phi^a$ & $\psi^a$ & $F^a$  \\ 
\hline
$Q_Y$ & 1 & 0 & 0 & 0 & 1 & $q_a$ & $q_a-1$ & $q_a -2$ \\
\hline
\end{tabular}
\caption{The $R$-charge $Q_Y$ of each field $Y$ appearing in the quiver model.}
\label{tab:rcharges}
\end{table}
For the Lagrangian to be invariant under equation \eqref{eq:Rsym}, the superpotential must satisfy a homogeneity condition so that it has overall $R$-charge $Q_W = 2$:
\begin{equation}
\label{eq:Wscaling}
W(\lambda^{q_a} \phi^a) = \lambda^2 \, W(\phi^a) \, ,
\end{equation}
or equivalently $\sum_a q_a \phi^a \partial_a W = 2 W$. If we redefine the fields, including $\xi$, by a time-dependent $R$-symmetry,
\begin{equation}
\label{eq:Rtransf} 
Y_{\rm old} = e^{-i \, Q_Y \frac{\rho}{2} \Omega \, t} \, Y_{\rm new} \, ,
\end{equation}
for some real parameter $\rho$, the only change to the Lagrangian and supersymmetry transformations is that all covariant derivatives are effectively shifted by constant $\mathrm{U}(1)_R$ background connections, because ${\cal D}_t Y_{\rm old} = e^{-i \, Q_Y \frac{\rho}{2} \Omega \, t} \lp \Dcal_t - i \, Q_Y \tfrac{\rho}{2} \Omega \rp Y_{\rm new}$.
Thus, all of the above expressions for the Lagrangian and the supersymmetry transformations remain unchanged provided we replace  
\begin{equation}
\label{eq:Dtransf}
\Dcal_t Y \to \tilde{\Dcal}_t Y \equiv \lp \Dcal_t - i \, Q_Y \tfrac{\rho}{2}\Omega \rp Y \, .
\end{equation}
In particular, picking $\rho = 1$ renders the supersymmetry parameter time independent. As we will see below, in an example in section \ref{sec:energydef}, and establish in general in appendix \ref{app:osp2bar4}, the appropriate value to identify the quiver Hamiltonian with the AdS global energy in ${\cal N}=2$ compactifications is
\begin{equation}
 \rho_{\rm AdS} = 2 \, .
\end{equation}



\section{Examples and physical interpretations}
\label{sec:examples}


The massive quiver matrix mechanics Lagrangian presented in section \ref{sec:Lag}  arises naturally in a number of string theoretic contexts. One is already well known; as we show in section \ref{sec:BMN}, the BMN matrix model \cite{Berenstein:2002jq}, describing the dynamics of D0-branes in a plane wave background, arises as a special case. In this paper we will, however, focus on a different interpretation; the nonrelativistic limit of massive particles living in global AdS$_4$, obtained by wrapping D$p$-branes on internal $p$-cycles in a flux compactification of string theory. Essentially, an AdS version of \cite{Denef:2002ru}. 

In this section we will substantiate and explore this interpretation by studying various simple examples.  
The interpretation of the harmonic potentials as the gravitational potential of AdS, and the corresponding identification $\Omega=1/\ell_{\rm AdS}$, is explained in section \ref{sec:interpads}. More interestingly, the model exhibits a number of smoking-gun phenomena usually associated with the presence of background fluxes, including the Myers effect (section \ref{sec:NAMyers}), magnetic trapping in the internal D-brane moduli spaces (section \ref{sec:internalmagnetic}) and flux-induced background charges on wrapped branes, forcing a nonzero number of confining strings to end on the branes (section \ref{sec:two_nodes}).
Finally, we will also use the examples to clarify the role of $R$-symmetry and subtleties associated with the existence of different $R$-frames (sections \ref{sec:energydef} and \ref{sec:nonzeroW}).

\subsection{One node, no arrows}


\begin{figure}[h!]
\centering
\includegraphics[width=10mm]{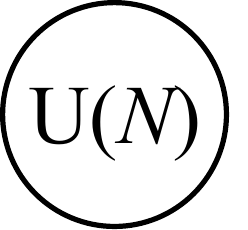}
\caption{A one-node quiver representing a matrix model with a single vector multiplet, with gauge group $\mathrm{U}(N)$.}
\label{fig:quiv0}
\end{figure}


\noindent The one-node quiver without arrows and dimension vector $(N)$, see figure \ref{fig:quiv0}, has just one vector multiplet with gauge group $\mathrm{U}(N)$ and no chiral multiplets. It describes $N$ identical D-particles in a 3+1-dimensional spacetime. 

\subsubsection{Lagrangian} 

The Lagrangian is
\begin{align}
\label{eq:ND0case}
\Lcal &= \Lcal_V^0 + \Lcal_{FI}^0 + \Lcal_V' + \Lcal_{FI}' \\
\Lcal_V^0 &= m \Tr\lp \tfrac{1}{2} (\Dcal_tX^i)^2 + \tfrac{1}{2} D^2 + \tfrac{1}{4}[X^i, X^j]^2  + \tfrac{i}{2}(\lambda^\dag \Dcal_t  \lambda - \Dcal_t  \lambda^\dag \lambda ) - \lambda^\dag\sigma^i[X^i,\lambda] \rp \nonumber \\
\Lcal_{FI}^0 &= - \theta \, \Tr D  \nonumber \\
\Lcal_V' &= - m  \Tr\lp \tfrac{1}{2} \, \Omega^2  (X^i)^2 + \tfrac{3}{2} \, \Omega \, \lambda^\dag \lambda + i \, \Omega \, \epsilon_{ijk}X^iX^jX^k \rp \nonumber \\
\Lcal_{FI}' &= \theta \, \Omega \, \Tr A  \ . \nonumber 
\end{align}
The covariant derivative 
acts in the adjoint of $\mathrm{U}(N)$, $\Dcal_t X = \partial_t X - i [A,X]$, so the diagonal $\mathrm{U}(1)$ part of $A$ does not couple to anything in the Lagrangian except to the constant $\theta \, \Omega$. Varying ${\Lcal}$ with respect to $\delta A = \delta a \, {\mathds 1}$ thus leads to the constraint $\Omega \, \theta = 0$, hence for $\Omega \neq 0$ we get the consistency condition $\theta=0$.
We will give an interpretation of this later in section \ref{sec:thetazero}, after we have studied examples of quivers in which $\theta$ can be nonzero. 


\subsubsection{Interpretation as a particle in AdS}
\label{sec:interpads}

For $N=1$ the Lagrangian \eqref{eq:ND0case} describes a nonrelativistic superparticle in an isotropic 3d harmonic oscillator potential with frequency $\Omega$:
\begin{equation}
\label{eq:LLLLL}
\Lcal = m \lp \tfrac{1}{2} (\dot{X}^i)^2  - \tfrac{1}{2} \, \Omega^2  (X^i)^2 + \tfrac{i}{2}(\bar \lambda  \dot \lambda - \dot {\bar \lambda} \lambda )  - \tfrac{3}{2} \, \Omega \, \bar\lambda \lambda  \rp \, .
\end{equation}
We can interpret this as a massive nonrelativistic superparticle near the bottom (origin) of global AdS$_4$.
Using isotropic coordinates $t, x_1, x_2, x_3$, with $x^2 \equiv x^2_1 + x^2_2 + x^2_3$,
an AdS$_4$ space of radius $\ell$ has a metric 
\begin{equation}
\label{eq:ads_met}
	ds^2 = \frac{-\left(1 + \frac{x^2}{4\ell^2}\right)^2 dt^2 + dx_1^2+ dx_2^2 + dx_3^2}{\left(1 - \frac{x^2}{4\ell^2}\right)^2} \ .
\end{equation}
The action of a massive particle in this background is 
\begin{equation}
S = -m\int dt \, \sqrt{\left(1+\tfrac{x^2}{4\ell^2}\right)^2\left(1-\tfrac{x^2}{4\ell^2}\right)^{-2} 
- \left(1-\tfrac{x^2}{4\ell^2}\right)^{-2} \dot{x}^2} \, .
\end{equation}
If $x \ll \ell$ and $\dot{x} \ll 1$ at any given time, the classical motion of such a particle remains nonrelativistic at all times.
In this regime, the action becomes
\begin{equation}
\label{eq:SAdS}
S \approx m \int dt \, \bigl( \tfrac{1}{2} \dot{x}^2 - \tfrac{1}{2} \Omega^2 x^2 - c^2 \bigr) \, , \quad \Omega = \frac{c}{\ell} \, .
\end{equation}
We have explicitly reinstated the speed of light $c$, to emphasize that $\Omega$ is most naturally viewed in this nonrelativistic setting as the universal oscillation frequency of a particle in the AdS gravitational potential well. 
The action obtained here reproduces the bosonic part of equation \eqref{eq:LLLLL}, confirming the interpretation of $\Omega$ announced earlier.


\subsubsection{Fermionic excitations, definitions of energy and $R$-frames}
\label{sec:energydef}

To extract the physics of the fermionic degrees of freedom, we have to quantize the system. 
This is straightforward in the case at hand, and done in general in appendix \ref{app:algebra}. 
We introduce canonical bosonic and fermionic annihilation operators,
$a_i \equiv \frac{\sqrt{m \Omega}}{\sqrt{2}} \, X^i + i \frac{1}{\sqrt{2 \, m \Omega}} P_i$ and $b_{\alpha} \equiv \sqrt{m} \, \lambda_\alpha$ respectively, together with their conjugate creation operators, which satisfy the algebra
\begin{equation}
 [a_i,a_j^\dagger] =  \, \delta_{ij} \, , \qquad \{ b_{\alpha},b_{\beta}^\dagger \} =  \, \delta_{\alpha \beta} \, ,
\end{equation}
where $i = 1,2,3$ and $\alpha = 1,2$ is a spinor index.
Then the Hamiltonian derived from \eqref{eq:LLLLL} and the generator of the 
$\mathrm{U}(1)_R$  symmetry $b_\alpha \to e^{i \vartheta} b_\alpha$ can be expressed as
\begin{equation}
\label{eq:HHHHHHH}
\hat{H} = \Omega \bigl( a^\dagger a + \tfrac{3}{2} \, b^\dagger b \bigr) \, , \qquad \hat{R} = - b^\dagger b \, .
\end{equation}
The normal ordering constants of the bosonic and fermionic parts cancel each other in $\hat H$, so the ground state energy is zero. The Hilbert space is spanned by the Fock eigenbasis built by acting with the $a_i^\dagger$ and $b_\alpha^\dagger$ on the ground state. 
Since $b^\dagger$ transforms in the spin $\frac{1}{2}$ representation of SO(3) and has $R$-charge $-1$, the fermionic sector consists of a spin zero state with $R$-charge 0, one spin $\frac{1}{2}$ doublet with $R$-charge $-1$ and another spin $0$ state with $R$-charge $-2$. Browsing through the tables of ${\rm OSp(2|4)}$ multiplets in, e.g., the appendix of \cite{Merlatti:2000ed}, one sees that our superparticle, together with its antiparticle, produces exactly the spin and $R$-charge content of a massive ${\cal N}=2$ hypermultiplet in AdS$_4$ (table 7).
However, at first sight, there seems to be a mismatch with the energy spectrum.
For our superparticle, the energy gaps for bosonic and fermionic excitations are, respectively,
\begin{equation}
\Delta E_B = \Omega, \quad \Delta E_F = \tfrac{3}{2} \Omega \ .
\end{equation}
Under the identification \eqref{eq:SAdS}, $\Delta E_B$ is exactly the scalar normal-mode energy gap in AdS, with respect to standard AdS global time.
However, according to table 7 in the appendix of \cite{Merlatti:2000ed}, 
we should then find $\Delta E_F = \frac{1}{2} \Omega $ instead of $\frac{3}{2} \Omega $. 

The solution to this puzzle is that our identification of $\hat H$ as ``the'' energy is ambiguous, since we can always shift to a different $R$-frame. 
In the Heisenberg picture, the transformation in equation \eqref{eq:Rtransf} becomes
\begin{equation}
Y_{\text{new}} = e^{i\frac{\rho}{2} R \Omega t}Y_{\text{old}} e^{-i\frac{\rho}{2}R \Omega t} \ .
\end{equation}
The equation of motion for $Y_{\text{new}}$ is then
\begin{equation}
-i\frac{d}{dt}Y_{\text{new}} = [\hat{H} + \tfrac{\rho}{2} \Omega \hat{R}, Y_{\text{new}}] \ .
\end{equation}
Thus, the operator $\hat{H}_\rho = H + \frac{\rho}{2}\Omega \hat{R}$ defines time translation in a different $R$-frame and is the energy operator in that frame.
The Hamiltonian $\hat{H}$ becomes
\begin{equation}
\label{eq:Hshiftt}
\hat{H}_{\rho} = \hat{H} + \tfrac{\rho}{2} \, \Omega \, \hat{R} %
= \Omega \lp a^\dagger a + \tfrac{3-\rho}{2} \, b^\dagger b \rp, 
\end{equation}
and the fermionic energy gap becomes
\beq
\Delta E_{F,\rho} = \tfrac{3-\rho}{2} \, \Omega \, .
\eeq
From the particle mechanics point of view, these different notions of energy $\hat{H}_\rho$ are all equally valid and just amount to a relabeling of conserved charges without affecting the bosonic part of the symmetry algebra (see appendix \ref{app:algebra}). 
However, only one choice of $\rho$ corresponds to the bulk gravitational AdS energy defined with respect to global AdS time, used in \cite{Merlatti:2000ed}. 
Matching energy gaps $\Delta E_F$, we see this is the case for $\rho=2$, so we are led to identify
\begin{equation}
 \hat H_{\rm AdS} = \hat H|_{\rho =2} = \hat{H}_0 + \Omega \, \hat R = \Omega \lp a^\dagger a + \tfrac{1}{2} \, b^\dagger b \rp \, .
\end{equation}
The relation $H_{\rm AdS} = H_0 + \Omega \, R$ turns out to hold in general, as we show in appendix \ref{app:osp2bar4}.
We reiterate that $\rho$ does not label truly different models the way, for example, different values of $\Omega$ do. 
Rather, it labels different reference frames, related to each other by the constant $R$-rotations given in equation \eqref{eq:Rtransf}.
This is similar to ordinary particle mechanics in a cylindrically symmetric potential described in different rotating frames.
The rotation shifts the Hamiltonian of the system by the corresponding angular momentum.
Physically, observers in different frames will perceive different centrifugal and Coriolis forces, but water remains water, and wine remains wine. 
One particular frame may be singled out if the system is part of a larger context, like a lab or distant stars.
This is the role played by AdS$_4$, which singles out the $\rho=2$ frame.


\subsubsection{Nonabelian model and Myers effect}
\label{sec:NAMyers}

When $N>1$, the Lagrangian in equation \eqref{eq:ND0case} contains interesting nonabelian interactions. The commutator-squared term already appeared in the flat-space quiver matrix models. 
It also universally appears in the worldvolume theories of $N$ coincident D-branes. Since we are interpreting $X$ as a position coordinate of a D-particle in AdS, we want to assign it the dimension of length. Similarly, we want $t$ to be time and $m$ to be a mass. 
However, then the commutator-squared term actually has the wrong dimension compared to, say, the kinetic term. This can be traced to the way the original flat-space quiver matrix model was obtained in \cite{Denef:2002ru}. It was essentially by dimensional reduction from a four dimensional gauge theory, in which the dimension of $X^i=A_i$ is naturally inverse length and $m=\frac{1}{g^2} {\rm Vol}_3$ is length cubed. To get the dimensions we want, we can rescale the fields and parameters by powers of a reference length $l$. This will produce explicit powers of $l$ at various places, including a factor $l^{-2}$ in front of the commutator term. Equivalently and more conveniently, we can pick units such that besides $c \equiv 1$ and $\hbar \equiv 1$, we also have $l \equiv 1$. 
Matching the kinetic and commutator terms with the standard expressions for D-brane worldvolume theories in the literature \cite{Myers:1999ps}, we see the appropriate length scale is the string length, or more precisely $l = \sqrt{2 \pi \, \alpha'}$. 
Thus, throughout we will be working in units with
\begin{equation}
\label{eq:units}
 2 \pi \alpha' \equiv 1 \, .
\end{equation}

Perhaps the most interesting new element in the massive quiver Lagrangian in equation \eqref{eq:ND0case} is the presence of a Myers term in ${\Lcal}'_V$,
\begin{equation}
\label{eq:Lmyersus}
\Lcal_{V}' = - i \, m \, \Omega \, \epsilon_{ijk} \, \Tr(X^iX^jX^k) + \cdots \, .
\end{equation}
Such terms arise in D-brane physics from the presence of background flux with legs in the directions transversal to the brane \cite{Myers:1999ps}. They allow multiple coincident branes to polarize into higher-dimensional dielectric branes. 
Specifically, for $N$ coincident D0-branes in a background R-R 4-form flux $F$, 
in the units of equation \eqref{eq:units} and the conventions of \cite{Myers:1999ps}, this term in the 
D0-brane Lagrangian is 
\begin{equation}
 {\Lcal}_{\rm Myers} =   i \, m \, \tfrac{1}{3} F_{tijk} \, \Tr(X^i X^j X^k) \, ,
\end{equation}
where $m=T_0 = \frac{1}{g_s \sqrt{\alpha'}}$ is the mass of the D0-brane. 
Comparing to equation \eqref{eq:Lmyersus}, we see the deformed quiver has a Myers coupling to an effective flux background
\begin{equation}
 F_{tijk} = - 3 \, \Omega \, \epsilon_{ijk} \, .
\end{equation}
Recalling that earlier we identified $\Omega = \frac{1}{\ell_{\rm AdS}}$ in equation \eqref{eq:SAdS}, this precisely agrees, including the coefficient,\footnote{The sign depends on a number of conventions for orientations and charges, which we did not try to sort out.}
with the 4-form flux supporting general Freund-Rubin compactifications of 11-dimensional supergravity \cite{Duff:1986hr}, %
such as, for example, the 11d AdS$_4 \times S^7/{\mathbb Z}_k$ or 10d AdS$_4 \times {\mathbb CP}^3$ compactifications \cite{Watamura:1983hj, Nilsson:1984bj} dual to ABJM theory \cite{Aharony:2008ug}.

The most striking consequence of the presence of such cubic terms is the Myers effect \cite{Myers:1999ps}; $N$ coincident D0-branes polarizing into a stable fuzzy sphere configuration, which, at large $N$, approximates a spherical D2-brane with $N$ units of worldvolume flux. 
In the case at hand, unlike the flat space case studied in \cite{Myers:1999ps}, the fuzzy sphere is supersymmetric. 
This can be seen from the supersymmetry variations of the gaugino given in section \ref{sec:susytransf}: 
\begin{equation}
 \delta \lambda = (\Dcal_tX^i)\sigma^i\xi  +\tfrac{1}{2}\epsilon^{ijk}[X^i,X^j]\sigma^k\xi + i D\xi - i \, \Omega \, X^i\sigma^i\xi \, .
\end{equation}
The configuration is supersymmeric provided $\delta \lambda =0$, that is to say, provided $X$ is time independent, $D=0$ (trivially the case here, by the equations of motion for $D$), and
\begin{equation}
 [X^i,X^j] = i \, \Omega \, \epsilon^{ijk} \, X^k \, .
\end{equation}
A maximally nonabelian solution to this equation is given by the $N$-dimensional representation of $\mathrm{SU}(2)$, yielding, at large $N$, a fuzzy sphere of radius \cite{Myers:1999ps}
\begin{equation}
 R_{\rm fuzzy} = \tfrac{1}{2} \Omega N  = \frac{\pi \alpha'}{\ell_{\rm AdS}} \, N \, .
\end{equation} 
Notice that R-R 4-form flux is sourced by D2-branes, or its uplift to M-theory by M2-branes. Thus, the fuzzy sphere can be interpreted as a membrane which has separated itself from the stack of membranes generating the AdS$_4$ Freund-Rubin compactification,  supported by its worldvolume flux. Such membranes in $\text{AdS}_4$, known as dual giant gravitons, were studied in, e.g., \cite{Nishioka:2008ib}.


\subsection{One node, $\kappa$ arrows}

\begin{figure}[h!]
\centering
\includegraphics[width=40mm]{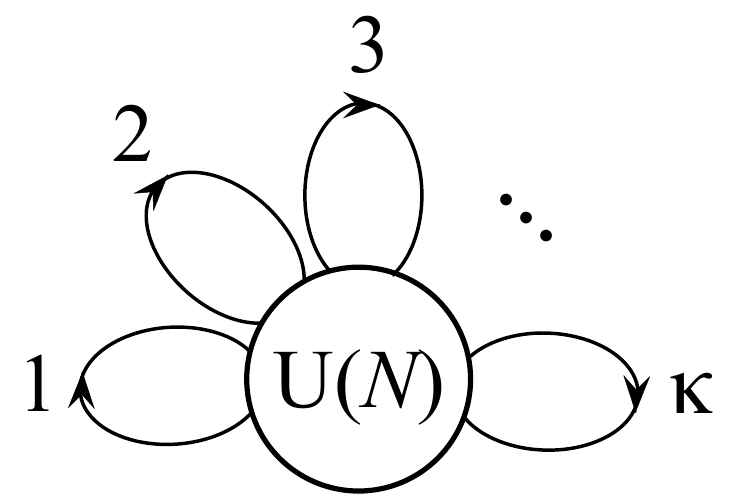}
\caption{A one-node quiver representing a matrix model with a single vector multiplet, with gauge group $\mathrm{U}(N)$, and $\kappa$ chiral multiplets.}
\label{fig:quiv1}
\end{figure}


\subsubsection{Lagrangian and consistent truncation}

To focus specifically on the new features brought by adding chiral multiplets and to simplify things as much as possible, let us consider a single-node quiver with $\kappa$ arrows from the node to itself, see figure \ref{fig:quiv1}, and let us put 
\begin{equation}
\label{eq:truncation}
 X = 0 \, , \qquad \lambda= 0 \, , \qquad \psi = 0 \, .
\end{equation}
Classically, this is a consistent truncation since there are no terms in the Lagrangian involving one of these fields coupled linearly to the other fields or to constants. 
Note that it is not possible to set $A$ or $D$ to zero in this way, since they do couple linearly to, for example, $\phi$. Thus, the truncated theory consists of a $\mathrm{U}(N)$ gauge field $A$ and auxiliary $D$ together with adjoint scalars $\phi^a$, $a = 1,\ldots,\kappa$. The covariant derivative 
acts as ${\cal D}_t \phi = \partial_t \phi - i[A,\phi]$, so the diagonal $\mathrm{U}(1)$ part does not couple to anything in the Lagrangian except to the constant $\theta \Omega$. Thus, as in the pure vector case discussed earlier, for $\Omega \neq 0$ we get the consistency condition $\theta=0$. Integrating out the auxiliary fields then puts
\begin{equation}
\label{eq:FDvals}
F^a=\partial_a W \, , \qquad D = \tfrac{1}{m} \mbox{$\sum_a$} [\phi^{a\dagger},\phi^{a}] \, ,
\end{equation}  
and the Lagrangian \eqref{eq:totalLag} becomes, 
\begin{align}
\label{eq:trunclag}
\Lcal &= \Lcal^0 + \Lcal' \\
\Lcal^0 &= \Tr\lp \, |\Dcal_t \phi^a|^2  -  |\partial_a W|^2 - \tfrac{1}{2m} \bigl([\phi^{a\dagger},\phi^a]\bigr)^{2} \, \rp  \nonumber  \\
\Lcal' &=  \, \tfrac{i}{2}  \Omega \Tr\bigl( \Dcal_t \phi^{a\dagger}  \phi^a - \phi^{a\dagger} \Dcal_t \phi^a \bigr) \, . \nonumber
\end{align}
The Gauss's law constraint obtained by varying $A$ is
\begin{equation}
i[\Dcal_t \phi^{a\dag},\phi^a] - i [\phi^{a\dag},\Dcal_t \phi^a] + \Omega \, [\phi^{a\dagger},\phi^a]  = 0 \, .
\end{equation}  
Having derived the Gauss's law constraint, we can pick a gauge $A \equiv 0$.


\subsubsection{Single particle: internal space magnetic and Coriolis forces}

\label{sec:internalmagnetic}

Let us first have a look at the $N=1$ case with $W=0$. Then the Lagrangian collapses to
\begin{equation}
\label{eq:Lphisimple}
\Lcal = |\dot \phi^a|^2  + \tfrac{i}{2} \Omega \bigl( \dot{\bar \phi}^a \, \phi^a - \bar\phi^a \, \dot\phi^a \bigr) \, .
\end{equation}
This describes a charged particle with complex position coordinates $\phi^a$ moving on a flat K\"ahler manifold ${\mathbb C}^\kappa$, 
in a background magnetic field $F$ proportional to the K\"ahler form: 
\begin{equation}
\label{eq:Fmagn}
F = \Omega \, \tfrac{i}{2} d\phi^a \wedge d\bar\phi^a \, .
\end{equation}
If the D-particle under consideration is a D0-brane in ten-dimensional IIA string theory, the $\phi^a$ are complex coordinates on the physical compactification space, and $F$ is then to be identified with an R-R 2-form flux. To more accurately describe such a situation, for example a D0-brane on the AdS$_4 \times {\mathbb CP}^3$ geometry dual to ABJM theory \cite{Aharony:2008ug}, we should consider a generalization of our quiver models to 
general curved-space K\"ahler potentials. This can be done, but we won't do it here. Sticking with our simple model, the classical solution to the equations of motion 
is a cyclotron motion of frequency $\Omega$ with arbitrary center $C$ and amplitude $A$:
\begin{equation}
\label{eq:phimotion0}
\phi^a(t) = C^a + A^a \, e^{-i \, \Omega \, t} \, .
\end{equation}
In particular, in the limit $|A|\to 0$ the effect of target-space curvature should become negligible, 
so we expect that in this limit the frequency of motion should be independent of the curvature. Comparing to the case ${\rm AdS}_4 \times {\mathbb CP}^3$ mentioned above and reviewed in detail below in section \ref{sec:ABJM}, it can be checked that the cyclotron frequency (with respect to global AdS time) of a D0 moving in the 2-form flux-carrying ${\mathbb CP}^3$ also equals $\Omega$.

Before jumping to conclusions, we should recall, however, that this is the motion in the 
original, $\rho=0$, $R$-frame, whereas the natural, ``inertial'' $R$-frame for D-particles in 
${\cal N}=2$ global AdS$_4$ corresponds to $\rho=2$. 
This was illustrated in section \ref{sec:energydef} and shown in general in appendix \ref{app:algebra}. 
Thus, for proper comparisons we should transform the motion in equation \eqref{eq:phimotion0} 
by a field redefinition, as in equation \eqref{eq:Rtransf}, with $\rho = 2$, yielding 
\begin{align}
\label{eq:chimotion}
\chi^{a}(t)  \equiv \phi_{\rm new}^a(t) &= e^{i \, \Omega \, q_a t} \, \phi^a_{\rm old}(t) \\
  &= C^a \, e^{i \, q_a \, \Omega \,  t} +  A^a \, e^{i (q_a-1) \Omega \,t} \, ,
\end{align} 
where $q_a$ is the $R$-charge of $\phi^a$. 
The Lagrangian in equation \eqref{eq:Lphisimple} becomes
\begin{equation}
\label{eq:Lphisimple2}
 {\Lcal} = |\dot \chi^a|^2  + \tfrac{i \, (1-2q_a)}{2}  \, \Omega  \bigl(\dot{\bar \chi}^a \chi^a  - \bar\chi^a \dot\chi^a \bigr)  + q_a(q_a-1) \Omega^2 \, |\chi^a|^2 \, .
\end{equation}
The new terms in the Lagrangian can be thought of as centrifugal (or electric) and Coriolis (or magnetic) forces due to the rotating frame transformation. An interesting symmetry can be observed between $q^a=0$ and $q^a=1$. In both cases the Lagrangian describes free motion in a magnetic field of equal magnitude, but of opposite sign. In line with this, the transformed motion in equation \eqref{eq:chimotion} is
\begin{equation}
  \chi^{a}(t)|_{q^a=0} = C^a +  A^a \, e^{- i \Omega \,t} \, , \qquad
  \chi^{a}(t)|_{q^a=1} = C^a \, e^{i \, \Omega \,  t} +  A^a \, , 
\end{equation}
so $\Omega \to -\Omega$ and the roles of amplitude and center get switched. More generally, there is a symmetry $q^a \leftrightarrow 1-q_a$. The fixed point $q_a=\frac{1}{2}$ is distinguished by the absence of any magnetic force.

In the context of an actual ${\cal N}=2$ flux compactification to AdS$_4$, the $R$-charge is identified with the $R$-charge of the ${\cal N}=2$ AdS$_4$ superalgebra ${\rm OSp(2|4)}$, which in turn typically arises as the Kaluza-Klein-charge of an isometry of the compactification manifold. 
Hence, in such a setup we could, in principle, determine the actual $q_a$. Here our discussion is more general and we do not know, a priori, the values of the $q_a$. 
In the present case we have assumed $W=0$, so any choice would give an $R$-symmetry.
On the other hand, if $W \neq 0$, the $q_a$ are not arbitrary, but constrained by the homogeneity condition in equation \eqref{eq:Wscaling}.
We consider this case next.


\subsubsection{Nonabelian case with superpotential}
\label{sec:nonzeroW}

A natural example of a system with a nonzero superpotential is the case of three arrows, $\kappa = 3$, and $N>1$, with superpotential
\begin{equation}
\label{eq:Wphicubed}
W = \frac{c}{3} \, \epsilon_{abc} \, \Tr \bigl( \phi^a \phi^b \phi^c \bigr) \, , 
\end{equation}
with $c$ some constant. This choice leads to a commutator-squared type potential:
\begin{equation}
 |\partial W|^2 = \frac{c^2}{2} \Tr \bigl( [\phi^a,\phi^b] \, [\phi^a,\phi^b]^\dagger \bigr) \, .
\end{equation}
This superpotential actually has an extended, nonabelian $R$-symmetry, but in line with the above discussion let us consider just the $\mathrm{U}(1)_R$ subgroup. 
The constraint in equation \eqref{eq:Wscaling} becomes
\begin{equation}
 q_1 + q_2 + q_3 = 2 \, .
\end{equation}
This has a two-dimensional family of solutions, all of which provide $R$-symmetries of the system. This is a manifestation of the presence of an extended $R$-symmetry. One possible choice is
\begin{equation} \label{eq:q1q2q3}
 (q_1,q_2,q_3)=(1,1,0) \, .
\end{equation}
After transforming to the $\rho = 2$ AdS frame by redefining, as before, $\chi^{a}(t)  \equiv  e^{i \, \Omega \, q_a t} \, \phi^a(t)$, 
the Lagrangian in equation \eqref{eq:trunclag} becomes (in $A=0$ gauge):
\begin{align}
\label{eq:Lchi}
\Lcal &=  \Tr\bigl( \, |\dot \chi^a|^2  -  \tfrac{c^2}{2}  [\chi^a,\chi^b] \, [\chi^a,\chi^b]^\dagger   - \tfrac{1}{2m} [\chi^{a\dagger},\chi^a] \, [\chi^{b\dagger},\chi^b] \, + \tfrac{i}{2}  s_a \Omega \bigl( \dot \chi^{a\dagger}  \chi^a - \chi^{a\dagger} \dot \chi^a  \bigr) \, \bigr)   \, , 
\end{align}
with $(s_1,s_2,s_3) = (-1,-1,+1) $.
To compare this to the nonabelian D0-brane Lagrangian in real coordinates,
we decompose the complex matrices $\chi^a$ into real and imaginary parts as $\chi^a = u^a + i v^a$. Then we have
\begin{align*}
  \Tr \bigl( [\chi^a,\chi^b] \, [\chi^a,\chi^b]^\dagger \bigr) &= -\Tr\bigl( 
   [u^a,u^b]^2 + [v^a,v^b]^2 + [u^a,v^b]^2 + [v^a,u^b]^2
  \bigr) \\
  & \quad + 2 \, \Tr \bigl( [u^a,v^a] [u^b,v^b] \bigr) \, , \\
  \Tr \bigl( [\chi^{a\dagger},\chi^a] \, [\chi^{b\dagger},\chi^b] \bigr) &= - 4 \, \Tr \bigl( [u^a,v^a] [u^b,v^b] \bigr) \ ,
\end{align*}
and thus, denoting $(y^1,y^2,y^3,y^4,y^5,y^6) \equiv (u^1,v^1,u^2,v^2,u^3,v^3)$,
\begin{align}
  \Tr \bigl( [\chi^a,\chi^b] \, [\chi^a,\chi^b]^\dagger + \tfrac{1}{2} [\chi^{a\dagger},\chi^a] \, [\chi^{b\dagger},\chi^b] \bigr) = -\Tr \, [y^m,y^n]^2 \, .
\end{align}
Hence we see that for this particular combination of complex field commutators, which corresponds to setting $ c \equiv \sqrt{2/m}$ in \eqref{eq:Lchi}, we obtain an SO(6)-symmetric commutator squared term. Finally, to bring the kinetic term into the same form as the kinetic term of the vector $X$ as in \eqref{eq:ND0case}, we rescale $y^n = \sqrt{\frac{m}{2}} \, Y^n$, so \eqref{eq:Lchi} becomes
\begin{equation}
\label{eq:Lint}
\Lcal = m \, \Tr \bigl( \tfrac{1}{2} \dot{Y}^2 + \tfrac{1}{4} [Y^m,Y^n]^2 + \tfrac{1}{2} \, \Omega \,\varepsilon_{mn} \,  Y^m \dot{Y}^n
 \bigr) \, ,
\end{equation}
where $\varepsilon_{mn}$ is a block-diagonal antisymmetric matrix with
$\varepsilon_{12}=-1$, $\varepsilon_{34}=-1$, and $\varepsilon_{56}=+1$. If we call the space parametrized by the 3-vector $X^i$ external and the space parametrized by the 6-vector $Y^n$ internal, then 
this is precisely the 6d internal-space part of the Lagrangian of a stack of D0-branes with a flat internal space threaded by the R-R 2-form field 
\begin{equation}
 F = \Omega \bigl(-dY^1 \wedge dY^2 - dY^3 \wedge dY^4 + dY^5 \wedge dY^6 \bigr) \, .
\end{equation}
Apart from the orientation reversal of the $(12)$ and $(34)$ planes, this is the same magnetic field as in equation \eqref{eq:Fmagn}, leading to cyclotron motions with frequency $\pm \Omega$ similar to those in equation \eqref{eq:phimotion0}.
As noted there, this is also the D0 cyclotron frequency in the ${\rm AdS}_4 \times {\mathbb CP}^3$ flux compactification dual to ABJM theory, so we see that this simple quiver model already captures quite accurately the dynamics of D0-branes in string flux compactifications. 
This could be improved further by generalizing the models to arbitrary target space K\"ahler potentials. 

This D0-model is just one of many possible one-node quiver models. Different values of $c$, for example, lead to models with the SO(6) symmetry of the Lagrangian in equation \eqref{eq:Lint} broken so some subgroup, modeling D-branes in flux compactifications with fewer isometries than AdS$_4 \times {\mathbb CP}^3$.
More generally, instead of D0-branes, we can model internally wrapped branes of different dimensions, possibly carrying worldvolume fluxes, and so on. The adjoint scalars $\phi^a$ will then correspond to geometric-deformation moduli of these brane configurations. From the general form of the deformed quiver Lagrangian it is clear that magnetic fields like \eqref{eq:Fmagn} will be a generic feature. However, in general these magnetic fields live on D-brane moduli space; they no longer have a direct physical-space interpretation as in the case of D0-branes. Nevertheless, their effect will be similar, causing oscillatory motion even in the absence of a potential on D-brane moduli space. Such dynamical features can presumably also be thought of as the result of the presence of background magnetic R-R-fluxes interacting with the D-branes.


\subsubsection{BMN matrix mechanics}
\label{sec:BMN}

The BMN matrix model describes D-branes or M-branes in a 
supersymmetric plane-wave background \cite{Berenstein:2002jq}. Although its interpretation is quite different from the nonrelativistic D-particles in AdS we have been considering so far, its Lagrangian is nevertheless a special case of our general massive quiver Lagrangian. It corresponds to a quiver with one node and three arrows and a cubic superpotential of the form in equation \eqref{eq:Wphicubed}, but now with 
\begin{equation}
  q_1 = q_2 = q_3 = \frac{2}{3} \, , \qquad \rho = \frac{3}{2} \, ,
\end{equation}
instead of $\rho=2$ and the $R$-charge assignments in equation \eqref{eq:q1q2q3}. Thus, the field redefinition of equation \eqref{eq:Rtransf} becomes $\phi^a = e^{- i \, \frac{1}{2} \Omega \, t} \, \chi^a$, which, for the chiral scalar field,
is effectively the same as the case $q_a = \frac{1}{2}$ in equation \eqref{eq:chimotion}. From equation \eqref{eq:Lphisimple2} we can then immediately read off that the magnetic interaction vanishes for the transformed Lagrangian for $\chi^a$ in this case, and that instead a harmonic oscillator potential $V(\chi) = \frac{1}{4} \Omega^2 |\chi^a|^2$ appears. Performing the same changes of variables as those leading up to equation \eqref{eq:Lint}, and combining this with the Langrangian in equation \eqref{eq:ND0case} for the vector multiplet scalars $X^i$, we obtain the Lagrangian 
\begin{align}
\label{eq:LintBMN}
\Lcal &= m \, \Tr \bigl(\tfrac{1}{2} \dot{X}^2 + \tfrac{1}{2} \dot{Y}^2 - \tfrac{1}{2}  \Omega^2 \, X^2  - \tfrac{1}{8} \Omega^2 \, Y^2 - i \, \Omega \, \epsilon_{ijk}X^iX^jX^k + \cdots \bigr) \, ,
\end{align} 
where the ellipsis denotes the fermionic and the commutator squared terms. (Some of these arise from the interaction part $\Lcal^0_I$ in equation \eqref{eq:totalLag}, which we have ignored so far in this section.) This correctly matches the BMN model, and it can be checked that the same holds for the fermions. 
The equality of the $R$-charges allows for an enhancement of the $R$-symmetry group from $\mathrm{U}(1)$ to $\mathrm{SO}(6)$.
Consequently, the superalgebra is also enhanced from $\mathfrak{su}(2|1)$ to $\mathfrak{su}(2|4)$.


\subsection{Two nodes, $\kappa$ arrows}
\label{sec:two_nodes}

\begin{figure}[h!]
\centering
\includegraphics[width=50mm]{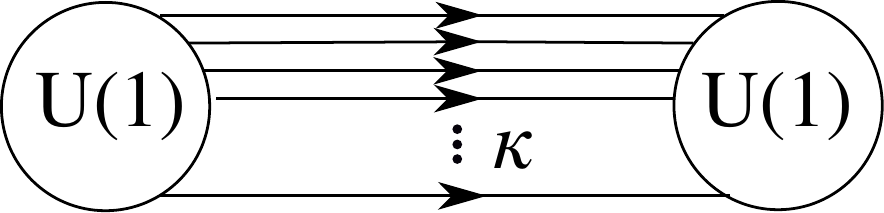}
\caption{A two-node quiver representing a matrix model with two vector multiplets, with gauge groups $\mathrm{U}(1)$, and $\kappa$ chiral multiplets.}
\label{fig:quiv2}
\end{figure}

Our next example is a quiver with two nodes and $\kappa$ arrows,
and dimension vector $(1,1)$, hence two $\mathrm{U}(1)$ vector multiplets containing the scalars $x_v^i$ and $\kappa$ charge $(-1,1)$ chiral multiplets containing the scalars $\phi^a$, see figure \ref{fig:quiv2}.

\subsubsection{Lagrangian}

Since we will not need the fermions in our discussion here, we just give the bosonic part of the Lagrangian, $\Lcal^b = \Lcal^b_{V} + \Lcal^b_{C}$, with 
\begin{align}
\label{eq:LV}
\Lcal^b_V &= \frac{m_1}{2} \bigl( \dot{x}_1^2 + D_1^2) + \frac{m_2}{2} \bigl( \dot{x}_2^2 + D_2^2) - \theta_1 D_1 - \theta_2 D_2 \nonumber \\ 
&\quad - \Omega^2 \, \bigl( \tfrac{1}{2} m_1 x_1^2 + \tfrac{1}{2} m_2 x_2^2\bigr) + \Omega \bigl( \theta_1 A_1 + \theta_2 A_2 \bigr) \, , \\
\Lcal^b_C &= |{\cal D}_t \phi^a|^2 - \bigl (|x_1 - x_2|^2 + D_1 - D_2 \bigr) |\phi^a|^2  \nonumber \\ 
\label{eq:LC}
&\quad - \tfrac{1}{2} \, \Omega \, i  \bigl(\bar \phi^a {\cal D}_t \phi^a - ({\cal D}_t \bar \phi^a)  \phi^a \bigr)  \, ,
\end{align}
where ${\cal D}_t \phi^a = \partial_t \phi^a + i (A_1-A_2) \phi^a$. 
Since a superpotential is forbidden by gauge invariance, the equations of motion for the auxiliary fields $F^a$ are trivial; that is $F^a = 0$.
Varying the gauge fields together, i.e., $\delta A_1=\delta A_2$, gives the consistency constraint 
\begin{equation}
 \Omega \, (\theta_1 + \theta_2) = 0 \, .
\end{equation}
This is analogous to the $\Omega \, \theta = 0$ consistency constraint we found for the single node quiver. In what follows we will therefore assume $\theta_1+\theta_2=0$.

As in the single node case, the $x_v^i$ may be thought of as the positions of two massive particles near the bottom of an AdS potential well, with $\Omega = 1/\ell_{\rm AdS}$. Despite the absence of translation invariance when $\Omega \neq 0$, the exact proportionality of potential and kinetic terms for the vector multiplet means it is still possible to separate the Lagrangian into a decoupled center-of-mass (c.o.m.)~part and a relative part, ${\cal L}={\cal L}_{\rm c.o.m.}+{\cal L}_{\rm rel}$. Given the gravitational interpretation of the potential, this can be interpreted as a consequence of the equivalence principle. Define c.o.m.~variables $Y_0$ and relative variables $Y$ for the vector multiplet as $Y_0 \equiv (m_1 Y_1 + m_2 Y_2)/(m_1 + m_2)$ and $Y \equiv Y_2-Y_1$, for $Y=x^i,A,D,\lambda$. Then the bosonic part of the c.o.m.~Lagrangian is given by
\beq
\Lcal^b_{\text{c.o.m.}} = \frac{m_1 + m_2}{2} \lp \dot{x}_0^2 - \Omega^2 x_0^2 + D_0^2 \rp \, ,
\eeq
and the bosonic part of the relative Lagrangians is 
\begin{align}
	\mathcal{L}^b_{\text{rel}} &= \frac{\mu}{2} \left( \dot{x}^2 - \Omega^2 x^2 
	+ D^2 \right) + \theta (\Omega A - D) \\
	 & \quad + |\mathcal{D}_t \phi^a|^2 - (x^2 -D)|\phi^a|^2 
	- \tfrac{1}{2} \, \Omega \, i  \bigl(\bar \phi^a {\cal D}_t \phi^a - ({\cal D}_t \bar \phi^a)  \phi^a \bigr) \ ,
\end{align}
where 
\begin{equation}
 \theta \equiv \theta_2 = -\theta_1 \, , \qquad
 \mu \equiv \frac{m_1 m_2}{m_1+m_2} \, , \qquad
  \mathcal{D}_t \phi^a \equiv (\partial_t - iA)\phi^a \, .
\end{equation}
We may also integrate out the auxiliary $D$ field, leading to the following potential:
\begin{equation} \label{Vphi}
  V(x,\phi) = \frac{\mu \, \Omega}{2} \, |x|^2 +  |x|^2 |\phi^a|^2 + \frac{1}{2\mu} \left(|\phi^a|^2 - \theta\right)^2 \, .
\end{equation}
Notice that if $\theta>0$, the potential attains its (zero) minimum at $x=0$, $|\phi^a|^2 = \theta$, whereas if $\theta < 0$, the potential reaches its (nonzero) minimum at $x=0$, $\phi^a=0$. If $x$ is held fixed at some sufficiently large fixed value, $V(\phi)$ is minimized at $\phi=0$. 

In the flat-space case, these observations essentially determine the bound state formation properties of the system \cite{Denef:2002ru}. In the massive case however, the Gauss's law constraint will significantly alter this analysis. We turn to this next.

\subsubsection{The Gauss's law constraint and confinement}

\label{sec:gaussconfines}

The Gauss's law constraint will turn out to have rather dramatic consequences; when $\Omega \theta \neq 0$, it causes the particles to be confined by fundamental strings. The Gauss's law constraint is obtained by varying the Lagrangian with respect to $A$. Working in the gauge $A \equiv 0$, we obtain:
\begin{equation}
\label{eq:gausslaw2}
 \Omega \bigl( \theta - |\phi|^2  \bigr) + i \bigl(\bar \phi^a  \dot \phi^a -  \dot{\bar \phi}^a  \phi^a \bigr)  = 0 \ ,
\end{equation}
where  $|\phi|^2 \equiv \sum_a |\phi^a|^2$.
One immediate consequence is that $\phi=0$ is inconsistent if $\Omega \theta \neq 0$, and that more generally, the Gauss's law constraint will force the $\phi^a$ to be in some excited state. Clearly then, the naive energy minimization analysis under (\ref{Vphi}) must be modified. 

In what follows we will interpret the constraint in string theory as a lower bound $n_{\rm min}$ on the number of physical open strings stretched between the two particles. More specifically, we will find $n_{\rm min}=\Omega \, \theta$.

Recall that in a 10d string context, the two particles in AdS correspond to two distinct internally wrapped D-branes, and the $\phi^a$ correspond to the lightest open-string modes that exist between the two branes. Roughly speaking, this means that if we hold $x$ fixed and quantize the $\phi^a$, the $n$-th energy level can be thought of as a state containing $n$ stretched strings, all in their string oscillator ground state. To check this, let us assume a large fixed separation $|x|$, so we can view the $\phi^a$ as simple harmonic oscillators with frequency $|x|$. Their $n$-th level excitation energy is then equal to $n |x|$. Keeping in mind our choice of units, equation \eqref{eq:units}, this is indeed the energy of $n$ fundamental strings stretched over a distance $|x|$. In the same vein, classical field excitations of $\phi^a$ may be thought of as quantum coherent states, which are superpositions of infinitely many different string number eigenstates. For large amplitudes however, the string number probability distribution will be sharply peaked, and we can infer the minimal number of strings by simply computing the minimal classical energy stored in the $\phi^a$ at large separation.

Since the Gauss's law constraint in equation \eqref{eq:gausslaw2} involves time derivatives, it will be necessary to include kinetic energy contributions in the energy budget, and take into account the dynamical equations of motion. It will be  convenient to keep the auxiliary fields explicit in the computation. The total  energy derived from the relative Lagrangian is 
\begin{align}
 {E}_{\text{rel}} &= \dot{x}^i \frac{\partial {\Lcal}_{\text{rel}}}{\partial \dot x^i} + \dot{\phi}^a \frac{\partial {\Lcal}_{\text{rel}}}{\partial \dot \phi^a} + \dot{\bar\phi}^a \frac{\partial {\Lcal}_{\text{rel}}}{\partial \dot{\bar\phi}^a} 
- {\Lcal}_{\text{rel}} \\
\label{eq:Erel}
&= \frac{\mu}{2} \left( \dot{x}^2 + \Omega^2 x^2 - D^2\right) + \theta D + |\dot \phi|^2 + (x^2 -D)|\phi|^2 \, ,
\end{align}
and here $|\dot\phi|^2 \equiv \sum_a |\dot{\phi}^a|^2$. The equations of motion derived from ${\Lcal}_{\text{rel}}$ are 
\begin{align}
\label{eq:xeomm}
\mu \bigl( \ddot{x} + \Omega^2 \, x \bigr) + 2 \, |\phi|^2 \, x &= 0\\
\label{eq:Deom}
\mu \, D - \theta + |\phi|^2 &= 0 \\
\label{eq:phi_eom}
\ddot \phi^a +  i \, \Omega \,  \dot \phi^a + (x^2-D) \phi^a  &= 0 \, .
\end{align}
If we assume for a moment that $x$ and $D$ are time-independent, the general solutions to equation \eqref{eq:phi_eom} are 
\begin{equation}
\label{eq:phi_soln}
\phi^a = \alpha^a \, e^{-i \omega_+ t} + \beta^a \, e^{-i \omega_- t}  \, , %
\qquad \omega_{\pm} = \tfrac{1}{2} \Omega \pm \sqrt{ \tfrac{1}{4} \Omega^2+x^2-D} \, .
\end{equation}
On this set of solutions, assuming the argument of the square root is positive, the Gauss's law constraint 
in equation \eqref{eq:gausslaw2} is equivalent to:
\begin{equation}
\label{eq:abcon}
|\beta|^2 - |\alpha|^2 = \frac{\Omega \, \theta}{\sqrt{\Omega^2+4(x^2-D)}}  \, ,
\end{equation}
where $|\alpha|^2 = \sum_a |\alpha^a|^2$ and $|\beta|^2 = \sum_a |\beta^a|^2$.
Notice in particular that if $\theta > 0$, the $\beta^a$ cannot all vanish, and if $\theta < 0$, the $\alpha^a$ cannot all vanish. The contribution to the total energy in equation \eqref{eq:Erel} from the motion of the $\phi^a$ is, taking into account the constraint in equation \eqref{eq:abcon},
\begin{equation}
 E_{\text{rel},\phi} \equiv |\dot \phi|^2 + (x^2 -D)|\phi|^2  = \tfrac{1}{2} \bigl( \Omega^2 + 4(x^2-D) \bigr) \bigl(|\alpha|^2+|\beta|^2\bigr) - \tfrac{1}{2} \Omega^2 \theta  \, .
\end{equation}
In view of the constraint in equation \eqref{eq:abcon}, we see that for $\theta > 0$ the energy is minimized when $|\alpha|^2=0$ (and $|\beta|^2 >0$), and for $\theta < 0$ when $|\beta|^2=0$ (and $|\alpha|^2 > 0$), and for $\theta = 0$ when $|\alpha|^2 = |\beta|^2 =0$. In each of these cases we have
\begin{equation}
\label{eq:phisqval}
 |\phi|^2 = \frac{\Omega |\theta|}{\sqrt{\Omega^2+4(x^2-D)}}
\end{equation}
and the corresponding minimal value of the energy stored in the oscillating $\phi^a$ fields is
\begin{equation}
 E_{\text{rel},\phi} = \tfrac{1}{2} \Omega |\theta| \sqrt{ \Omega^2 + 4(x^2-D) }  - \tfrac{1}{2} \Omega^2 \theta  \, .
\end{equation}
In particular, in the limit $|x| \to \infty$, with everything else fixed, this becomes
\begin{equation}
 E_{\text{rel},\phi} \approx  \Omega |\theta|  |x| \, ,
\end{equation}
which, taking into account our units, equation \eqref{eq:units}, is the characteristic confining linear energy of $\Omega |\theta|$ fundamental strings. Hence, we conclude that the minimal number of stretched open strings is 
\begin{equation} \label{nminstring}
  n_{\rm min} = \Omega |\theta| \, .
\end{equation}

We end this part with a more careful analysis of the regime of validity of our analysis.
In actual solutions to the full system of equations of motion, $x$ and $D$ will not be time independent, and moreover $\phi$ will back-react onto $x$ and $D$, leading to a complicated nonlinear system of equations. 
However, if $|x|^2 - D \gg \frac{1}{4}\Omega^2$ and $|\phi|^2 \ll \mu \Omega^2$, the oscillatory motion of $x$ will be much slower than that of the $\phi^a$, hence to a good approximation the $\phi^a$ will oscillate as stated above, with slowly varying amplitude and frequency. To describe this regime more explicitly, notice from \eqref{eq:phisqval} that $|x|^2 - D \gg \frac{1}{4}\Omega^2$ implies  $|\phi|^2 \ll |\theta|$, and therefore by \eqref{eq:Deom} that $D \approx \theta/\mu$. 
Thus self-consistency requires $|x|^2 - \frac{\theta}{\mu} \gg \frac{1}{4} \Omega^2$, and moreover we have from \eqref{eq:phisqval} that $|\phi|^2 \approx \frac{\Omega |\theta|}{\sqrt{\Omega^2+4(x^2-\frac{\theta}{\mu})}} \approx \frac{\Omega |\theta|}{2 \, \sqrt{x^2-\frac{\theta}{\mu}}}$. Our second condition $|\phi|^2 \ll \mu \Omega^2$ thus becomes $\frac{|\theta|}{\mu} \ll 2 \, \Omega \, \sqrt{x^2-\frac{\theta}{\mu}}$.  Since we have already assumed $x^2-\frac{\theta}{\mu} \gg \frac{1}{4}\Omega^2$, this condition is satisfied if $\frac{|\theta|}{\mu} \ll \Omega^2$. Thus, we conclude that the above separation into fast and slow variables is valid in the regime
\begin{equation}
 |x| \gg \Omega \gg \sqrt{\tfrac{|\theta|}{\mu}} \, .
\end{equation}
Moreover, in this regime we have
\begin{equation}
 D \approx \frac{\theta}{\mu} \, , \qquad |\phi|^2 \approx \frac{\Omega |\theta|}{2 |x|} \, , \qquad E_{\rm min, \phi} \approx \Omega |\theta| |x| \, ,
\end{equation}
so that the equation of motion effectively becomes
\begin{equation}
 \mu \bigl( \ddot{x} + \Omega^2 \, x \bigr) + \Omega |\theta| \frac{x}{|x|} = 0 \, ,
\end{equation}
which is the equation of motion for the relative position of two nonrelativistic particles in AdS$_4$ connected by $\Omega |\theta|$ fundamental strings.

\subsubsection{Ground state structure}

\label{sec:groundstates}

In the flat-space case, $\Omega=0$, the ground state structure of the model can easily be inferred from a static potential analysis, as was done under equation (\ref{Vphi}). In particular, one finds the system has a classical supersymmetric ground state at $x=0$, $|\phi|^2 = \theta$ if $\theta>0$, and no supersymmetric ground state if $\theta<0$. In the former case, the relative $\mathrm{U}(1)$ is Higgsed and we obtain a bound state, which can be thought of as a fusion of the two constituent branes into one \cite{Kachru:1999vj}. In the latter case, the minimal energy configuration has $\phi=0$ but $x$ arbitrary, so the relative $\mathrm{U}(1)$ remains unbroken and classically there is no bound state. 

Interestingly, the supersymmetric bound state at $\theta > 0$ survives the $\Omega \neq 0$ mass deformation entirely unaffected, at least classically. It is easy to check that $D=\dot{x}=\dot{\phi}=0$, $|\phi|^2=\theta$ solves the equations of motion while satisfying the Gauss's law constraint. The  energy \eqref{eq:Erel} is minimized at zero on this configuration. Furthermore, this solution is supersymmetric, 
as follows  by checking $\delta \lambda = \delta \psi = 0$. 

For $\theta<0$, the situation is more complicated. The Gauss's law constraint forbids the undeformed solution $\phi=0$, and the presence of confining strings means that the ground state remains a bound state, but one with broken supersymmetry. It is possible to derive analytic expressions for the energy as a function of $\theta$, but as this will be of no further use to us in this paper, we will not do this here.

\subsubsection{Concluding comments}
\label{sec:thetazero}

The interpretation of a nonzero $\theta_v \Omega$ as a minimum number of confining strings attached to a particle also suggests an interpretation of the ``overall part'' of the Gauss's law constraint, i.e., $\Omega \, (\theta_1+\theta_2)=0$ for the two-node quiver, and $\Omega \, \theta = 0$ for the one-node quiver. When there are two particles, it is possible to have finite energy configurations with nonzero $\theta_v$ for the individual nodes, because the strings emanating from one particle can terminate on the other one. The number of strings emanating from the first particle is $\Omega |\theta_1|$ and the number of strings terminating on the other one $\Omega |\theta_2|$. If $\theta_1 + \theta_2 = 0$, there are as many strings coming out of the first particle as there are strings going into the second. However if $\theta_1 + \theta_2 \neq 0$, some strings have nowhere to go and hence must stretch all the way to the conformal boundary of AdS, leading to an infinite energy configuration. It is reassuring that this situation is excluded by consistency from these low energy models, which only include string modes stretched between the particles.

A more complete analysis of the two node quiver system requires the use of quantum mechanics. We provide the quantum Hamiltonian and other generators of the supersymmetry algebra in appendix \ref{app:algebra}, but will not perform the analysis here. A quantum treatment also requires integral quantization of $n_{\rm min} = \Omega |\theta|$, thus imposing a quantization condition on the FI parameter. This may seem obscure from our treatment above, but becomes clear upon recalling that the Gauss's law constraint is a charge cancellation constraint. Explicitly, the quantum Gauss's law constraint is given by equation \eqref{eq:quantumgausslaw}, which requires the sum of the diagonal $\mathrm{U}(1)$ gauge charge and the background charge $\Omega \theta_v$ to cancel for each node $v$. Quantization of $\theta_v$ is therefore a consequence of Gauss's law and the integral quantization of this $\mathrm{U}(1)$ charge. 

Another interesting quantum question, raised in \cite{Anninos:2013mfa}, is whether the ``Coulomb branch'' bound states of \cite{Denef:2002ru} persist in the presence of confining strings. We leave this to future work.

The appearance of confining strings ending on wrapped branes is a phenomenon also universally observed in the context of flux compactifications. 
In fact, this was one motivation for this work, prompted by the related difficulties encountered in \cite{Anninos:2013mfa}. In the next section we will review this for the AdS$_4 \times {\mathbb CP}^3$ flux compactification of type IIA string theory, and compare in some detail the properties of the two-node quiver derived here with properties of wrapped branes in this compactification.


\section{Comparison to wrapped branes on AdS$_4 \times {\mathbb CP}^3$}
\label{sec:ABJM}

The various examples of massive matrix models we have considered all 
describe some features of wrapped branes in general AdS$_4$ flux compactifications. 
To make the comparison more precise, we consider 
the specific AdS$_4 \times {\mathbb CP}^3$ compactification of type IIA string theory, with $N$ units of magnetic R-R 6-form flux and $k$ units of magnetic R-R 2-form flux \cite{Watamura:1983hj, Nilsson:1984bj}, holographically dual to the ABJM theory \cite{Aharony:2008ug}. 
We will focus, in particular, on the confining fundamental strings between internally wrapped D-branes and the role of the FI parameters $\theta_v$, which we identify in terms of the parameters of this compactification.


\subsection{Notation and conventions}

As before, we work in the notation and conventions of \cite{Myers:1999ps}. This means, in particular, the type IIA supergravity action contains the terms 
\begin{equation}
\label{eq:IIAaction}
 S_{\rm IIA} = \frac{2\pi}{g_s^2 L_s^8}  \int d^{10} x \, \sqrt{-g} \, \bigl( e^{-2 \phi} R - \tfrac{1}{2} F^{(2)} \wedge * F^{(2)} - \tfrac{1}{2} F^{(4)} \wedge * F^{(4)} + \cdots \bigr) \, ,
\end{equation}
where $F^{(2)} = d C^{(1)}$ and $F^{(4)}=d C^{(3)} + dB^{(2)} \wedge C^{(1)}$, and
\begin{equation}
 L_s = 2 \pi \sqrt{\alpha'} \, 
\end{equation}
is our convention for the string length in this section.\footnote{In \cite{Myers:1999ps} other versions of the string scale are introduced, namely $\ell_s \equiv \sqrt{\alpha'}$ as well as $\lambda \equiv 2 \pi \alpha'$. The massive quiver Lagrangian in equation \eqref{eq:totalLag} is given in units with $\lambda = 1$, i.e.\ $L_s = \sqrt{2\pi}$.}
The abelian D$p$-brane action is 
\begin{equation}
\label{eq:Dpaction}
 S_{\text{D}p} = -\frac{2 \pi}{g_s L_s^{p+1}} \int d^{p+1} x \, e^{-\phi} \sqrt{-\det(g + L_s^2 {\cal F})} + \frac{2 \pi}{g_s L_s^{p+1}} \int \mbox{$\sum_n$} C^{(n)} \wedge e^{L_s^2 {\cal F}}  \, ,
\end{equation}
with ${\cal F} = \frac{1}{L_s^2} B^{(2)}+\frac{1}{2\pi} dA^{(1)}$, $A^{(1)}$ being the $\mathrm{U}(1)$ gauge field living on the brane worldvolume.


\subsection{AdS$_4 \times {\mathbb CP}^3$ flux compactification}

With these conventions, the AdS$_4 \times {\mathbb CP}^3$ flux compactification, originally described in \cite{Watamura:1983hj, Nilsson:1984bj} and more recently studied in \cite{Aharony:2008ug,Gutierrez:2010bb} and many other works, consists of the following nontrivial IIA supergravity field excitations. The metric is of product form
\begin{equation}
 ds^2 = ds^2_{{\rm AdS}_4(\ell)} + ds^2_{{\mathbb CP}^3(L)} \, , \qquad L = 2 \, \ell \, ,
\end{equation}
where $ds^2_{{\rm AdS}_4(\ell)}$ denotes the metric on AdS$_4$ with radius $\ell$, i.e., the hyperboloid $\eta_{IJ} X^I X^J=-\ell^2$ embedded in ${\mathbb R}^5$ with signature $(-,-,+,+,+)$ metric $ds_{{\mathbb R}^5}^2 = \eta_{IJ} dx^I dx^J$, while $ds^2_{{\mathbb CP}^3(L)}$ denotes the metric on ${\mathbb CP}^3$ with radius $L$, i.e., the sphere $\delta_{ij} Z^i \bar Z^j = L^2$ embedded in ${\mathbb C}^4$ with Euclidean metric $ds^2_{{\mathbb C}^4} = \delta_{ij} dZ^i d{\bar Z}^j$, quotiented by the $\mathrm{U}(1)$ action $Z^i \to e^{i \phi} Z^i$. The quotient space has three complex dimensions and is a K\"ahler manifold with K\"ahler form $J_{{\mathbb CP}^3(L)}$. The R-R 4-form and 2-form field strengths are
\begin{equation}
F^{(4)} = \frac{3}{\ell} \, dV_{\rm AdS_4(\ell)} \, , \qquad %
F^{(2)} = \frac{1}{\ell} J_{{\mathbb CP}^3(2 \ell)} \, .
\end{equation}
The NS-NS 2-form $B^{(2)}=0$ and the dilaton $\phi=0$.  The dual magnetic flux to $F^{(4)}$ is $F^{(6)} = *F^{(4)} = -\frac{3}{\ell} dV_{{\mathbb CP}^3} = -\frac{3}{\ell} \frac{1}{6} J_{{\mathbb CP}^3}^3$. The minus sign arises because of the Lorentzian signature of the 10d metric.\footnote{This sign was missed in \cite{Gutierrez:2010bb}, which is the origin of the sign discrepancy with that work we will find later, in the formula for the number of confining strings terminating on the brane, equation \eqref{NFformula}.} The complex surface $Z^1=0$ defines a ${\mathbb CP}^2$ submanifold of  ${\mathbb CP}^3$, and the complex curve $Z^1=Z^2=0$ similarly defines a ${\mathbb CP}^1$ submanifold. The fluxes through these cycles are integrally quantized as
\begin{equation}
\label{eq:fluxquant}
 \frac{1}{g_s L_s} \int_{{\mathbb CP}^1} F^{(2)} = k  \, , \qquad \frac{1}{g_s L_s^5} \int_{{\mathbb CP}^3}  F^{(6)} = -N \, ,
\end{equation}
with $k$ and $N$ positive integers (positivity follows from $F^{(2)} \propto + J$, $F^{(6)} \propto - J^3$ and positivity of the K\"ahler form).
Using some basic algebraic geometry (see, e.g., \cite{Denef:2008wq} section 5.4 and equation (5.16)), the volume of these ${\mathbb CP}^n$ subspaces, $n=1,2,3$, are found to be 
\begin{equation}
\label{eq:CPvolumes}
\text{vol}({\mathbb CP}^n) = \int_{{\mathbb CP}^n} \frac{1}{n!} J_{{\mathbb CP}^3(2 \ell)}^n= \frac{1}{n!} (4 \pi \ell^2)^n \, .
\end{equation} 
Using $F^{(2)}=\frac{1}{\ell} J_{{\mathbb CP}^3}$ and $F^{(6)} =  -\frac{3}{\ell} \frac{1}{6} J_{{\mathbb CP}^3}^3$, together with \eqref{eq:fluxquant} and \eqref{eq:CPvolumes}, we obtain the relation
$\frac{1}{g_s L_s} \frac{1}{\ell} (4 \pi \ell^2) = \frac{4 \pi \ell}{g_s L_s} = k$ and $\frac{1}{g_s L_s^5} \frac{3}{\ell} \frac{1}{6} (4 \pi \ell^2)^3 = \frac{2^5 \pi^3 \ell^5}{g_s L_s^5} = N$.
Solving these for $g_s$ and $\ell$, we get
\begin{equation}
\label{eq:gsellsols}
 g_s = \left(\frac{32 \, \pi^2 N}{k^5}\right)^{\frac{1}{4}} \, , \qquad 
 \ell =  \left(\frac{N}{8 \, \pi^2 k}\right)^{\frac{1}{4}} L_s  \ , 
\end{equation}
where we recall we define $L_s=2\pi \sqrt{\alpha'}$ and work in units where $L_s = \sqrt{2\pi}$ in the quiver Lagrangian in equation \eqref{eq:totalLag}. The four dimensional Planck length is defined as $\ell_\text{P} \equiv \sqrt{G_\text{N}}$, with the four dimensional Newton constant $G_\text{N}$ obtained by dimensional reduction of equation \eqref{eq:IIAaction}, i.e., $\frac{1}{16 \pi G_\text{N}} = \frac{2\pi}{g_s^2 L_s^8} {\rm vol}({\mathbb CP}^3)$.
In combination with equation \eqref{eq:CPvolumes} this yields
\begin{equation}
\label{eq:lplanck}
 \ell_\text{P} = \left(\frac{3 }{8 \pi k N} \right)^{\frac{1}{2}} L_s \ .
\end{equation}


\subsection{Probe brane masses}

Type IIA string theory has 0-, 2-, 4- and 6-branes, which we can wrap on non-contractible cycles of the same dimension in ${\mathbb CP}^3$ to obtain charged particles in AdS$_4$. The bare masses of these particles can be obtained from the D-brane action in equation \eqref{eq:Dpaction}. This was done in \cite{Gutierrez:2010bb}. We review and streamline their computations here. 

From \eqref{eq:Dpaction} together with \eqref{eq:CPvolumes} and \eqref{eq:gsellsols}, it follows that a D($2n$)-brane wrapped on ${\mathbb CP}^n$ ($n=0,1,2,3$) with ${\cal F} = 0$ looks like a particle in AdS$_4$ with mass 
\begin{equation}
\label{eq:mD2n}
 m_{D(2n)} 
 = \frac{v^n}{n!}  \, m_{D0} \, , \qquad v \equiv \frac{4 \pi \ell^2}{L_s^2} = \sqrt{\frac{2 N}{k}} \, , \qquad m_{D0} = \frac{2 \pi}{g_s L_s} =\frac{k}{2 \, \ell} \, .
\end{equation}
The case with nonzero ${\cal F}$ can be considered with minimal additional effort. It is characterized by a quantized flux $w \equiv \int_{{\mathbb CP}^1} {\cal F}$;\footnote{Roughly speaking, $w$ is integrally quantized. In general, this may be shifted by non-integral amounts due to curvature effects and $B$-fields. We will ignore these subtleties here, since they do not matter for our purposes. Furthermore, in the regime $k,N \to \infty$, $\frac{k}{N} \gg 1$, in which 10d supergravity is reliable, the ratio $w/v$ effectively becomes a continuous parameter, and we will treat it as such here.}
equivalently, we can write ${\cal F} = \frac{w}{4 \pi \ell^2} J_{{\mathbb CP}^3}$. As for any K\"ahler manifold, the Born-Infeld action in equation \eqref{eq:Dpaction} now becomes proportional to the absolute value of the \emph{complexified} K\"ahler volume, rather than the real K\"ahler volume we had before. The complexified K\"ahler volumes are obtained as in \eqref{eq:CPvolumes} after substituting 
\begin{equation}
  J \to J_{\mathbb C} \equiv J + i L_s^2 {\cal F} = J + i \, \frac{w}{v} \, J \, .
\end{equation}
This changes \eqref{eq:mD2n} to
\begin{equation}
\label{eq:mDnvk}
 m_{D(2n)(w)} = \left| \frac{(v+i w)^n}{n!} \right| \, m_{D0} = \frac{(v^2+w^2)^{\frac{n}{2}}}{n!} \, m_{D0} \, .
\end{equation}
From the second part of the D-brane action in equation \eqref{eq:Dpaction}, we can furthermore infer that turning on  ${\cal F}$ induces lower dimensional D-brane charges; for example, on a singly wrapped D6-brane ($q_{D6} = 1$), we get an induced $q_{D4} = w$, $q_{D2} = \frac{w^2}{2}$ and $q_{D0} = \frac{w^3}{6}$. Expanding $\frac{1}{6}|(v+i w)^3|=\frac{1}{6}|(i v - w)^3| = |\frac{1}{6}(i v)^3 - \frac{1}{2} w (i v)^2 + \frac{1}{2} w^2 (iv) - \frac{1}{6} w^3|$, we can alternatively write the mass formula in equation \eqref{eq:mDnvk} in terms of a complexified mass $\mu_Q$ that depends linearly on the charge $Q=(q_{D6},q_{D4},q_{D2},q_{D0})$ as
\begin{equation}
\label{eq:muQ}
 m_Q = |\mu_Q|  , \qquad \mu_Q \equiv \left(\tfrac{1}{6} q_{D6} \, (iv)^3 - \tfrac{1}{2} q_{D4} \,(iv)^2 + q_{D2} \, (iv) - q_{D0} \right) m_{D0} \, .
\end{equation}
In terms of the four-dimensional Planck mass in equation \eqref{eq:lplanck} this becomes 
\begin{equation}
\label{eq:ZQ}
 m_Q = \tfrac{1}{\ell_\text{P}} |Z_Q| \, , \qquad Z_Q \equiv \left(\tfrac{3}{4 v^3}\right)^{\frac{1}{2}} \left(\tfrac{1}{6} q_{D6} \, (iv)^3 - \tfrac{1}{2} q_{D4} \,(iv)^2 + q_{D2} \, (iv) - q_{D0} \right) \, .
\end{equation}

In asymptotically flat ${\cal N} = 2$ supergravity theories, the energy of a BPS bound state is given by a central charge formula $E=\frac{1}{\ell_\text{P}}|Z|$.
Recall the central charge of a BPS particle is a complex parameter, determined as a function of its electric and magnetic charges and of the background scalars of the supergravity theory. 
In the asymptotically flat-space context, it is defined as the eigenvalue of the state under the action of the central charge operator that appears in the $\mathcal{N}=2$ flat-space superalgebra \cite{Freedman:2012zz}. 
Its phase parametrizes the $\mathcal{N}=1$ subspace inside the ${\cal N} = 2$ superalgebra preserved by the BPS state \cite{Moore:1998pn}. 
In the context of geometric ${\cal N} = 2$ string compactifications to flat space, the central charge of a BPS particle produced by wrapping a brane on a compact cycle has a geometrical interpretation and can be computed as such \cite{Billo:1999ip}.

In the context of the AdS$_4$ compactifications we are considering, though, there is no corresponding central charge operator in the superalgebra $\mathfrak{osp}(2|4)$. 
It is thus remarkable that we find equation \eqref{eq:ZQ}, which has the same form as a BPS central charge formula.

In \cite{Anninos:2013mfa}, a simple gauged ${\cal N}=2$ supergravity model was considered in an attempt at capturing the low-energy bulk dynamics of AdS$_4$ flux compactifications such as the ${\mathbb CP}^3$ compactification under consideration here. The simple model was chosen because it provides the minimal amount of ingredients needed to lift the multicentered bound states of \cite{Denef:2000nb,Denef:2002ru,Denef:2007vg,Anninos:2011vn} to AdS$_4$ while staying within an ${\cal N}=2$ supergravity framework. Based on the close similarity of this supergravity theory to ungauged ${\cal N}=2$ supergravity, where BPS extremal black holes have mass $m=\frac{1}{\ell_\text{P}} |Z|$, together with the fact that the mass of an extremal black hole becomes insensitive to the AdS curvature when the black hole is much smaller than the AdS radius, a mass formula was postulated for charged probe particles in this model. Interestingly, this formula (equation (2.10) in \cite{Anninos:2013mfa}) coincides exactly with equation \eqref{eq:ZQ} above, upon identifying $y=v$ and $(p^0,p^1,q_1,q_0)=(q_{D6},q_{D4},q_{D2},-q_{D0})$, with the difference in sign for the D0-charge being due to a trivial difference in conventions. 

However, as was pointed out already in \cite{Anninos:2013mfa}, this model misses an important feature of all actual flux compactifications of string theory, namely the fact that a subset of charges obtained by wrapping branes on internal cycles are in fact confined; they necessarily come with fundamental strings attached. In the following section we will review this phenomenon in more detail for AdS$_4 \times {\mathbb CP}^3$.


\subsection{Confining strings}

In the AdS$_4 \times {\mathbb CP}^3$ flux compactification under consideration, wrapped D2- and D6-branes necessarily come with confining fundamental strings attached. From a low-energy effective field theory point of view, this is due to the fact that these have a nonzero magnetic charge with respect to a Higgsed $\mathrm{U}(1)$ \cite{Aharony:2008ug}. The Higgs condensate forces the magnetic flux lines into flux tubes, which act as confining strings.  
This can also be seen directly from the Gauss's law constraint for the D-brane worldvolume gauge field. 
Consider a D6-brane wrapped on ${\mathbb CP}^3$. Varying the action in equation \eqref{eq:Dpaction} with respect to the gauge field $A$ leads to the Born-Infeld generalization of the Maxwell equations of motion 
\begin{equation}
 d *_7 {\cal G}^{(2)} = F^{(6)} + \tfrac{L_s^4}{2} F^{(2)} \wedge {\cal F} \wedge {\cal F} \, .
\end{equation}
The 2-form ${\cal G}^{(2)}$ is the Born-Infeld field strength, the detailed form of which does not matter. The terms on the right hand side arise from varying the second part of the action with respect to $A$ and integrating by parts, using $F^{(n)}=d C^{(n)}$. Integrating both sides of the equation over the ${\mathbb CP}^3$ and using \eqref{eq:fluxquant} gives $0 = -N + k \, \frac{w^2}{2}$, or more generally $-N q_{D6} + k q_{D2} = 0$. The brane configuration is only considered consistent when this equation is satisfied. 
In all other cases, the background fluxes induce an electric tadpole for the worldvolume $\mathrm{U}(1)$ gauge field that must be canceled by other sources. The only other such sources are fundamental string endpoints terminating on the D-brane, for those indeed couple as unit point charges to the gauge field, with sign depending on the orientation of the string. Thus, in general, the Gauss's law constraint forces a nonzero net number $N_F = N_{\rm in} - N_{\rm out}$ of fundamental strings to terminate on the D-brane, given by
\begin{equation} \label{NFformula}
 N_F = N \, q_{D6} - k \, q_{D2} \, . 
\end{equation}
The overall sign is a matter of conventions and we will not try to be precise about it, but the relative sign is important. It disagrees with \cite{Gutierrez:2010bb} but agrees with \cite{Aharony:2008ug}. 


\subsection{Comparison}
\label{sec:comparison}

The brane-confining strings appearing here and the quiver-confining strings appearing in section \ref{sec:gaussconfines} have essentially identical origins and properties, so it seems clear they must be referring to the same things. Nevertheless, this leads to some remarkable predictions, one of them being that, when scanning over parameter space (e.g., over brane worldvolume fluxes), supersymmetric bound states of two wrapped D-branes cease to exist exactly when the confining strings disappear. In the quiver model this is an immediate consequence of sections \ref{sec:gaussconfines} and \ref{sec:groundstates}. 

To make a more refined comparison, let us consider a 2-particle system consisting of a wrapped D6 with worldvolume $\mathrm{U}(1)$ flux $w$ and the corresponding anti-brane, a wrapped D6 with flux $-w$. The D$p$-charges of these two branes are $Q_1 \equiv (1,w,\frac{1}{2} w^2,\frac{1}{6} w^3)$ and $Q_2 \equiv (-1,w,-\frac{1}{2} w^2,\frac{1}{6} w^3)$, respectively. The total charge of the system is $Q_{\rm tot} = Q_1 + Q_2 = (0,2 w,0,\frac{1}{3} w^3)$. From  (\ref{NFformula}) we see these two branes have $N_F = N - k \, \frac{w^2}{2}$ strings stretched between them, and none running off to infinity. 

We want to model this two-particle system as the 2-node quiver of section \ref{sec:two_nodes}. According to (\ref{nminstring}), we have $|N_F| = |\theta|  \Omega = |\theta|/\ell$, with $\theta=\theta_2=-\theta_1$, in units with $2\pi\alpha' \equiv 1$, and recall $\ell$ is the AdS$_4$ radius. This suggests we should identify $\theta$ as follows for this system:\footnote{We picked the + sign because this leads to the existence of supersymmetric bound states with D4-D0 charge in the limit in which the the ${\mathbb CP}^3$ is much larger than the string scale, which is the limit $N/k \to \infty$, according to (\ref{eq:gsellsols}). This is consistent with geometric considerations.}
\begin{equation}
\label{eq:theta2}
  \theta 
 = \ell \, (N - \tfrac{w^2}{2} \, k) \, .
\end{equation}
More generally, this suggests identifying the FI parameter $\theta_v$ of a node corresponding to a wrapped brane with charge $Q_v = (q_{D6,v},q_{D4,v},q_{D2,v},q_{D0,v})$ as 
\begin{equation}
 \theta_v = \ell \, (N\, q_{D6,v} - k\, q_{D2,v}) \, .
\end{equation}
In the case of flat-space quivers, considerations of binding energies, perturbative string spectra and comparison to supergravity bound states leads one to identify the FI parameter $\theta_p$ of a node with charge $Q_p$ as $\theta_p = {\rm Im} \left( e^{-i\alpha} \mu_{Q_p} \right)$, where $\mu_Q = \frac{1}{\ell_\text{P}} Z_Q$ is the complexified mass 
and $Z_Q$ the usual ${\cal N}=2$ central charge, and $\alpha$ is the phase of $Z_{\rm tot}=\sum_p Z_{Q_p}$ \cite{Denef:2002ru}. 
This identification is expected to be valid at least in the regime of small mutual phase differences of the constituents. 
For the charges considered here, when $w$ is not too large, we have $\alpha=0$, so in the flat-space case we would be led to $\theta_p = {\rm Im} \, \mu_p$. 
One may wonder whether a similar identification holds in the AdS case at hand. 

To check this, we compute the imaginary part of the complexified mass defined in \eqref{eq:muQ}, again with $2\pi\alpha' \equiv 1$:
\begin{equation}
 {\rm Im} \, \mu_{Q_2} = \tfrac{1}{6} v^3 \, m_{D0} - \tfrac{w^2}{2} \, v \, m_{D0} =   
 \tfrac{1}{3} \ell N - \tfrac{w^2}{2} \, \ell k \, .
\end{equation}
This is almost the same as equation \eqref{eq:theta2}, but differs from it in the extra factor $\frac{1}{3}$ multiplying the first term. 
To appreciate the physical differences with the flat-space case, and to think about whether we should have expected something like this or not on physical grounds, let us consider the 
case in which the phases of the complexified masses $\mu_1 = \frac{1}{6} (-w + i v)^3 \, m_{D0}$ and $\mu_2 = -\frac{1}{6} (w + i v)^3 \, m_{D0}$ line up. This happens when $w=\frac{1}{\sqrt{3}} v = \sqrt{\frac{2 N}{3 k}}$, since then $\mu_1=\mu_2=\frac{8 v^3}{3 \sqrt{3}}$. In the flat space case this would correspond to $\theta=0$ and marginal stability,  because $|\mu_1|+|\mu_2|=|\mu_{\rm tot}|$. For slightly larger $w$, a stable supersymmetric bound state would no longer exist. In the AdS case at hand, however, when $w=\frac{1}{\sqrt{3}} v$, we have $N_F = \frac{2N}{3}$, and $\theta = \frac{2N\ell}{3} > 0$, hence according to the quiver picture we do still get a stable supersymmetric bound state at $w=\frac{1}{\sqrt{3}} v$. This is not in contradiction with physical expectations. Whereas it is true that $|\mu_1|+|\mu_2|=|\mu_{\rm tot}|$, in AdS this does not mean the bound state is just marginal, since the attached strings  produce an additional binding force; unlike the flat-space case, when we pull the two branes apart the total energy of the system is much larger than the sum of their masses. Because of this, supersymmetric bound states may be expected even for values of $w$ substantially larger than $\frac{1}{\sqrt{3}} v$. This is indeed what we see here, for as long as $\theta$ remains positive.

Eventually though, $\theta$ will become zero, and subsequently flip sign. This happens when $w=v=\sqrt{\frac{2N}{k}}$, since then $N_F = N - \frac{w^2}{2} \, k = 0$. At this point in parameter space no strings are left stretched between the two branes, and according to the quiver picture this is also the point where a supersymmetric bound state ceases to exist. 
Notice that $\mu_1 \propto (1+i)$ and $\mu_2 \propto (1-i)$, so in contrast to the flat-space case, the complexified masses do {\it not} line up at this stability wall; the mass of the bound state is a factor $\sqrt{2}$ smaller than the mass of its constituents.  
Upon further increasing $w$, confining strings reappear, but with opposite orientation. Hence the two particles will still form a bound state, albeit a nonsupersymmetric one. 

To summarize, the sign of $\theta$ determines whether supersymmetric bound states exist or not. In flat space, this is controlled by the relative phases of the complexified masses, since when these phases line up, bound states become energetically marginally stable. In AdS flux vacua, this kinematical argument fails, due to the presence of confining strings and the AdS potential well. It is natural to expect this to facilitate the formation of bound states, including supersymmetric bound states. Thus we expect the sign of $\theta$ to ``lag behind'' with respect to the sign of the relative phases, changing sign only deep in what, in flat space, would be the regime without bound states. This is indeed what we see here.

It would be very interesting to check the bound state predictions of the quiver picture against dual CFT predictions or geometric D-brane considerations, and to compare (\ref{eq:theta2}) to independent identifications of the FI parameter $\theta$, such as perturbative string spectra and brane binding energies, in analogy with the arguments reviewed in \cite{Denef:2002ru}. 
However, this is outside the scope of this paper.


\section{Derivation from dimensional reduction on $\mathbb{R} \times S^3$}
\label{sec:reduction}

In section \ref{sec:Lag} we presented the Lagrangian and supersymmetry transformations for massive quiver matrix mechanics with four supersymmetries. In appendix \ref{app:susy} we sketched how these can be obtained by brute force: add to the flat-space, ``massless'' quiver mechanics Lagrangian of \cite{Denef:2002ru} the most general set of terms compatible with gauge and $\mathrm{SO}(3)$ invariance, keep the target-space metric flat, and fix the coefficients by imposing supersymmetry. However, even just verifying this result is rather tedious.
Hence, it would clearly be desirable to have a more systematic way of obtaining the Lagrangian, with the potential for generalizations in mind.

Indeed, for massless quivers, the arduous brute force route can be avoided by making use of textbook results \cite{Wess:1992cp}. The Lagrangian and supersymmetries can be obtained by simple dimensional reduction of 3+1-dimensional ${\cal N}=1$ quiver gauge theories to $0+1$ dimensions, truncating the Fourier expansion of the fields to the zero-mode sector, i.e.\ to spatially constant fields. In this section we will explain how massive quiver matrix mechanics can be obtained from an analogous reduction of ${\cal N}=1$ quiver gauge theories on ${\mathbb R} \times S^3$. 

Let us first recall why the truncation to the zero-mode sector works in the massless case. The key property that guarantees this is translation invariance of the four-dimensional theory. 
This implies that classically, fields with translationally invariant initial conditions will remain translationally invariant at all times. For indeed, if they did not, there would be a solution with translationally invariant initial conditions at $t=0$ that evolved to a non-translationally-invariant configuration at some later time $t$. We could then translate this solution to obtain a {\it different} solution, evolving from the {\it same} initial conditions. Classically, this is impossible, so we have arrived at a contradiction.\footnote{We assume here that all gauge redundancy has been fixed in a translationally invariant way.} Thus, the truncation to translationally invariant fields is classically consistent.\footnote{Quantum mechanically, things are more complicated. However, since our goal is simply to obtain a reduced Lagrangian that is guaranteed to be supersymmetric by virtue of the symmetries of its parent theory, we do not need to concern ourselves with this here.} A second key fact is that the four-dimensional ${\cal N}=1$ supersymmetry generators commute with the translation generators and hence the supersymmetry transformations likewise map translationally invariant fields to translationally invariant fields. Thus, dimensional reduction by restriction to translationally invariant fields is guaranteed to produce an ${\cal N}=4$ supersymmetric quiver matrix mechanics theory, inheriting its symmetries from the parent four-dimensonal quiver gauge theory. A final element in the story is the string theory interpretation of the dimensionally reduced theory, as describing internally wrapped branes localized in space. In this setup, this follows directly from the interpretation of the original quiver gauge theory as describing internally wrapped branes extending over all of space \cite{Douglas:1996sw,Brunner:1999jq,Douglas:2000ah}, making use of the close relation of D-brane actions wrapping different numbers of flat directions. This, in turn, can be traced to T-duality, as extensively reviewed and applied in \cite{Myers:1999ps}. 

The massive quiver matrix models we are after cannot be obtained by this kind of elementary dimensional reduction of a Poincar\'e invariant four dimensional gauge theory. The main issue is that we want to obtain mass terms for the vector multiplets such as $\frac{\mu}{2} \, {\rm Tr} \, (X^i)^2$, but since the $X^i$ descend from the spatial components of the gauge fields $A_i$ in four dimensions, mass terms of this kind would break gauge invariance in the parent theory. Moreover, we want the time component $A_t$ of the gauge field to remain massless, so the parent theory would have to break Lorentz invariance as well. Although this seems discouraging, there is a version of the idea that does in fact work. In \cite{Kim:2003rza}, Kim, Klose and Plefka pointed out that the massive BMN matrix model can be obtained by dimensionally reducing ${\cal N}=4$ super-Yang-Mills (SYM) theory on ${\mathbb R} \times S^3$, generalizing the way its massless cousin, the BFSS matrix model \cite{Banks:1996vh}, can be obtained by elementary dimensional reduction on flat space. We will use their ideas in combination with the general description of ${\cal N}=1$ theories on ${\mathbb R} \times S^3$ given in \cite{Sen:1985ph,Festuccia:2011ws} to obtain the  massive quiver matrix mechanics of section \ref{sec:Lag} from ${\cal N}=1$ quiver gauge theories in four dimensions. In this construction, the mass deformation scale $\Omega$ is proportional to the inverse radius of the three-sphere. 

\subsection{Reduction on ${\mathbb R} \times S^3$: general idea} \label{sec:genidea}

We will describe the reduction in detail in the sections below, but let us first sketch the general idea. The theory on ${\mathbb R} \times S^3$ is invariant under the spatial isometry group $\mathrm{SO}(4) \simeq \mathrm{SU}(2)_l \times \mathrm{SU}(2)_r$, so, in analogy with the translationally invariant case, we may consistently truncate to, say, $\mathrm{SU}(2)_r$-invariant field configurations. 
By consistent truncation, we mean that a classical solution to the dimensionally reduced theory is also a solution of the full four-dimensional theory.
Indeed, $\mathrm{SU}(2)_r$-invariant initial conditions will evolve into $\mathrm{SU}(2)_r$-invariant field configurations, for if they did not, there would  be multiple classical solutions corresponding to the same initial conditions. Furthermore, in appropriate conventions, the supersymmetries commute with $\mathrm{SU}(2)_r$ rotations (though not with $\mathrm{SU}(2)_l$ rotations) \cite{Sen:1985ph}, hence they map the $\mathrm{SU}(2)_r$-invariant sector into the $\mathrm{SU}(2)_r$-invariant sector. Thus, in complete analogy with the translationally invariant case, the truncated theory inherits all four supersymmetries from its parent theory.

To identify the  field configurations invariant under $\mathrm{SU}(2)_r$, first recall we can parametrize $S^3$ by group elements $u \in \mathrm{SU}(2)$.\footnote{Explicitly, $u={\,\,\,\, z_1 \,\, z_2 \choose -\bar z_2 \,\, \bar z_1}$ with $(z_1,z_2) \in {\mathbb C}^2$ and $|z_1|^2+|z_2|^2 = 1$.} The action of $\mathrm{SU}(2)_l \times \mathrm{SU}(2)_r$ on a scalar field $\phi(u)$ is then given by $\phi(u) \to \phi'(u) = \phi(g_l^{-1} u g_r)$. From this we see that an $\mathrm{SU}(2)_r$ invariant scalar is actually constant on $S^3$; the truncation simply sets
\begin{equation}
 \phi(t,u)=\phi(t) \, .
\end{equation}

For 1-form fields such as the gauge field, things are more subtle. There exist nontrivial $\mathrm{SU}(2)_r$-invariant 1-forms, namely the three right-invariant 1-forms $V^{(i)}$, $i=1,2,3$, defined by 
$du \, u^{-1} = \frac{i}{\rsp} \sum_i V^{(i)} \sigma^i$,
where $\sigma^i$ are the Pauli matrices and $\rsp$ is a normalization factor that we take to be equal to the radius of the sphere, for reasons explained below. Thus, for a gauge field $A$ on ${\mathbb R} \times S^3$ the truncation to the $\mathrm{SU}(2)_r$ singlet sector means it is of the form 
\begin{equation} \label{AtuVdef}
  A(t,u) = A_t(t) \, dt + X^i(t) \, V^{(i)}(u) \, , \qquad du \, u^{-1} = \frac{i}{\rsp} \, V^{(i)} \sigma^i \, .
\end{equation}
Since $d(du \, u^{-1})=du \, u^{-1} \wedge du \, u^{-1}$, we have that $dV^{(i)} = -\frac{1}{\rsp} \, \epsilon^{ijk} V^{(j)} \wedge V^{(k)}$, and so we can write the gauge field strength as 
\begin{equation} \label{eq:FeqV}
F = dA - i \, A \wedge A = D_t X^i \, dt \wedge V^{(i)} - \bigl( \tfrac{i}{2} [X^i,X^j] + \tfrac{1}{\rsp} \epsilon^{ijk} X^k \bigr) \, V^{(i)} \wedge V^{(j)} ,
\end{equation}
where $D_t X^i = \partial_t X^i - i[A,X^i]$. To obtain the reduced theory, we have to substitute this in the Yang-Mills action 
\begin{equation}
 S = - \frac{1}{2g_{\rm YM}^2} \int_{{\mathbb R} \times S^3} {\rm Tr} \, F \wedge * F \, ,
\end{equation} 
and integrate over $S^3$. 
For this we need a metric. The metric on ${\mathbb R} \times S^3$, with $S^3$ the round sphere of radius $\rsp$, can be expressed in terms of the right-invariant 1-forms as follows:
\begin{equation}
 ds^2=-dt^2+\tfrac{1}{2} \rsp^2 \, {\rm tr}\,(i \, du \, u^{-1})^2 = -dt^2 + V^{(i)} V^{(i)} \, .
\end{equation} 
Thus, with the normalization we chose,  the $V^{(i)}$ form an orthonormal frame, and 
\begin{equation}
 *(dt \wedge V^{(i)})=-\tfrac{1}{2} \epsilon^{ijk} \, V^{(j)} \wedge V^{(k)} \, , \qquad
 *(V^{(i)} \wedge V^{(j)}) = \epsilon^{ijk} \, dt \wedge V^{(k)} \, .
\end{equation}
The reduced action is now easily obtained:
\begin{equation}
 S = m \int dt \, {\rm Tr} \, \bigl( \, \tfrac{1}{2}(D_t X^i)^2 
 - \bigl( \tfrac{i}{2} [X^i,X^j] +  \tfrac{1}{\rsp}  \epsilon^{ijk} X^k \bigr)^2 \, \bigr) \, , \qquad m =\frac{2 \pi^2 \rsp^3}{g_{\rm YM}^2} \, .
\end{equation}
The factor $2 \pi^2 \rsp^3$ is the volume of the 3-sphere. 

Comparing to the massive quiver matrix mechanics Lagrangian in equation \eqref{eq:totalLag}, we recognize the mass and Myers terms for $X^i$, with mass deformation parameter 
\begin{equation} \label{OmegaRidentification}
 \Omega = \frac{2}{\rsp} \, .
\end{equation} 
Other terms in the Lagrangian arise from different terms in the four dimensional action. In particular, the coupling of the FI parameter $\theta$ to the gauge field $A_t$ in equation (\ref{eq:Lprm}), causing the quiver confinement described in section \ref{sec:gaussconfines}, descends from a term of the same form in the four dimensional SYM Lagrangian on $\mathbb{R} \times S^3$, whose presence is required by supersymmetry in this curved background \cite{Sen:1985ph,Sohnius:1982fw}. Similarly, the terms linear in time derivatives of the chiral multiplet scalars in equation \eqref{eq:Lprm} are direct descendants of such terms in the four dimensional ${\cal N}=1$ chiral multiplet Lagrangian on $\mathbb{R} \times S^3$, again required by supersymmetry \cite{Sen:1985ph,Festuccia:2011ws}. Also, the peculiar time dependence of the supersymmetry parameter is inherited from the time dependence already present in the parent theory \cite{Festuccia:2011ws}.  

We provide some of the details of the reduction below. We will not spell out all calculations, but will present all the necessary ingredients and present some sample computations. 
The remainder of this section is as follows: in section \ref{sec:susylag} we collect the different terms in the Lagrangian of ${\cal N}=1$ SYM on ${\mathbb R} \times S^3$ coupled to chiral multiplets as well as the supersymmetry transformations of the four dimensional fields; in section \ref{sec:lagrangian_reduction} we perform the dimensional reduction of the Lagrangian; in section \ref{sec:dimredsusy} we discuss the reduction of the supersymmetry transformations; and in section \ref{sec:reduccomparison} we compare our results to other models and mention possible generalizations.

\subsection{Interpretation of the dimensionally reduced model}

Unlike the flat-space case, the interpretation of the $S^3$-reduced gauge theory as describing localized particles in AdS is not immediately clear. While in the flat-space case, space-filling D-branes and localized D-branes are directly related by T-duality, there is no duality map we know of that would transform a D-brane wrapping $S^3$ to a D-branes localized in AdS$_4$. Nevertheless, as we have seen in the previous sections, the reduced theory does have the appropriate physical properties to be consistent with this interpretation. More generally, we show in appendix \ref{app:algebra} that the reduced theory has superalgebra $\mathfrak{su}(2|1)\rtimes \mathfrak{u}(1)_R$, and in appendix \ref{app:osp2bar4} that is precisely the superalgebra expected for a superparticle in an ${\cal N}=2$ AdS$_4$ compactification. This provides an algebraic proof of the consistency of our interpretation. A better physical insight into why this had to be the case would nevertheless be useful. 

Another point which is not manifest from the derivation by dimensional reduction is the conjecture, stated in section \ref{sec:genlagandsusy}, 
that for connected quivers, the only possible mass deformation preserving $\mathrm{SO}(3)$-invariance and ${\cal N}=4$ supersymmetry (with a flat metric on the target space) is the one we wrote down in equation \eqref{eq:totalLag}. In other words, that the mass deformation depends on just one parameter, $\Omega$. 
Within the dimensional reduction scheme, in view of the identification in equation \eqref{OmegaRidentification}, a massive quiver matrix model with more than one mass deformation parameter $\Omega_I$ would have to arise from a compactification of fields living on spheres with different radii $R_I$. There is of course no obstruction to doing so for a quiver with more than one connected component, for example, two nodes with no arrows between the nodes. Each connected component can live on a different sphere in a trivial way, since the corresponding degrees of freedom are completely decoupled from each other. However, vector multiplets associated with two nodes connected by arrows cannot live on different spheres, since in order to write down the four dimensional supersymmetric theory, each of them has to live on the same sphere as the bifundamental chiral multiplets corresponding to those arrows. Of course, this does not prove that ${\cal N}=4$ matrix mechanics with multiple mass parameters does not exist for connected quivers; it merely shows that they cannot be constructed by dimensional reduction along the lines detailed in this section. Nevertheless, it may be viewed as evidence in favor of the conjecture.


\subsection{${\cal N}= 1$ supersymmetric Lagrangians on ${\mathbb R}\times S^3$}

\label{sec:susylag}

Here we will put together the different terms of the action
\beq
\label{actionfour dimensional}
S = \int_{{\mathbb R} \times S^3} d^4 x \, \sqrt{-g} \, \Lcal_{\rst}
\eeq
for ${\cal N}=1$ supersymmetric gauge theories coupled to chiral multiplets, together with its supersymmetries %
\footnote{See appendix \ref{app:reductiondetails} for our index conventions for the spacetime $\rst$.}.
Supersymmetric gauge theories on $\rst$ were first constructed in \cite{Sen:1985ph}.  Modern techniques for putting supersymmetric field theories on curved backgrounds, with explicit results for $\rst$, are discussed in \cite{Festuccia:2011ws}.
In that work, they considered supersymmetric theories on arbitrary curved backgrounds.
The idea is to couple supergravity multiplets to other supersymmetric field theory multiplets, set the gravitino to zero, and then freeze the bosonic fields of the supergravity multiplet by taking $G_\text{N}\to 0$.
Enforcing the vanishing of the gravitino variation results in a set of constraints on the spacetime and auxiliary fields of the supergravity multiplet.
We use their results for the `new minimal' supergravity formulation \cite{Sohnius:1982fw}.

We will begin with the four-dimensional Lagrangian coupled to the `new minimal' supergravity multiplet and then substitute background field values to get a $\Ncal=1$ supersymmetric theory on $\rst$.
The ingredients of our Lagrangian are 
\begin{itemize}
\item The supergravity fields $(e_\mu^i, \zeta_\alpha, A_\mu, a_{\mu\nu})$ with $e_\mu^i$ the veilbein, $\zeta_\alpha$ the gravitino, $A_\mu$ an auxiliary $\mathrm{U}(1)_R$ gauge field, and $a_{\mu\nu}$ an auxiliary abelian field.
Our Lagrangians will actually make use of the dual of the flux of $a_{\mu\nu}$ defined by $V^\mu = \tfrac{1}{4}\epsilon^{\mu\nu\lambda\sigma}\partial_\nu a_{\lambda\sigma}$.
As usual we define $e\equiv \det(e_\mu^i)$.
\item The vector multiplet fields $(B_\mu, \lambda_\alpha, D)$ with $B_\mu$ a $\mathfrak{u}(n)$ valued gauge field, $\lambda_\alpha$ a Weyl fermion, and $D$ an auxiliary field.
The fermion and auxiliary fields live in the adjoint of $\mathrm{U}(N)$.
The coupling constant will be denoted by $\cc$.
\item A set of chiral multiplet fields $(\phi^i, \psi_\alpha^i, F^i)$ with $\phi^i$ a complex scalar, $\psi_\alpha^i$ a Weyl fermion, and $F^i$ a complex auxiliary field.
The multiplet is accompanied with a gauge charge $c_i = \pm 1$ denoting whether it lives in the fundamental ($c_i = +1$) or anti-fundmental ($c_i = -1$) of $\mathrm{U}(N)$.
Each chiral multiplet is charged under the $\mathrm{U}(1)_R$ gauge field $A_\mu$ and the charge is denoted by $q_i$.
Derivatives with respect to the complex scalars will be denoted by subscripts $i,~j,\dots$ and with respect to their conjugates by $\bar{i},~\bar{j},\dots$.
For example,
\beq
\bar{W}_{\bar{i}} = \pderiv{}{\phi^{i\dag}}\bar{W}(\phi^\dag) \ .
\eeq
\end{itemize}
Spinor and gauge indices will almost always be suppressed.

A generic $\Ncal = 1$ theory is characterized by a real scalar function $K(\phi, \phi^\dag)$ (the K\"ahler potential), a holomorphic function $W(\phi)$ (the superpotential), and a holomorphic symmetric tensor $f_{AB}(\phi)$ (the kinetic gauge function).
For the models we are interested in we set $K(\phi, \phi^\dag) = \sum_{i} \phi^{i\dag} \phi^i$, $f_{AB} = \delta_{AB}$, and leave the superpotential arbitrary.
We will comment at the end on how the dimensional reduction generalizes to arbitrary K\"ahler potentials and kinetic gauge functions.

When writing down the Lagrangian coupled to the supergravity backgrounds we implicitly set the gravitino to zero, $\zeta_\alpha\to 0$.
The chiral Lagrangian with the superpotential terms can be grabbed directly from equation (6.5) of \cite{Festuccia:2011ws} %
\footnote{We use a definition of the Ricci scalar that gives positive curvature for $\rst$.}.
\begin{align}
\Lcal_{\text{chiral}} &= \lp -\frac{1}{2}\Rcal - 3V_\mu V^\mu\rp\lp \frac{1}{2}q_i\phi^{i\dag}\phi^i\rp + \lp F^{i\dag}F^i - \hat{\Dcal}_\mu\phi^{i\dag}\hat{\Dcal}^\mu\phi^i\rp \nonumber \\
&\quad + iV^\mu\lp \phi^{i\dag} \hat{\Dcal}_\mu \phi^i - (\hat{\Dcal}_\mu\phi^{i\dag})\phi^i\rp - i\psi^{i\dag}\bar{\sigma}^\mu \hat{\Dcal}_\mu\psi^i  \nonumber \\
\label{eq:lrstchiral}
&\quad + F^i W_i - \frac{1}{2}W_{ij}\psi^i\psi^j + F^{i\dag}\bar{W}_{\bar{i}} - \frac{1}{2}\bar{W}_{\bar{i}\bar{j}}\psi^{i\dag}\psi^{j\dag} \, ,
\end{align}
where
\begin{align}
\hat{\Dcal}_\mu\phi^i &= (\partial_\mu - iq_i A_\mu - ic_i B_\mu)\phi^i, \quad \hat{\Dcal}_\mu\phi^{i\dag} = (\partial_\mu + iq_i A_\mu + ic_i B_\mu)\phi^{i\dag} \\
\hat{\Dcal}_\mu\psi^i &= \lp \nabla_\mu - i(q_i - 1)A_\mu - \frac{i}{2}V_\mu - ic_i B_\mu\rp\psi^i \ .
\end{align}
The vector Lagrangian can be pulled from equation (5.9) of \cite{Sohnius:1982fw}, suitably generalized to the nonabelian case.
It can also be found in \cite{Nishioka:2014zpa}, equation (3.27):%
\begin{align}
\label{eq:sohniusvector}
\Lcal_{\text{vector}} &= \frac{1}{\cc^2} \Tr\lp -\frac{1}{4}F_{\mu\nu}F^{\mu\nu} - i\lambda^\dag\bar{\sigma}^\mu \hat{\Dcal}_\mu\lambda + \frac{1}{2}D^2\rp, 
\end{align}
where
\begin{align}
F_{\mu\nu} &= \partial_\mu B_\nu - \partial_\nu B_\mu - i[B_\mu, B_\nu] \\
\hat{\Dcal}_\mu\lambda &= \nabla_\mu\lambda - i(A_\mu - \tfrac{3}{2}V_\mu)\lambda - i[B_\mu, \lambda] \ .
\end{align}
The remaining interaction terms between a nonabelian vector multiplet and a chiral multiplet in the fundamental of the gauge group with flat Kahler potential can be taken from equation (5.17) of \cite{Sohnius:1982fw}. These are reproduced in equation (3.24) of \cite{Nishioka:2014zpa}.
They are identical to the flat space expressions: 
\begin{equation}
\Lcal_{\text{int}} = c_i[\sqrt{2}i(\phi^{i\dag}\lambda\psi^i - \psi^{i\dag}\lambda^\dag\phi^{i}) + \phi^{i\dag}D\phi^{i}]  \ .
\end{equation}
We are also interested in Fayet-Iliopoulos (FI) terms, which are treated in equation (3.1) of \cite{Sohnius:1982fw}.
This piece of the Lagrangian was also written down in \cite{Sen:1985ph}, but here we show how one can get it directly from supergravity.
\begin{equation}
\Lcal_{\text{FI}} = \theta \, \Tr(D + 2V^\mu B_\mu) \ ,
\end{equation}
with $\theta$ some real number.

Now we have to substitute in the background supergravity fields.
The constraint that the gravitino variation vanish can be satisfied by setting the metric to that of $\rst$, $V_a = A_a = 0$, and $V_0 = 1/\rsp$ \cite{Festuccia:2011ws}.
The scalar curvature is given by $\Rcal = 6 / \rsp^2$.
Lastly there is a choice for the field $A_0$ related to the strength of the presence of the $R$-current in the Lagrangian.
Setting $A_0 = \rho / \rsp$ will correspond to different $R$-frames as described in section \ref{sec:R-symm}.
Since different $R$-frames are physically equivalent in the matrix model, we will simply set $\rho = 0$.
The resulting four-dimensional Lagrangian density we are now working with is 
\begin{align}
\label{eq:lrst}
\Lcal_{\rst} &= \Lcal_{\text{chiral}} + \Lcal_{\text{vector}} + \Lcal_{\text{int}} + \Lcal_{\text{FI}} \\
%
\Lcal_{\text{chiral}} &= - \hat{\Dcal}_\mu\phi^{i\dag}\hat{\Dcal}^\mu\phi^i - i\psi^{i\dag}\bar{\sigma}^\mu \hat{\Dcal}_\mu\psi^i + F^{i\dag}F^i \nonumber \\
&\quad - \frac{i}{\rsp}\lp \phi^{i\dag} \hat{\Dcal}_t \phi^i - (\hat{\Dcal}_t\phi^{i\dag})\phi^i\rp  \nonumber \\
&\quad + F^i W_i - \frac{1}{2}W_{ij}\psi^i\psi^j + F^{i\dag}\bar{W}_{\bar{i}} - \frac{1}{2}\bar{W}_{\bar{i}\bar{j}}\psi^{i\dag}\psi^{j\dag} \\
%
\Lcal_{\text{vector}} &= \frac{1}{\cc^2} \Tr\lp -\frac{1}{4}F_{\mu\nu}F^{\mu\nu} - i\lambda^\dag\bar{\sigma}^\mu \hat{\Dcal}_\mu\lambda + \frac{1}{2}D^2\rp \\
%
\Lcal_{\text{int}} &= c_i[\sqrt{2}i(\phi^{i\dag}\lambda\psi^i - \psi^{i\dag}\lambda^\dag\phi^{i}) + \phi^{i\dag}D\phi^{i}] \\
%
\Lcal_{\text{FI}} &= \theta\Tr\lp D - \frac{2}{\rsp}B_0 \rp \ , 
\end{align}
with the following additional definitions for covariant derivatives
\begin{align}
\hat{\Dcal}_\mu\phi^i &= (\partial_\mu - ic_i B_\mu)\phi^i, \quad \hat{\Dcal}_\mu\phi^{i\dag} = (\partial_\mu + ic_i B_\mu)\phi^{i\dag} \\
\hat{\Dcal}_t\psi^i &= \lp \partial_t - \frac{i}{2\rsp} - ic_i B_0\rp\psi^i, \quad \hat{\Dcal}_a\psi^i = \lp \nabla_a - ic_i B_a\rp\psi^i \\
\hat{\Dcal}_t\lambda &= \partial_t\lambda + \frac{3i}{2\rsp}\lambda - i[B_0, \lambda] \\
\hat{\Dcal}_a\lambda &= \nabla_a\lambda - i[B_a, \lambda] \\
F_{\mu\nu} &= \partial_\mu B_\nu - \partial_\nu B_\mu - i[B_\mu, B_\nu] \ .
\end{align}
The supersymmetry transformations with parameter $\xi$ that leave this Lagrangian invariant are
\begin{align}
\delta B_\mu &= i\lambda^\dag\bar{\sigma}_\mu \xi - i\xi^\dag \bar{\sigma}_\mu \lambda \\
\delta \lambda_\alpha &= (-\sigma^{\mu\nu}F_{\mu\nu} + iD)_\alpha^{\phantom{\alpha}\beta}\xi_\beta \\
\delta D &= \xi^\dag\bar{\sigma}^\mu\hat{\Dcal}_\mu \lambda + (\hat{\Dcal}_\mu\lambda^\dag)\bar{\sigma}^{\mu}\xi \\
\delta \phi^i &= \sqrt{2}\xi\psi^i \\
\delta \psi^i_\alpha &= i\sqrt{2} (\sigma^\mu\xi^\dag)_\alpha \hat{\Dcal}_\mu \phi^i + \sqrt{2} \xi_\alpha F^i\\
\delta F^i &= i\sqrt{2}\xi^\dag\bar{\sigma}^\mu\hat{\Dcal}_\mu \psi^i + 2ic_i\xi^\dag\lambda^\dag \phi^{i} \ .
\end{align}
The supersymmetry parameter $\xi$ satisfies the differential equation (obtained by the vanishing of the gravitino variation)
\beq
0 = \nabla_\mu \xi_\alpha + i(\sigma_{\mu\nu}\xi)_\alpha V^\nu + i(V_\mu - A_\mu)\xi_\alpha \ . 
\eeq
The temporal and spatial components of this equation are given by
\begin{align}
0 &= \partial_t \xi_\alpha + \frac{i}{\rsp}\xi_\alpha \\
0 &= \nabla_a \xi_\alpha - \frac{i}{2\rsp} (\sigma_a\bar{\sigma}_0 \xi)_\alpha \ .
\end{align}
The solution to these equations is given by
\beq
\xi_\alpha = S_\alpha^{\hat{\alpha}+}\xitt_{\hat{\alpha}}(t) = e^{-\frac{i}{\rsp}t}S_\alpha^{\hat{\alpha}+}\xitt_{\hat{\alpha}0} \ ,
\eeq
where $\xitt_{\hat{\alpha}}(t) = \exp(-\tfrac{i}{\rsp}t)\xitt_{\hat{\alpha}0}$, $\xitt_{\hat{\alpha}0}$ is an arbitrary two-component complex vector, and $S_\alpha^{\hat{\alpha}+}$ is defined in appendix \ref{app:reductiondetails}.


\subsection{Dimensional reduction of the Lagrangian}
\label{sec:lagrangian_reduction}

Here we reduce the four dimensional action in equation \eqref{actionfour dimensional} to one dimension by performing the integral over $S^3$. We truncate to modes which are singlets under $\mathrm{SU}(2)_r$. As we have seen already in section \ref{sec:genidea}, for scalars this means we truncate to constant modes, while for vectors it means we truncate to the right-invariant 1-forms. Similarly, for spin 1/2 fields there is a finite number of $\mathrm{SU}(2)_r$ invariant modes. It is possible to go beyond this invariant sector and consider general mode expansions of the fields in $S^3$ spherical harmonics, classified by $\mathrm{SU}(2)_l \times \mathrm{SU}(2)_r$ representations. See appendix \ref{app:reductiondetails} for details and conventions; in particular, table \ref{tab:harmonics} lists all scalar, spinor and vector harmonics. The $\mathrm{SU}(2)_r$-invariant sector corresponds to $\ell=0$ in this table.  

The truncated mode expansions we substitute are thus
\begin{gather}
\label{eq:modeex}
\phi^i(t, \vec{x}) 
= \phitt^i(t) \, , %
\qquad \psi_\alpha^i(t, \vec{x}) = \sum_{\hat{\alpha}=1}^2 \psitt_{\hat{\alpha}}^i(t) \, S_\alpha^{\hat{\alpha}+}(\vec{x}) \, , 
\qquad F^i 
= \Ftt^i(t) \\
B_0(t,\vec{x}) 
= \Btt_0(t), %
\qquad B_a(t,\vec{x}) = \sum_{\hat{a}=1}^3 \Btt_{\hat{a}}(t) \, V_a^{\hat{a}+}(\vec{x}) \\
\lambda_\alpha(t, \vec{x}) = \sum_{\hat{\alpha}=1}^2 \lambdatt_{\hat{\alpha}}(t) \, S_\alpha^{\hat{\alpha}+}(\vec{x}), %
\qquad D 
= \Dtt(t) \ .
\end{gather}
The mode expansion implies we are in Coulomb gauge since $\nabla^a B_a = \Btt_{\hat{a}}\nabla^a V_{a}^{\hat{a}+} = 0$ (using \eqref{eq:divergenceless}) and is relevant to other places in the reduction.

An immediate simplification arises from this expansion.
Any terms in the Lagrangian which contain the complex scalars, auxiliary fields, or time component of the gauge field are constant over the $S^3$ and can immediately be pulled out of the spatial integral.
If these are the only fields present in these terms, then these terms retain their form in the reduced Lagrangian.
In particular, we have the zero-mode action 
\begin{align}
S_0 &= 2\pi^2 \rsp^3\int dt\, \hat{\Dcal}_t\phitt^{i\dag}\hat{\Dcal}_t\phitt^{i} - \frac{i}{\rsp}\lp \phitt^{i\dag} \hat{\Dcal}_t \phitt^i - (\hat{\Dcal}_t\phitt^{i\dag})\phitt^i\rp + \Ftt^{i\dag}\Ftt^i \nonumber \\
&\quad\hphantom{2\pi\rsp^3\int } + \Ftt^i W_i(\phitt) + \Ftt^{i\dag} \bar{W}_{\bar{i}}(\phitt^\dag) + \frac{1}{2}\Tr(\Dtt^2) + \phitt^{i\dag}\Dtt\phitt^i + \theta\Tr\lp \Dtt - \frac{2}{\rsp}\Btt_0 \rp \ .
\end{align}
Next let us consider terms with fermion bilinears.
Due to \eqref{eq:spinors0} and \eqref{eq:spinorep}, fermion bilinears also reduce to the zero mode on the sphere.
In particular,
\begin{align}
&\frac{1}{2\pi^2\rsp^3}\int_{S^3} \epsilon^{\beta\alpha}\psi^i_\alpha\psi^j_\beta = \frac{1}{2\pi^2\rsp^3}\int_{S^3} \epsilon^{\hat{\beta}\hat{\alpha}}\psitt_{\hat{\alpha}}^i \psitt_{\hat{\beta}}^j = \epsilon^{\hat{\beta}\hat{\alpha}}\psitt_{\hat{\alpha}}^i \psitt_{\hat{\beta}}^j \equiv \psitt^i\psitt^j, \\
&\frac{1}{2\pi^2\rsp^3}\int_{S^3} -i\psi^{i\dag}\bar{\sigma}^0\hat{\Dcal}_t\psi^{i} = \frac{1}{2\pi^2\rsp^3}\int_{S^3} i\duind{\delta}{\hat{\alpha}}{\hat{\beta}}\psitt^{i\dag\hat{\alpha}}\hat{\Dcal}_t\psitt_{\hat{\beta}}^{i} = i\duind{\delta}{\hat{\alpha}}{\hat{\beta}}\psitt^{i\dag\hat{\alpha}}\hat{\Dcal}_t\psitt_{\hat{\beta}}^{i} \equiv i\psitt^{i\dag}\hat{\Dcal}_t\psitt^{i} \ .
\end{align}
Thus, we can add the following fermion terms to our reduced action
\begin{align}
S_{1/2} &= 2\pi^2\rsp^3\int dt\, i\psitt^{i\dag}\hat{\Dcal}_t\psitt^{i} - \frac{1}{2}W_{ij}(\phitt)\psitt^i\psitt^j - \frac{1}{2}\bar{W}_{\bar{i}\bar{j}}(\phitt^\dag)\psitt^{i\dag}\psitt^{j\dag} \nonumber \\
&\quad \hphantom{2\pi\rsp^3\int dt\, } + \frac{1}{\cc^2}i\Tr(\lambdatt^{\dag}\hat{\Dcal}_t\lambdatt) + \sqrt{2}i(\phitt^{i\dag}\lambdatt\psitt^i - \psitt^{i\dag}\lambdatt^\dag\phitt^i) \ .
\end{align}
The only terms left to integrate over are those containing spatial derivatives, spatial components of the gauge field, or both.
We do not show the remaining calculations here, although we note that the reduction of the bosonic part of the vector Lagrangian was already given in section \ref{sec:genidea}.

The final steps are to get the Lagrangian of the massive quivers in equation \eqref{eq:totalLag} are few in number.
All chiral multiplets are coupled to two vector multiplets, one with $(c_w)_{i} = +1$ and the other with $(c_v)_{i} = -1$, so that it is in a bifundamental represenation of the gauge group.
This amounts to replacing the single spatial gauge field by a difference of gauge fields $B_a \to B_a^1 - B_a^2$ and contracting the gauge indices appropriately.
The parameters from the four-dimensional theory are related to the quiver by
\beq
\Omega = \frac{2}{\rsp}, \quad m_v = \frac{1}{\cc^2}, \quad \theta_v = -\theta
\eeq
and for the fields
\begin{gather}
\label{eq:replacements}
\Btt_{0} \rightarrow A_v, \quad \Btt_{\hat{a}}\rightarrow -X_v^i, \quad \lambdatt\rightarrow \lambda_v, \quad \Dtt\rightarrow D_v \\
\phitt^i\rightarrow \phi^a, \quad \phitt^{i\dag} \rightarrow \phi^{a\dag},\quad \psitt^{i}\rightarrow \psi^{a}, \quad \psitt^{i\dag}\rightarrow \psi^{a\dag},\quad \Ftt^i\rightarrow F^a, \quad \Ftt^{i\dag} \rightarrow F^{a\dag} \ .
\end{gather}
Note the minus sign on the spatial components of the gauge field.
The reasons for this minus sign are to identify with the flat space quivers when we take $\Omega\to 0$.
This convention is due to how one chooses to implement gauge invariance along with various other conventions for the Pauli matrices and fermion kinetic term.


\subsection{Dimensional reduction of the supersymmetry transformations} 

\label{sec:dimredsusy}

Here we give one example of how to obtain the supersymmetry tranformation for the field $\Btt_{\hat{a}}$.
It is performed by projecting onto the proper spherical harmonic.
One has
\begin{align}
\delta \Btt_{\hat{a}} &= \frac{1}{2\pi^2 \rsp^3} \delta_{\hat{a}\hat{b}}\int_{S^3} V^{\hat{b}+a}(\delta B_a) \\
&= \frac{1}{2\pi^2 \rsp^3} \delta_{\hat{a}\hat{b}}\int_{S^3} (i\lambdatt^{\dag\hat{\alpha}} \xitt_{\hat{\beta}} - i\xitt^{\dag\hat{\alpha}} \lambdatt_{\hat{\beta}})(S_{\hat{\alpha}}^+)^\dag_{\dot{\alpha}}\bar{\sigma}_a^{\dot{\alpha}\beta}S_\beta^{\hat{\beta}+}V^{\hat{b}+a} \\
&= \frac{1}{2\pi^2 \rsp^3} \delta_{\hat{a}\hat{b}}\int_{S^3} (i\lambdatt^\dag_{\hat{\alpha}} \xitt_{\hat{\beta}} - i\xitt^\dag_{\hat{\alpha}} \lambdatt_{\hat{\beta}})(-1)\duind{(\sigma^{\hat{b}})}{\hat{\alpha}}{\hat{\beta}} \\
&= -(i\lambdatt^\dag \sigma_{\hat{a}} \xitt - i\xitt^\dag \sigma_{\hat{a}} \lambdatt) \ .
\end{align}
After making the replacements \eqref{eq:replacements} we have
\beq
\delta X_v^i = i\lambda_v^\dag \sigma^i \xi - i\xi^\dag \sigma^i \lambda_v \ ,
\eeq
which matches \eqref{eq:susyX}.
For scalar fields, one integrates against the mode $Y_{(0)}^{00} = 1$.
For the fermions, one integrates against the mode $(S_{\hat{\alpha}}^+)^\dag_{\dot{\alpha}} \bar{\sigma}_0^{\dot{\alpha}\beta}$.
For example
\beq
\delta\lambda_{\hat{\alpha}} = \frac{1}{2\pi^2 \rsp^3}\int_{S^3} (S_{\hat{\alpha}}^+)^\dag_{\dot{\alpha}} \bar{\sigma}_0^{\dot{\alpha}\beta} (\delta\lambda_\alpha) \ .
\eeq 


\subsection{Comparison to other models and generalizations}
\label{sec:reduccomparison}

The superalgebra for the massive quiver matrix models is $\mathfrak{su}(2|1)\rtimes \mathfrak{u}(1)_R$ in the $\rho = 0$ frame, as noted in appendix \ref{app:algebra}.
If no $R$-symmetry is present, one is restricted to the $\rho = 0$ frame and simply drops the $\mathfrak{u}(1)_R$ factor and the $R$-generator.
Quantum mechanics with $\mathfrak{su}(2|1)$ symmetry has been studied before \cite{Ivanov:2014ima,Ivanov:2013ova}.
The algebra listed in equation (2.1) of \cite{Ivanov:2014ima} is our algebra in the $\rho = 1$ frame.\footnote{One must also make the identifications $m = \Omega$, $F = \frac{1}{2}R$, and $I_{\alpha}^{\beta} = J^k\sigma^{k\beta}_{\alpha}$.}
The supersymmetries are time-independent and consequently the superalgebra is a centrally extended $\mathfrak{su}(2|1)$ where the Hamiltonian plays the role of the central charge.
They also comment on the ability to shift the Hamiltonian by ``$m F$," which in our language is accounted for by different $R$-frames.

They also study different multiplets.
In particular, their $(\mathbf{2},\mathbf{4},\mathbf{2})$ multiplet corresponds to our chiral multiplets, but they have Lagrangian expressions for an arbitrary K\"ahler potential.
After changing to the $\rho = 1$ $R$-frame and removing all instances of the vector multiplets, our chiral Lagrangian agrees with equation (5.11) of \cite{Ivanov:2014ima} with the choice of a flat K\"ahler potential.
On the other hand, our superpotential is arbitrary up to $R$-symmetry constraints.
To generalize our results to an arbitrary K\"ahler potential, we can perform the dimensional reduction on the chiral Lagrangian \eqref{eq:lrstchiral} with abitrary K\"ahler potential, that is, equation (6.5) of \cite{Festuccia:2011ws}, with the background supergravity fields set to the appropriate values.
The calculation is simplified by the fact that the K\"ahler potential, a function of the scalars, will only contain the zero mode scalar harmonic of the $S^3$.
Furthermore, the identities of appendix \ref{app:reductiondetails} show that fermion bilinears (e.g., $\psi\psi$) reduce to the scalar mode on the sphere even before integration.
The result matches equation (5.11) of \cite{Ivanov:2014ima}.
Since the reduction preserves the $\mathrm{SU}(2|1)$ symmetry, the dimensional reduction remains classically consistent for arbitrary K\"ahler potential as well.

In contrast to other $\mathfrak{su}(2|1)$ quantum mechanical systems, we have gauge invariance.
The vector multiplet acts as the $(\mathbf{3},\mathbf{4},\mathbf{1})$ multiplet coupled to the `gauge' multiplet, as described in \cite{Delduc:2006yp}, except that our models take into account the mass deformation parameter $\Omega$.
Without chiral multiplets, the abelian vector multiplet Lagrangian is just that of a supersymmetric harmonic oscillator, as described in section \ref{sec:interpads}.
Nonabelian gauge invariance allows for greater complexity in the form of cubic and quartic terms.
We should be able to obtain even more general Lagrangians using an abitrary K\"ahler potential and kinetic gauge function in the dimensional reduction.
Gauge invariance for an arbitrary K\"ahler potential is possible with constraints relating the K\"ahler potential, kinetic gauge function, the structure constants, and moment maps \cite{Freedman:2012zz}.
The reduction of an $\Ncal=1$ theory on $\rst$ with arbitrary K\"ahler potential, superpotential, and kinetic gauge function, obtained directly from the ``new minimal" supergravity \cite{Sohnius:1982fw}, would yield some very interesting $\mathfrak{su}(2|1)$ quantum mechanical models.



\section{Outlook}
\label{sec:concl}

We have discussed the significance of our results in the introduction and throughout, so we conclude with a number of interesting questions that we leave for future work. One key question, raised in \cite{Anninos:2013mfa}, is whether the molecule-like quantum Coulomb branch bound states of \cite{Denef:2002ru} persist even in the presence of confining strings. Another set of questions is related to generalizations. We imposed SO(3) symmetry and ${\cal N}=4$ supersymmetry, and as in \cite{Denef:2002ru} we restricted to a flat target-space for simplicity. Relaxing any of these will lead to new models. As mentioned before, generalization to arbitrary K\"ahler metrics for the chiral multiplets can be obtained by dimensional reduction on $\mathbb{R} \times S^3$ of the theories in \cite{Festuccia:2011ws}. For the vector multipet scalars more work will be needed; the $(\mathbf{3}, \mathbf{4},\mathbf{1})$ multiplet of \cite{Ivanov:2013ova} and the deformed $S^3$-reductions of \cite{Assel:2015nca} suggest some possible directions.

In further explorations of the model itself, it would be useful to check, for example, the bound state predictions discussed in section \ref{sec:comparison} against independent results. 
In the flat-space case it is known that quiver predictions can be unreliable if there is no point in the physical moduli space where the FI parameters all vanish \cite{Denef:2007vg}; analogous subtleties may arise in the present case. 
Finally, it would be very interesting to apply this model to understand aspects of the microscopic dynamics of black holes, generalizing some of the successes of quiver matrix mechanics in flat space. For example, some of the fuzzy membrane ideas explored in, e.g., \cite{Gaiotto:2004ij,Denef:2007yt,Bena:2013dka,Raeymaekers:2015sba}, might find a more natural home in the current setup, in view of the  existence of the supersymmetric, nonabelian, fuzzy membrane ground states of massive quiver matrix mechanics, in contrast to the flat-space case. 
Additionally, all of the above can be reinterpreted as questions in a holographically dual CFT. 



\acknowledgments

We thank D.~Anninos, T.~Anous, D.~Berenstein, R.~Monten and C.~Toldo for helpful discussions.
CA and FD were supported in part by a grant from the John Templeton Foundation, and are supported in part by the U.S.~Department of Energy (DOE) under DOE grant DE-SC0011941. 
ED is supported by the DOE Office of Science Graduate Fellowship Program (DOE SCGF), made possible in part by the American Recovery and Reinvestment Act of 2009, administered by ORISE-ORAU under contract no. DE-AC05-06OR23100.


\appendix


\section{Definitions and conventions}
\label{app:Def}

We use the following index conventions for the quivers: %
quiver nodes, $v,~w$; %
quiver arrows, $a,~b,~c,\dots$; %
$\mathrm{SO}(3)$ vector indices, $i,~j,~k,\dots$; %
$\mathrm{SO}(3)$ spinor indices, $\alpha,~\beta,~\gamma,\dots$; %
gauge indices, $A,~B,~C,\dots$

The field content of an $\Ncal=4$, $d=1$ quiver matrix model $Q$ is described by a graph with nodes $v\in V$ and directed edges (arrows) $a\in A$, and dimension vector $N=(N_v)_{v\in V}$.
To each node is assigned a linear, or vector, multiplet $(A_v^i,~X_v^i,~{\lambda_v}_\alpha,~D_v)$ with $i=1,~2,~3$.
The field $A_v$ is the gauge field for the group $\mathrm{U}(N_v)$.
The fields $X_v^i$, ${\lambda_v}_\alpha$, $D_v$ live in the adjoint of $\mathrm{U}(N_v)$.
To each edge $a:v\rightarrow w$ is assigned a chiral multiplet $(\phi^a,~\psi^a_\alpha,~F^a)$.
All fields of a chiral multiplet live in the bifundmental $(N_w, \bar{N}_v)$ of $\mathrm{U}(N_w)\times \mathrm{U}(N_v)$.

The quiver model is described by the additional real scalar parameters: %
the mass $m_v$ of the particle represented by each node; %
an FI parameter for each node $\theta_v$; 
the mass deformation parameter or coupling constant $\Omega$. %
The quiver may also have a superpotential $W(\phi^a)$, a holomorphic function of the chiral multiplet scalars $\phi^a$.
If the superpotential satisfies the homogeneity condition $W(\lambda^{q_a} \phi^a) = \lambda^2 W(\phi^a)$, then the quiver model has an $R$-symmetry, with $R$-charge $q_a$ for $\phi^a$ and for the other fields as in table \ref{tab:rcharges}.

Gauge transformations exist for each node individually and act on gauge fields, adjoint fields $g_v\in(X_v^i,~{\lambda_v}_{\alpha},~D_v)$, and bifundamental fields $h^a\in(\phi^a,~\psi^a_\alpha,~F^a)$ as
\begin{equation}
A_v\rightarrow U_v(A_v + i\partial_t)U_v^\dag,%
\quad g_v\rightarrow U_v g_v U_v^\dag,%
\quad h^a\rightarrow U_w h^a U_v^\dag
\end{equation}
with $a:v\to w$.
The covariant derivatives are given by 
\begin{align}
\Dcal_t X_v^i &= \partial_t X_v^i - i[A_v, X_v^i], \\
\Dcal_t\lambda_v &= \partial\lambda_v - i[A_v, \lambda_v], \quad \Dcal_t\lambda_v^\dag = \partial\lambda_v^\dag - i[\lambda_v^\dag, A_v], \\
\Dcal_t \phi^a &= \partial_t \phi^a - iA_w \phi^a + i\phi^a A_v, \quad \Dcal_t \phi^{a\dag} = \partial_t \phi^{a\dag} + i\phi^{a\dag}A_w - iA_v\phi^{a\dag} \\
\Dcal_t\psi^a &= \partial_t \psi^a - iA_w\psi^a + i\psi^a A_v, %
\quad \Dcal_t\psi^{a\dag} = \partial_t \psi^{a\dag} + i\psi^{a\dag} A_w - iA_v \psi^{a\dag}\, .
\end{align}

The fermions are Grassmann valued $\mathrm{SO}(3)$ Weyl spinors.
The complex conjugate of a spinor is denoted with a dagger and has upper indices $\psi^{\dag\alpha} = (\psi_\alpha)^*$
We always explicitly write out the Levi-Civita symbol, i.e., $\psi\psi = \psi^\alpha\psi_\alpha = \epsilon^{\beta\alpha}\psi_\alpha\psi_\beta = -\psi_\alpha\epsilon^{\alpha\beta}\psi_\beta$.
Fermion bilinears are formed by contracting with the Levi-Civita symbol or the Pauli matrices
\begin{align}
\epsilon^{\alpha\beta} &= -\epsilon^{\beta\alpha}, %
\quad \epsilon^{12} = \epsilon_{21} = 1 \\
(\epsilon\psi)^\alpha &= \epsilon^{\alpha\beta}\psi_\beta, \quad (\psi^\dag\epsilon)_{\alpha} = \psi^{\dag\beta}\epsilon_{\beta\alpha} \\
\chi^\dag\sigma^i\psi &= \chi^{\dag\alpha}\sigma^{i\beta}_\alpha\psi_\beta \\
\sigma^1 &= \begin{pmatrix} 0 & 1 \\ 1 & 0 \end{pmatrix} %
\quad \sigma^2 = \begin{pmatrix} 0 & -i \\ i & 0 \end{pmatrix} %
\quad \sigma^3 = \begin{pmatrix} 1 & 0 \\ 0 & -1 \end{pmatrix}
\end{align}

Complex conjugation on a vector multiplet only affects the fermion indices as those fields are represented as Hermitian matrices.
On chiral multiplets, complex conjugation changes the representation from $(N_w, \bar{N}_v)$ of $\mathrm{U}(N_w)\times \mathrm{U}(N_v)$ to the $(\bar{N}_w, N_v)$.
It additionally acts on fermion indices.
Thus, when we consider an arbitrary field $Y^\dag$, it is unambiguous to have the dagger act on the gauge indices of fields as matrix Hermitian conjugation and spinor indices as complex conjugation, should any be present.


\section{Matrix model Lagrangian}
\label{app:Lag}


For convenience, we restate the Lagrangian and supersymmetry transformations given in section \ref{sec:Lag}.
The Lagrangian of massive quiver mechanics with deformation parameter $\Omega$ is given by
\begin{equation}
\label{eq:app_totalLag}
\Lcal_\Omega = \Lcal^0 + \Lcal'_\Omega
\end{equation}
where $\Lcal^0$ is the original, undeformed, flat-space quiver Lagrangian,
given in appendix C of \cite{Denef:2002ru}, and $\Lcal'_\Omega$ is the mass deformation:
\begin{align}
\label{eq:app_Lag1}
\Lcal^0 &= \Lcal_V^0 + \Lcal_{FI}^0 + \Lcal_C^0 + \Lcal_I^0 + \Lcal_W^0 \\
\Lcal_V^0 &= \sum_v  m_v \Tr\lp \tfrac{1}{2} (\Dcal_tX_v^i)^2 + \tfrac{1}{2} D_v^2 + \tfrac{1}{4}[X^i_v, X^j_v]^2  + \tfrac{i}{2}(\lambda^\dag_v \Dcal_t  \lambda_v - (\Dcal_t  \lambda^\dag_v) \lambda_v ) - \lambda^\dag_v\sigma^i[X^i_v,\lambda_v] \rp \nonumber \\
\Lcal_{FI}^0 &= - \sum_v \theta_v \Tr D_v  \nonumber \\
\Lcal_C^0 &= \sum_a \Tr\lp |\Dcal_t \phi^a|^2 + |F^a|^2 + \tfrac{i}{2} (\psi^{a\dag} \Dcal_t \psi^a  - (\Dcal_t \psi^{a\dag}) \psi^a  ) \rp \nonumber \\
\Lcal_W^0 &= \sum_a \Tr\lp \partial_a W F^a + \text{h.c.}\rp + \sum_{ab}\Tr\lp  \tfrac{1}{2} \partial_a \partial_b W \psi^b\epsilon\psi^a + \text{h.c.}\rp \nonumber \\
\Lcal_I^0 &= -\sum_{a:v\rightarrow w} \Tr \bigl( (\phi^{a\dag}X_w^i - X_v^i\phi^{a\dag})(X_w^i\phi^a - \phi^aX_v^i) - \phi^{a\dag}(D_w\phi^a - \phi^aD_v)  \nonumber \\
&\qquad  + \psi^{a\dag}\sigma^i(X^i_w\psi^a - \psi^aX^i_v) %
+ i\sqrt{2}\lp (\phi^{a\dag}\lambda_w - \lambda_v\phi^{a\dag})\epsilon\psi^a %
- \psi^{a\dag}\epsilon(\lambda_w^\dag\phi^a - \phi^a\lambda^\dag_v)\rp \bigr) \nonumber \\
\label{eq:app_Lp}
\Lcal'_\Omega  &= \Lcal_V' + \Lcal_{FI}' + \Lcal_C' \\
\Lcal_V' &= -\sum_v m_v  \Tr\lp \tfrac{1}{2} \, \Omega^2  (X_v^i)^2 + \tfrac{3}{2} \, \Omega \, \lambda^\dag_v\lambda_v + i \, \Omega \, \epsilon_{ijk}X_v^iX_v^jX^k_v \rp \nonumber \\
\Lcal_{FI}' &= \sum_v \theta_v \, \Omega \, \Tr A_v  \nonumber \\
\Lcal_C' &= \sum_a \Tr\lp \tfrac{i}{2} \, \Omega \bigl( \Dcal_t \phi^{a \dagger}  \phi^a - \phi^{a \dagger} \Dcal_t \phi^a \bigr)  - \tfrac{1}{2} \, \Omega\,\psi^{a\dag}\psi^a \rp \, . \nonumber
\end{align}


The action is supersymmetric with respect to the transformations
\begin{align}
\delta A_v &= i\lambda^\dag_v\xi - i\xi^\dag\lambda_v \\
\label{eq:susyX}
\delta X_v^i &= i\lambda^\dag_v\sigma^i\xi - i\xi^\dag\sigma^i\lambda_v \\
\delta \lambda_v &= (\Dcal_tX^i_v)\sigma^i\xi  +\tfrac{1}{2}\epsilon^{ijk}[X^i_v,X^j_v]\sigma^k\xi + iD_v\xi - i \, \Omega \, X^i_v\sigma^i\xi \\
\delta D_v &= -(\Dcal_t\lambda_v^\dag)\xi  - i[X_v^i,\lambda^\dag_v]\sigma^i\xi - \xi^\dag(\Dcal_t\lambda_v)  - i\xi^\dag\sigma^i[X^i_v,\lambda_v] \,
+ \tfrac{3 i}{2} \, \Omega \, \lambda_v^\dagger \xi 
- \tfrac{3 i}{2} \, \Omega \, \xi^\dagger \lambda_v 
\\
\delta \phi^a &= -\sqrt{2} \,\xi\epsilon\psi^a \\
\delta \psi^a &= -i\sqrt{2} \,\xi^\dag\epsilon(\Dcal_t\phi^a) - \sqrt{2} \,\sigma^i(\xi^\dag\epsilon)(X^i_w\phi^a - \phi^aX_v^i) + \sqrt{2} \,\xi F^a  \\
\delta F^a &= -i\sqrt{2} \,\xi^\dag(\Dcal_t\psi^a) + \sqrt{2} \,\xi^\dag\sigma^i(X^i_w\psi^a - \psi^aX_v^i) - 2i\xi^\dag\epsilon(\lambda^\dag_w\phi^a - \phi^a\lambda^\dag_v) + \tfrac{\sqrt{2}}{2} \,  \Omega \, \xi^\dag\psi^a \\
\xi(t) & = e^{- \tfrac{i}{2} \Omega \, t} \, \xi_0 \, .
\end{align}


\section{Checking supersymmetry}
\label{app:susy}

We obtained the Lagrangian and its supersymmetries given in section \ref{sec:genlagandsusy} by dimensional reduction as described in section \ref{sec:reduction}, but also by more direct methods; that is to say, by considering the most general SO(3)-symmetric and gauge invariant mass deformations and inferring, through direct computation, the parameter constraints and additional terms needed to preserve ${\cal N}= 4$ supersymmetry. Our analysis along those lines leads us to conjecture that the Lagrangian in equation (\ref{eq:totalLag}) is the most general ${\cal N}=4$ mass deformation preserving SO(3)-symmetry, assuming a flat target-space metric for both vector and chiral scalars. In this appendix we will demonstrate in some detail how supersymmetry constrains the Lagrangian,  using direct methods. Since the computation is lengthy and not particularly illuminating, we will not reproduce it in its entirety here, focusing instead on clarifying the supersymmetric origin of the terms responsible for some of the most interesting physical features of the model, as explored in section \ref{sec:examples}. 

\subsection{Vector multiplet}

The supersymmetry of the FI Lagrangian $\Lcal_{FI} = \theta \, \Tr \lp \Omega A - D\rp $ is easy to check:
\begin{align*}
\delta_\xi \Lcal_{FI} = \theta \, \Tr\lp \partial_t \lambda^\dagger \xi - \tfrac{i}{2}\Omega\lambda^\dagger \xi \rp %
= \theta \, \partial_t \Tr \lp \lambda^\dagger \xi \rp \, .
\end{align*}
Here the commutator term $-i[X_v^i, \lambda_v^\dag]\sigma^i \xi$ in $\delta D_v$ vanishes once one takes the trace.
The result is a total derivative, and similarly for $\delta_{\xi^\dag} \Lcal_{FI}$.
Hence, the FI action is separately invariant. Notice that the mass deformation requires  the FI parameter to couple to the gauge connection. As discussed in section \ref{sec:two_nodes}, it is this coupling that leads to particle confinement.

To check the supersymmetry of the remainder of the vector multiplet Lagrangian, it is useful to exploit its $R$-symmetry and first redefine the fields as in \eqref{eq:Rtransf} with $\rho=1$, so that the supersymmetry parameter becomes time independent. We can then save ourselves some effort by making use of the fact that we already know the undeformed ($\Omega=0$) model is supersymmetric under time-independent supersymmetry transformations. 
Denote $\Lcal = \Lcal^0 + \Lcal'$, where $\Lcal^0$ is $\Lcal$ evaluated at $\Omega=0$, and similarly $\delta = \delta^0 + \delta'$ with $\delta^0$ the supersymmetry variation at $\Omega=0$ and $\delta'$ the additional variation due to the extra terms  proportional to $\Omega$ in the supersymmetry variations.
Then we already know 
\begin{equation}
\delta^0 \Lcal^0 = 0 + \text{total derivative},
\end{equation}
so what remains to be shown is
\begin{equation}
\delta \Lcal_{\Omega}' + \delta' \Lcal^0 = 0 + \text{total derivative} \, .
\end{equation}
If $\xi$ did depend on time, then we could not separate the variations order by order in $\Omega$ because time derivatives would yield addition factors of $\Omega$.
This is why we consider here a frame in which $\xi$ is time independent. 
Performing the substitutions in equation \eqref{eq:Dtransf} with $\rho=1$ we obtain
\begin{align*}
\delta' A_v &= 0 & \delta' \phi^a &= 0 \\
\delta' X_v^i &= 0 & \delta' \psi^a &= -\tfrac{\sqrt{2}}{2} \, \Omega \, q_a \, \phi^a  \xi^\dag\epsilon  \\
\delta' \lambda_v &= - i \, \Omega \, X^i_v\sigma^i\xi 
&\delta' F^a &= 
-\tfrac{\sqrt{2}}{2} \, \Omega  (q_a-2)  \xi^\dag \psi^a  \\
\delta' D_v &= i \Omega \lambda_v^\dag \xi  - i \Omega \xi^\dag \lambda_v  
\end{align*}
Without loss of generality, we may consider the case of a single vector multiplet, say of mass $m_v=1$. Then after the substitutions in equation \eqref{eq:Dtransf} with $\rho=1$ we have
\begin{align*}
\Lcal_V^0 &= \Tr\lp \, \tfrac{1}{2} (\Dcal_tX^i)^2 + \tfrac{1}{2} D^2 + \tfrac{1}{4}[X^i, X^j]^2  + \tfrac{i}{2}(\lambda^\dag \Dcal_t  \lambda - (\Dcal_t  \lambda^\dag) \lambda ) - \lambda^\dag\sigma^i[X^i,\lambda] \, \rp \\
\Lcal'_V &= \Tr\lp \, - \tfrac{1}{2} \, \Omega^2  (X^i)^2 - \, \Omega \, \lambda^\dag\lambda - i \, \Omega \, \epsilon_{ijk}X^iX^jX^k \, \rp.
\end{align*}
Notice that the coefficient of $\lambda^\dagger \lambda$ is different from the one in \eqref{eq:Lprm}, due to the shift, 
in equation \eqref{eq:Dtransf}, of the $\Dcal_t \lambda$ terms in the full Lagrangian. 
The relevant variations are now relatively easy to compute, using the cyclic property 
of the trace and other algebraic manipulations: 
\begin{align*}
\delta_\xi \Lcal_V' &= \Tr \lp \, -\Omega \lambda^\dagger (\mathcal{D}_t X^i) \sigma^i \xi + \Omega \lambda^\dagger [X^i,X^j] \epsilon_{ijk} \sigma^k \xi - i \Omega D \lambda^\dagger \xi \, \rp , 
\end{align*}
and $\delta'_\xi \Lcal^0_V$ turns out to be minus this, up to a total derivative.
As the sum of these is therefore a total derivative, this establishes the supersymmetry of the vector multiplet sector. 
Notice, in particular, that the mass deformation necessitated the addition of a Myers-like term cubic in the $X^i$. The physical implications of this term were reviewed in section \ref{sec:NAMyers}.

\subsection{Chiral multiplet and vector-chiral interaction terms}

\label{app:chiralsusy}

Checking the supersymmetry of the full Lagrangian is long and elaborate, but it can be organized by, for example, collecting all terms in the variation with the same fields in the same degree, which must cancel among each other. 
As an example, in the full variation of $\mathcal{L}_I$, which we give below in equation \eqref{eq:LI_calc}, one can see that there are terms with the same fields in the same degree among the non-quadratic terms in $\delta_\xi\Lcal^0_C$ and $\delta_\xi \Lcal'_C$.
We give all these terms in equation \eqref{eq:LI_calc}, then collect all terms proportional to $\Omega$. We then show how these collected terms cancel with other terms in the variation, up to a total derivative.
The remaining terms (those that are not in boxes in equation \eqref{eq:LI_calc}) are exactly those one would obtain 
in the variation of the flat-space, $\Omega = 0$ quiver Lagrangian, which are known to cancel \cite{Denef:2002ru}.

For simplicity, we first integrate by parts, so the chiral fermion kinetic term is $i\psi^{a\dag} \Dcal_t \psi^a$ and we start with the variation with respect to $\xi$.
We discuss the variation with respect to $\xi^\dag$ afterwards. 
The computation is implicitly under a trace of color indices; 
the adjoint of $\mathrm{U}(N_v)$ for the vector multiplets and the bi-fundamental for the chiral multiplets.
Blue text color in these portions indicate the variation of the fields.
Terms are groups by colored boxes.
Except for the groups immediately after the variation, the terms in the following lines have the same text color as the boxes they came from.
\begin{align}
\label{eq:LI_calc}
&\delta_\xi\Lcal^\text{(non-quadratic)}_C + \delta_\xi\Lcal_I  \\
&= \tggreen{\Dcal_t \phi^{a\dag}[-i(\var{i\lambda^\dag_w\xi})\phi^{a} + i\phi^a(\var{i\lambda^\dag_v\xi})] + [i\phi^{a\dag}(\var{i\lambda^\dag_w\xi}) - i(\var{i\lambda^\dag_v\xi})\phi^{a\dag}]\Dcal_t \phi^a} \nonumber \\
&\qquad + \tggreen{ i\Omega [i\phi^{a\dag}(\var{i\lambda^\dag_w\xi}) - i(\var{i\lambda^\dag_v\xi})\phi^{a\dag}] \phi^a} \nonumber \\
&\qquad + \tgorange{i\lp\var{-\sqrt{2}((\epsilon\xi)\sigma^i)(\phi^{a\dag}X_w^i - X_v^i\phi^{a\dag})} \Dcal_t \psi^a \rp}
	+ {i\psi^{a\dag}(-i(\var{i\lambda^\dag_w\xi})\psi^{a} + i\psi^a(\var{i\lambda^\dag_v\xi}))} \nonumber \\
&\qquad	+ \tgorange{-\tfrac12 \Omega \var{[-\sqrt{2}\sigma^i (\epsilon\xi)(\phi^{a\dag}X^i_w - X^i_v \phi^{a\dag})]} \psi} \nonumber \\
&\qquad + \lp{\var{\sqrt{2}(\psi^{a\dag}X_w^i - X_v^i\psi^{a\dag})(\sigma^i\xi)}} + {\var{2i(\phi^{a\dag}\lambda_w - \lambda_w\phi^{a\dag})\epsilon\xi}}\rp F^a \nonumber \\
&\qquad {- (\phi^{a\dag}X_w^i - X_v^i\phi^{a\dag})(X_w^i(\var{-\sqrt{2}\xi\epsilon\psi^a}) - (\var{-\sqrt{2}\xi\epsilon\psi^a})X_v^i)} \nonumber \\
&\qquad\qquad {- (\phi^{a\dag}(\var{i\lambda^\dag_w\sigma^i\xi}) - (\var{i\lambda^\dag_v\sigma^i\xi})\phi^{a\dag})(X_w^i\phi^a - \phi^a X_v^i)} \nonumber \\
&\qquad\qquad {- (\phi^{a\dag}X_w^i -  X_v^i\phi^{a\dag})((\var{i\lambda^\dag_w\sigma^i\xi})\phi^a - \phi^a(\var{i\lambda^\dag_v\sigma^i\xi}))} \nonumber \\
&\qquad + \phi^{a\dag}\lp (\tggreen{\var{-(\Dcal_t\lambda^\dag_w)\xi}}  \var{{- i[X_w^i,\lambda^\dag_w\sigma^i\xi]}} 
	+ \tggreen{\var{\tfrac{3i}{2}\Omega \lambda_w^\dag \xi}})\phi^a \rr \nonumber \\
&\qquad\qquad \lr - \phi^a(\tggreen{\var{-(\Dcal_t \lambda^\dag_v)\xi}} \var{{-i[X_v^i,\lambda^\dag_v\sigma^i\xi]}}
	+ \tggreen{\var{\tfrac{3i}{2}\Omega \lambda_v^\dag \xi}} )\rp \\
&\qquad	 + {\phi^{a\dag}\lp D_w(\var{-\sqrt{2}\xi\epsilon\psi^a}) - (\var{-\sqrt{2}\xi\epsilon\psi^a})D_v\rp} \nonumber \\
&\qquad - \lp \tgorange{\var{i\sqrt{2}(\Dcal_t\phi^{a\dag})(\epsilon\xi)}} {\var{-\sqrt{2}(\phi^{a\dag}X_w^i - X_v^i\phi^{a\dag})((\epsilon\xi)\sigma^i)}} \rp \sigma^j(X_w^j\psi^a - \psi^aX_v^j) \nonumber \\
&\qquad\qquad - { \psi^{a\dag}\sigma^i\lp(\var{i\lambda^\dag_w\sigma^i\xi})\psi^a - \psi^a(\var{i\lambda^\dag_v\sigma^i\xi})\rp} 
	- {\psi^{a\dag}\sigma^i\lp X_w^i(\var{\sqrt{2}\xi F^a}) - (\var{\sqrt{2}\xi F^a})X_v^i\rp} \nonumber \\
&\qquad - i\sqrt{2}\lb \phi^{a\dag}\lp \tgorange{\var{(\Dcal_tX_w^i - i\Omega X_w^i)\sigma^i\xi}} + {\var{\tfrac{1}{2}\epsilon^{ijk}[X_w^i, X_w^j]\sigma^k\xi}} + {\var{iD_w\xi}} \rp \rr \nonumber \\
&\qquad\qquad \lr - \lp \tgorange{\var{(\Dcal_tX_v^i - i\Omega X_v^i)\sigma^i\xi}} + {\var{\tfrac{1}{2}\epsilon^{ijk}[X_v^i, X_v^j]\sigma^k\xi}} {\var{iD_v\xi}} \rp\phi^{a\dag}\rb \epsilon\psi^{a} \nonumber \\
&\qquad\qquad - {i\sqrt{2}(\phi^{a\dag}\lambda_w - \lambda_v\phi^{a\dag})\epsilon(\var{\sqrt{2}\xi F^a})} \nonumber \\
&\qquad + i\sqrt{2}\lp \tggreen{\var{i\sqrt{2}(\Dcal_t\phi^{a\dag})(\epsilon\xi)}} - {\var{\sqrt{2}(\phi^{a\dag}X_w^i - X_v^i\phi^{a\dag})((\epsilon\xi)\sigma^i)}} \rp \epsilon(\lambda^\dag_w\phi^{a} - \phi^{a}\lambda^\dag_v) \nonumber \\
&\qquad\qquad + { i\sqrt{2}\psi^{a\dag}\epsilon(\lambda^\dag_w(\var{-\sqrt{2}\xi\epsilon\psi^a}) - (\var{-\sqrt{2}\xi\epsilon\psi^a})\lambda^\dag_v)} \ . \nonumber \\
\tgorange{\phantom{X}} &= \tgpurple{-i\sqrt{2}(\phi^{a\dag}X_w^i - X_v^i\phi^{a\dag})(\epsilon\xi)\sigma^i(\Dcal_t\psi^a)} \nonumber \\
&\qquad + \tgred{\sqrt{2}\tfrac12 \Omega (\epsilon\xi) \phi^{a\dag}\sigma^i(X_w^i\psi^a - \psi^aX_v^i)} \\
&\qquad - \sqrt{2}\lp \tgpurple{i(\Dcal_t\phi^{a\dag})} \rp (\epsilon\xi)\sigma^j(X_w^j\psi^a - \psi^aX_v^j)
 \nonumber \\
&\qquad - i\sqrt{2}\lp \phi^{a\dag}\lp \tgpurple{(\Dcal_tX_w^i)} + \tgcyan{- i\Omega X_w^i}\rp - \lp \tgpurple{(\Dcal_tX_v^i)} + \tgcyan{- i\Omega X_v^i}\rp \phi^{a\dag}\rp (\epsilon\xi)(\sigma^i\psi^{a}) \nonumber \\
&= \ttpurple{-\Dcal_t\lp i\sqrt{2}(\phi^{a\dag}X_w^i - X_v^i\phi^{a\dag})(\epsilon\xi)(\sigma^i\psi^a)\rp} \nonumber \\
&\qquad + \sqrt{2}\lb \ttpurple{i\partial_t(\epsilon\xi)} + (\ttred{\tfrac12 \Omega} - \ttcyan{\Omega}) (\epsilon\xi) \rb\phi^{a\dag}\sigma^i(X_w^i\psi^a - \psi^aX_v^i) \nonumber \\
\label{eq:intvar_orange}
&= \Dcal_t\lp i\sqrt{2}(\phi^{a\dag}X_w^i - X_v^i\phi^{a\dag})(\epsilon\xi)(\sigma^i\psi^a)\rp \\
\tggreen{\phantom{X}} &= \tgorange{\Dcal_t \phi^{a\dag}}((\lambda^\dag_w\xi)\phi^{a} - \phi^a(\lambda^\dag_v\xi)) - (\phi^{a\dag}(\lambda^\dag_w\xi) - (\lambda^\dag_v\xi) \phi^{a\dag})\tgorange{\Dcal_t \phi^a} \nonumber \\
&\qquad \tglime{-i\Omega \phi^{a\dag} ((\lambda_w^\dag \xi)\phi^a - \phi^a (\lambda_v^\dag\xi)) } \\
&\qquad + \phi^{a\dag}\lp (\tgpurple{-(\Dcal_t\lambda^\dag_w)\xi} + \tglime{i\tfrac{3}{2}\Omega\lambda^\dag_w\xi})\phi^a - \phi^a(\tgpurple{-(\Dcal_t\lambda^\dag_v)\xi} + \tglime{i\tfrac{3}{2}\Omega\lambda^\dag_v\xi})\rp \nonumber \\
&\qquad + 2i\lp \tgorange{i(\Dcal_t\phi^{a\dag})}\rp ((\lambda^\dag_w\xi)\phi^{a} - \phi^{a}(\lambda^\dag_v\xi)) \nonumber \\
&= \ttorange{-(\Dcal_t\phi^{a\dag})((\lambda^\dag_w\xi)\phi^{a} - \phi^a(\lambda^\dag_v\xi)) - (\phi^{a\dag}(\lambda^\dag_w\xi) - (\lambda^\dag_v\xi) \phi^{a\dag})(\Dcal_t\phi^a)} \nonumber \\
&\qquad \ttpurple{-\phi^{a\dag}((\Dcal_t\lambda^\dag_w)\xi\phi^{a} - \phi^{a}(\Dcal_t\lambda^\dag_v)\xi)} + \ttlime{\lp -i\Omega + i\tfrac{3}{2}\Omega \rp
\phi^{a\dag}((\lambda^\dag_w\xi)\phi^{a} - \phi^{a}(\lambda^\dag_v\xi))} \nonumber \\
&= \Dcal_t\lp -\phi^{a\dag}((\lambda^\dag_w\xi)\phi^{a} - \phi^a(\lambda^\dag_v\xi))\rp + \phi^{a\dag}((\lambda^\dag_w\partial_t\xi)\phi^{a} - \phi^a(\lambda^\dag_v\partial_t\xi)) \nonumber \\
&\qquad + i\tfrac{1}{2}\Omega\phi^{a\dag}((\lambda^\dag_w\xi)\phi^{a} - \phi^{a}(\lambda^\dag_v\xi)) \nonumber \\
\label{eq:intvar_green}
&= \Dcal_t\lp -\phi^{a\dag}((\lambda^\dag_w\xi)\phi^{a} - \phi^a(\lambda^\dag_v\xi))\rp  \ .
\end{align}
For the variation with respect to $\xi^\dag$, because 
the total Lagrangian is hermitian, we only need to consider those terms which are asymmetric in time derivatives.
The variation of all the other terms will be conjugate to those above.
Thus, we only need to redo the $\tgorange{\phantom{x}}$ variation.
\begin{align}
& \tgpurple{i\psi^{a\dag}\Dcal_t\lp \var{-\sqrt{2}\sigma^i(\xi^\dag\epsilon)(X_w^i\phi^a - \phi^aX_v^i)}\rp} +
	\tgmagenta{-\tfrac12 \Omega \psi^{a\dag} \var{[-\sqrt{2}\sigma^i(\xi^\dag \epsilon)(X_w^i\phi^a - \phi^aX_v^i)]}} \\
& -\psi^{a\dag}\sigma^i\lp X_w^i\lp \tgpurple{\var{-i\sqrt{2}(\xi^\dag\epsilon)(\Dcal_t\phi^a)}}\rp - \lp \tgpurple{\var{-i\sqrt{2}(\xi^\dag\epsilon)(\Dcal_t\phi^a)}} \rp X_v^i\rp \\
& +i\sqrt{2}\psi^{a\dag}\epsilon \lp \lp \tgpurple{\var{(\xi^\dag\sigma^i)(\Dcal_tX_w^i)}} + \tgcyan{\var{i\Omega(\xi^\dag\sigma^i)X_w^i}}\rp \phi^a - \phi^a\lp \tgpurple{\var{(\xi^\dag\sigma^i)(\Dcal_tX_v^i)}} + \tgcyan{\var{i\Omega(\xi^\dag\sigma^i)X_v^i}}\rp \rp \\
&= \sqrt{2}(\psi^{a\dag}\sigma^i)\lb\ttpurple{i\partial_t(\xi^\dag\epsilon)} + \lp \ttcyan{-\Omega} + \ttmagenta{\tfrac12 \Omega}\rp(\xi^\dag\epsilon)\rb (X_w^i\phi^a - \phi^aX_v^i) \\
&= 0 \ .
\end{align}


\section{Quantum mechanics and operator algebra}
\label{app:algebra}

The fields satisfy the (anti-)commutation relations
\begin{align}
[\fduind{X^i_v}{A}{B}, \fduind{P^j_w}{C}{D}] &= i\delta^{ij}\delta_{vw}\duind{\delta}{A}{D}\duind{\delta}{C}{B} \\
\{\fduind{\lambda_{w\alpha}}{A}{B}, \fduind{\lambda_v^{\dag\beta}}{C}{D}\} &= \frac{1}{m_v}\delta_{vw}\duind{\delta}{\alpha}{\beta}\duind{\delta}{A}{D}\duind{\delta}{C}{B} \\
[\fduind{\phi^a}{A}{B}, \fduind{\pi^b}{C}{D}] &= i\delta^{ab}\duind{\delta}{A}{D}\duind{\delta}{C}{B} \\
\{\fduind{\psi^a_{\alpha}}{A}{B}, \fduind{\psi^{b\dag\beta}}{C}{D}\} &= \delta^{ab}\duind{\delta}{\alpha}{\beta}\duind{\delta}{A}{D}\duind{\delta}{C}{B}
\end{align}
The adjoint bosonic fields obey the reality condition $(\fduind{X^i_v}{A}{B})^{\dag} = \fduind{X^i_v}{B}{A}$.

The supersymmetry algebra for the massive quivers is spanned by the Hamiltonian $H$, the generators of the internal $\mathrm{SO}(3)$ symmetry $J^i$, the $\mathrm{U}(1)$ $R$-symmetry generator $R$, and the supercharges $Q_\alpha$ and $Q^{\dag\alpha}$.
They are given in terms of the fields by 
\begin{align}
H_\rho &= \sum_v \Tr\lb \frac{1}{2m_v}P_{vi}^2 + \frac{1}{2}m_v\Omega^2(X_v^{i})^2 - \frac{m_v}{4}[X_v^i, X_v^j]^2 + im_v\Omega\epsilon^{ijk}X_v^i X_v^jX_v^k  \phantom{\sum_a^2}\rr \\
&\qquad \lr + m_v \frac{3}{4}(\lambda^\dag_v\lambda_v - \lambda_v\lambda_v^\dag) + m_v\lambda^\dag_v\sigma^i[X_v^i,\lambda_v] + \frac{1}{2m_v} \lp \theta_v - \sum_{a:*\rightarrow v}\phi^a\phi^{a\dag} + \sum_{b:v\rightarrow *} \phi^{b\dag}\phi^b \rp^2 \rb \\
&\qquad + \sum_{a} \Tr \lp \pi_a^\dag\pi_a + (\tfrac{1}{2}\Omega)^2\phi^{a\dag}\phi^a + \tfrac{i}{2}\Omega(\phi^{a\dag}\pi^{a\dag} - \pi^a\phi^a) + \tfrac{1}{2}\Omega \psi^{a\dag}\psi^a + \rr \\
&\qquad \hphantom{+ \sum_{a} \Tr (} + |X_w^i\phi^a - \phi^aX_v^i|^2 + \psi^{a\dag}\sigma^i (X_w^i\psi^a - \psi^a X_v^i) \\
&\qquad \hphantom{+ \sum_{a} \Tr (}\lr + i\sqrt{2}\lp (\phi^{a\dag}\lambda_w - \lambda_v\phi^{a\dag})\epsilon\psi^{a} - \psi^{a\dag}\epsilon(\lambda^\dag_{w}\phi^a - \phi^a\lambda^\dag_v) \rp \rp \\
&\qquad + \sum_{a}\Tr(|W_a|^2) - \frac{1}{2} \sum_{ab} \Tr(W_{ab}\psi^a\epsilon\psi^b + \Wbb_{\abb \bbb}\psi^{a\dag}\epsilon\psi^{b\dag}) + \frac{\rho}{2}\Omega R \\
M^{ij} &= \sum_v \Tr\lp X_v^iP_{vj} - P_{vi}X_v^j + \frac{1}{2}m_v\epsilon^{ijn}\lambda^\dag_v\sigma^n\lambda_v \rp + \sum_{a}\Tr\lp \frac{1}{2}\epsilon^{ijn}\psi^{a\dag}\sigma^n\psi^a \rp \\
J^i &= \frac{1}{2}\epsilon^{ijk}M^{jk} \\
R &= - \Tr\lp \sum_v m_v \lambda^\dag_v \lambda_v + \sum_a \lb iq_a(\phi^{a\dag}\pi^{a\dag} - \pi^a\phi^a) + (q_a - 1)\psi^{a\dag} \psi^a\rb \rp\\
Q &= \sum_v \Tr\lb \lp iP_{vk} + m_v\Omega X_v^k + \tfrac{i}{2}m_v\epsilon^{ijk}[X_v^i, X_v^j]\rp(\lambda^\dag_v\sigma^k\epsilon) - \theta_v(\lambda^\dag_v\epsilon)\rb \nonumber \\
&\qquad + \sqrt{2}\sum_a \Tr\lb \vphantom{\frac{1}{\sqrt{2}}}\lp -\pi_a - \frac{i}{2}\Omega\phi^{a\dag}\rp \psi^a - i\Wbb_{\abb}(\psi^{a\dag}\epsilon) - i(\phi^{a\dag}X_w^i - X_v^i\phi^{a\dag})(\sigma^i\psi^a) \rr \nonumber \\
&\quad \hspace{3cm} \lr + \frac{1}{\sqrt{2}}\phi^{a\dag}\lp(\lambda^\dag_w\epsilon)\phi^a - \phi^a(\lambda^\dag_v\epsilon)\rp\rb \\
Q^\dag &= \sum_v \Tr\lb (\epsilon\sigma^k\lambda_v)\lp -iP_{vk} + m_v\Omega X_v^k + \frac{i}{2}m_v\epsilon^{ijk}[X_v^i, X_v^j]\rp - \theta_v(\epsilon\lambda_v)\rb \nonumber \\
&\qquad + \sqrt{2}\sum_a \Tr\lb \vphantom{\frac{1}{\sqrt{2}}} \psi^{a\dag}\lp -\pi_a^\dag + \frac{i}{2}\Omega\phi^{a}\rp + iW_{a}(\epsilon\psi^a) + i(\psi^{a\dag}\sigma^i)(X_w^i\phi^{a} - \phi^{a}X_v^i) \rr \nonumber \\
&\quad \hspace{3cm} \lr + \frac{1}{\sqrt{2}}\lp\phi^{a\dag}(\epsilon\lambda_w) - (\epsilon\lambda_v)\phi^{a\dag}\rp\phi^{a}\rb
\end{align}
They satisfy the following algebra
\begin{align}
\label{eq:alg1}
[J^i, H_\rho] &= [J^i, R] = [H_\rho, R] = 0 \\
[J^i, J^j] &= i\epsilon^{ijk}J^k \\
[J^i, Q_\alpha] &= -\frac{1}{2}\sigma^{i\beta}_\alpha Q_\beta, \quad [J^i, Q^{\dag\alpha}] = \frac{1}{2}Q^{\dag\beta} \sigma^{i\alpha}_\beta \\
\{Q_\alpha, Q^{\dag\beta}\} &= 2\delta_\alpha^\beta \lp H_\rho - \frac{\rho}{2} \Omega R\rp + 2\Omega J^i\sigma^{i\beta}_\alpha \\
[R, Q_\alpha] &= -Q_\alpha \quad [R, Q^{\dag\alpha}] = +Q^{\dag\alpha} \\
\label{eq:alg6}
[H_\rho, Q_\alpha] &= \frac{1}{2}(1 - \rho) \Omega Q_\alpha, \quad [H, Q^{\dag\alpha}] = -\frac{1}{2}(1 - \rho) \Omega Q^{\dag\alpha}
\end{align}
All other (anti-)commutators vanish.
For convenience, the algebra listed here accounts for all $R$-frames, characterized by $\rho$ in the presence of an $R$-symmetry, as discussed in section \ref{sec:R-symm}.
In the $\rho=0$ frame, the algebra is easily recognizable as $\mathfrak{su}(2|1)\rtimes \mathfrak{u}(1)_R$. 

There are also the generators of $\mathrm{U}(N_v)$ gauge transformations $\fduind{\hat{G}_v}{A}{B}$, one for each node of the quiver.
We write them in normal ordered form 
\begin{align}
\fduind{\hat{G}_v}{A}{B} &= -i\sum_{i} \lp \fduind{X_v^i}{A}{C}\fduind{P_v^i}{C}{B} - \fduind{X_v^i}{C}{B}\fduind{P_v^i}{A}{C}\rp \nonumber \\
&\qquad - m_v\lp\fduind{\lambda^{\dag\alpha}}{A}{C}\fduind{\lambda_\alpha}{C}{B} - \fduind{\lambda^{\dag\alpha}}{C}{B}\fduind{\lambda_\alpha}{A}{C}\rp \nonumber \\
&\qquad + \sum_{a:*\to v} i\lp\fduind{\phi^{a\dag}}{C}{B}\fduind{\pi^{a\dag}}{A}{C} - \fduind{\phi^{a}}{A}{C}\fduind{\pi^{a}}{C}{B}\rp + \fduind{\psi^{a\dag\alpha}}{C}{B}\fduind{\psi^{a}_\alpha}{A}{C} \nonumber \\ 
\label{eq:gaugeop}
&\qquad + \sum_{b:v\to *} i\lp\fduind{\phi^{b}}{C}{B}\fduind{\pi^{b}}{A}{C} - \fduind{\phi^{b\dag}}{A}{C}\fduind{\pi^{b\dag}}{C}{B}\rp - \fduind{\psi^{b\dag\alpha}}{A}{C}\fduind{\psi^{b}_\alpha}{C}{B} \ .
\end{align}
A state $\ket{\chi}$ is physical if it satisfies all $\mathrm{U}(N_v)$ gauge constraints
\begin{equation} \label{eq:quantumgausslaw}
\fduind{\hat{G}_v}{A}{B}\ket{\chi} = -\theta_v \Omega\, \duind{\delta}{A}{B}\ket{\chi} \ .
\end{equation}
The supersymmetry algebra closes up to gauge terms and thus only closes on physical states.


\section{AdS superalgebras}
\label{app:osp2bar4}

The superalgebra of AdS$_4$, with its fermionic counterpart, and $\Ncal$ supersymmetries is $\mathfrak{osp}(\Ncal|4)$.
The spacetime algebra is $\mathfrak{sp}(4)\simeq \mathfrak{so}(3,2)$ and the $R$-symmetry algebra is $\mathfrak{so}(\Ncal)$.

Let $A,~B,~C,\dots =0,~1,~2,~3,-1$ denote $\mathfrak{so}(3,2)$ indices; %
$\alpha,~\beta,~\gamma,\dots =1,\dots,4$ denote spinor indices of $\mathfrak{so}(3,1)$; %
$\mu,~\nu,~\rho,~\sigma=0,\dots,3$ denote $\mathfrak{so}(3,1)$ vector indices; %
and $I,~J,~K,~L=1,\dots,\Ncal$ denote $\mathfrak{so}(\Ncal)$ vector indices.
Define $\eta^{AB} = \diag(-1,~1,~1,~1,-1)$, gamma matrices satisfying the Clifford algebra $\{\gamma^\mu, \gamma^\nu\} = 2\eta^{\mu\nu}I_4$, and symbols $\Sigma^{AB}$ by
\begin{equation}
\Sigma^{\mu\nu} \equiv \gamma^{\mu\nu} = \frac{1}{2}[\gamma^\mu,\gamma^\nu],\quad %
\Sigma^{\mu,-1} \equiv \gamma^\mu
\end{equation}
The superalgebra is spanned by Lorentz generators $M^{AB} = -M^{BA}$, $R$-symmetry generators $T_{IJ} = -T_{JI}$, and Majorana supercharges $Q_{I\alpha}$.
The supercommutation relations are given by
\begin{align}
[M^{AB}, M^{CD}] &= -i(\eta^{AD}M^{BC} - \eta^{BD}M^{AC} - \eta^{AC}M^{BD} + \eta^{BC}M^{AD}), \\
[T_{IJ}, T_{KL}] &= -i(\delta_{JK}T_{IL} - \delta_{JL}T_{IK} - \delta_{IK}T_{JL} + \delta_{IL}T_{JK}), \\
[T_{IJ}, Q_{K\alpha}] &= i(\delta_{IK}Q_{J\alpha} - \delta_{JK}Q_{I\alpha}), \quad [M^{AB}, Q_{I\alpha}] = \frac{i}{2}\duind{(\Sigma^{AB})}{\alpha}{\beta}Q_{I\beta}, \\
[T_{IJ}, \bar{Q}_{K}^{\alpha}] &= i(\delta_{IK}\bar{Q}_{J}^{\alpha} - \delta_{JK}\bar{Q}_{I}^{\alpha}), \quad [M^{AB}, \bar{Q}_{I}^{\alpha}] = -\frac{i}{2}\bar{Q}_{I}^{\beta}\duind{(\Sigma^{AB})}{\beta}{\alpha}, \\
\{Q_{I\alpha}, \bar{Q}_{J}^{\beta}\} &= i\delta_{IJ}\duind{(\Sigma^{AB})}{\alpha}{\beta}M_{AB} - 2i\duind{\delta}{\alpha}{\beta}T_{IJ}
\end{align}
with the Dirac bar defined as $\bar{\Psi} = i\Psi^\dag\gamma^0$.
All other supercommutators vanish.
We can write this in a more recognizable four-dimensional notation by defining momentum generators $P^\mu \equiv M^{\mu,-1}$.
To connect the algebra with the physical spacetime we introduce the AdS length $\ell$ and rescale the momenta and supercharges as $Q\to\sqrt{\ell}Q$ and $P\to \ell P$.
The supercommutation relations become
\begin{align}
\label{eq:ads_alg1}
[M^{\mu\nu}, M^{\rho\sigma}] &= -i(\eta^{\nu\rho}M^{\mu\sigma} - \eta^{\nu\sigma}M^{\mu\rho} - \eta^{\mu\rho}M^{\nu\sigma} + \eta^{\mu\sigma}M^{\nu\rho}), \\
[M^{\mu\nu}, P^\lambda] &= -i(\eta^{\nu\lambda}P^\mu - \eta^{\mu\lambda}P^\nu), \quad [P^\mu, P^\nu] = -\frac{i}{\ell^2}M^{\mu\nu} \\
[T_{IJ}, T_{KL}] &= -i(\delta_{JK}T_{IL} - \delta_{JL}T_{IK} - \delta_{IK}T_{JL} + \delta_{IL}T_{JK}), \\
[M^{\mu\nu}, Q_{I\alpha}] &= \frac{i}{2}\duind{(\gamma^{\mu\nu})}{\alpha}{\beta}Q_{I\beta},\quad [P^\mu, Q_{I\alpha}] = \frac{i}{2\ell}\duind{(\gamma^{\mu})}{\alpha}{\beta}Q_{I\beta} \\
[M^{\mu\nu}, \bar{Q}_{I}^{\alpha}] &= -\frac{i}{2}\bar{Q}_{I}^{\beta}\duind{(\gamma^{\mu\nu})}{\beta}{\alpha}, \quad [P^\mu, \bar{Q}_{I}^{\alpha}] = -\frac{i}{2\ell}\bar{Q}_{I}^{\beta}\duind{(\gamma^\mu)}{\beta}{\alpha} \\
[T_{IJ}, Q_{K\alpha}] &= i(\delta_{IK}Q_{J\alpha} - \delta_{JK}Q_{I\alpha}), \quad [T_{IJ}, \bar{Q}_{K}^{\alpha}] = i(\delta_{IK}\bar{Q}_{J}^{\alpha} - \delta_{JK}\bar{Q}_{I}^{\alpha})\\
\{Q_{I\alpha}, \bar{Q}_{J}^{\beta}\} &= \delta_{IJ}\lp -2i\duind{(\gamma^\mu)}{\alpha}{\beta}P_\mu + \frac{i}{\ell}\duind{(\gamma^{\mu\nu})}{\alpha}{\beta}M_{\mu\nu}\rp - \frac{2i}{\ell}\duind{\delta}{\alpha}{\beta}T_{IJ}
\end{align}
As we take $\ell \to\infty$ we obtain the flat space super Poincar\'e algebra.
If we additionally scale the $R$-charges $T\to \ell T$, then the algebra contracts to the super Poincar\'e algebra with central charges.

From our interpretation of the massive quiver matrix models as descriptions of wrapped branes in AdS$_4$, we expect the above algebra to be related to the operator algebra we gave in appendix \ref{app:algebra}. 
We describe that relationship here.
Even though the worldline action of a superparticle in AdS$_4$ is symmetric under the full $\mathfrak{osp}(2|4)$, gauge fixing will leave only some of the symmetry manifest, e.g., reparametrization invariance and $\kappa$-symmetry \cite{Berenstein:2002jq,Billo:1999ip}.
For example, the action of a bosonic particle in flat space with coorindates $(t(\tau),~\vec{x}(\tau))$ is given by
\begin{equation}
S = -m\int d\tau \, \sqrt{\dot{t}^2 - \dot{\vec{x}}^2} \ .
\end{equation}
This action is invariant under the full Poincar\'e group.
One can remove the reparameterization invariance by setting $t(\tau) = \tau$.
The action becomes
\begin{equation}
S = -m\int d\tau \, \sqrt{1 - \dot{\vec{x}}^2} \ .
\end{equation}
The remaining symmetry of the action is some reparametrization invariance $\tau\to \tau + \alpha$, translation invariance, and rotation invariance.
The symmetry $\tau\to \tau + \alpha$ is related to time translations since $t(\tau) = \tau$ and thus should be identified with the Hamiltonian.
Indeed, the quantum Hamiltonian of a free particle $\hat{H} = \frac{1}{2m}\hat{p}^i\hat{p}^i$ commutes with the momenta and angular momentum generators.
In quantizing the worldline of the free bosonic particle, we have lost boost invariance.

The same story is true for the superparticle, except we have an additional fermionic $\kappa$-symmetry.
The general action for superparticles in an arbitrary $\Ncal=2$ (asymptotically flat) supergravity background was written down in \cite{Billo:1999ip}.
We do not possess the superparticle action for an AdS supergravity background, but we expect it to have some of the same features.
In particular, gauge fixing the $\kappa$-symmetry should remove half of the fermionic degrees of freedom.
The resulting algebra of the worldline should then only have four supersymmetries.
For the bosonic part of the action, we choose to write the AdS metric in isotropic coordinates
as in equation \eqref{eq:ads_met} and fix the reparametrization invariance $t(\tau) = \tau$.
Once we gauge fix, the only remaining symmetries are time translations and rotations; we lose spatial translations and boosts.
Thus, we seek a subalgebra of $\mathfrak{osp}(2|4)$ that contains time translations, rotations, and four of the supersymmetries.
Supergravity in AdS has a background $\mathrm{U}(1)$ $R$-symmetry gauge field which we expect to couple to the superparticle.
The subalgebra we want should then also have an $R$-generator.

We begin with the superalgebra $\mathfrak{osp}(2|4)$ listed above and form a Dirac supercharge $S_\alpha = (Q_{1\alpha} - iQ_{2\alpha})/\sqrt{2}$.
We additionally define boosts and angular momenta generators $K^i = M^{0i}$, $J^i = \tfrac{1}{2}\epsilon^{ijk}M^{jk}$.
The $R$-charge and anti-commutation relations for the supercharges are
\begin{align}
[T_{12}, S] &= -S, \quad [T_{12}, \bar{S}] = \bar{S} \\
\{S_\alpha, S_\beta\} &= \{\bar{S}^\alpha, \bar{S}^\beta\} = 0 \\
\{S_\alpha, \bar{S}^\beta\} &= -2i\duind{(\gamma^\mu)}{\alpha}{\beta} P_\mu + \frac{i}{\ell}\gamma^{\mu\nu}M_{\mu\nu} + \frac{2}{\ell}\duind{\delta}{\alpha}{\beta}T_{12} \ .
\end{align}
We now switch to a Weyl representation of the gamma matrices
\begin{align}
\sigma^\mu_{\alpha\dot{\alpha}} &= (-I_2, \sigma ^i), \quad \bar{\sigma}^{\mu\dot{\alpha}\alpha} = (-I_2, -\sigma^i), \\
\duind{(\sigma^{\mu\nu})}{\alpha}{\beta} &= \frac{1}{2}\duind{(\sigma^\mu\bar{\sigma}^\nu - \sigma^\nu\bar{\sigma}^\mu)}{\alpha}{\beta}, %
\quad \udind{(\bar{\sigma}^{\mu\nu})}{\dot{\alpha}}{\dot{\beta}} = \frac{1}{2}\udind{(\bar{\sigma}^\mu\sigma^\nu - \bar{\sigma}^\nu\sigma^\mu)}{\dot{\alpha}}{\dot{\beta}} \\
\gamma^\mu &= i\begin{pmatrix} 0 & \sigma^\mu \\ \bar{\sigma}^\mu & 0 \end{pmatrix}, %
\quad \gamma^{\mu\nu} = -\begin{pmatrix} \sigma^{\mu\nu} & 0 \\ 0 & \bar{\sigma}^{\mu\nu} \end{pmatrix} \ .
\end{align}
We parametrize the Dirac spinor in terms of two Weyl spinors as $S^T = (\eta_{-\alpha},~\eta_{+}^{\dag\dot{\alpha}})$, where the $\pm$ subscript denotes the $R$-charge.
Conjugation negates the value of the $R$-charge.
The Dirac conjugate is then given by $\bar{S} = (\eta_{+}^\alpha,~\eta_{-\dot{\alpha}}^\dag)$.
The anti-commutation relations become
\begin{align}
\{\eta_{-\alpha},\eta_{-\beta}\} &= \{\eta_{+}^{\alpha}, \eta_{+}^{\beta}\} = \{\eta_{-\alpha}, \eta_{+}^{\dag\dot{\beta}}\} = 0 \\
\lc \begin{pmatrix}\eta_{-\alpha} \\ \eta_{+}^{\dag\dot{\beta}}\end{pmatrix}, %
\begin{pmatrix}\eta_{+}^{\beta} & \eta_{-\dot{\alpha}}^\dag\end{pmatrix}\rc &= %
\begin{pmatrix} -\tfrac{i}{\ell}\duind{(\sigma^{\mu\nu})}{\alpha}{\beta}M_{\mu\nu} + \tfrac{2}{\ell}\duind{\delta}{\alpha}{\beta} T_{12} & 2\sigma^\mu_{\alpha\dot{\alpha}} P_\mu \\ %
2\bar{\sigma}^{\mu\dot{\beta}\beta} P_\mu & -\tfrac{i}{\ell}\udind{(\bar{\sigma}^{\mu\nu})}{\dot{\beta}}{\dot{\alpha}}M_{\mu\nu} + \tfrac{2}{\ell}\udind{\delta}{\dot{\beta}}{\dot{\alpha}} T_{12} \end{pmatrix} \ .
\end{align}
Lastly, we define the spinors 
\begin{align}
\chi_{1\alpha} &= \frac{1}{\sqrt{2}}(\eta_{-\alpha} + \sigma^0_{\alpha\dot{\beta}}\eta_{+}^{\dag\dot{\beta}}), %
\quad \chi_{2\alpha} = \frac{1}{\sqrt{2}}(\eta_{-\alpha} - \sigma^0_{\alpha\dot{\beta}}\eta_{+}^{\dag\dot{\beta}}) \\
\chi_{1\dot{\alpha}}^\dag &= \frac{1}{\sqrt{2}}(\eta_{-\dot{\alpha}}^\dag + \eta_{+}^{\beta}\sigma^0_{\beta\dot{\alpha}}), %
\quad \chi_{2\dot{\alpha}}^\dag = \frac{1}{\sqrt{2}}(\eta_{-\dot{\alpha}}^\dag - \eta_{+}^{\beta}\sigma^0_{\beta\dot{\alpha}}) 
\end{align}
and the non-vanishing anti-commutation relations are
\begin{align}
\{\chi_{1\alpha}, \chi_{1\dot{\alpha}}^\dag\} &= (\sigma^\mu_{\alpha\dot{\alpha}} + \sigma^0_{\alpha\dot{\beta}}\bar{\sigma}^{\mu\dot{\beta}\beta}\sigma^0_{\beta\dot{\alpha}})P_\mu - \frac{i}{2\ell}(\sigma^0_{\alpha\dot{\beta}}\udind{(\bar{\sigma}^{\mu\nu})}{\dot{\beta}}{\dot{\alpha}} + \duind{(\sigma^{\mu\nu})}{\alpha}{\beta}\sigma^0_{\beta\dot{\alpha}})M_{\mu\nu} + \frac{2}{\ell}\sigma^0_{\alpha\dot{\alpha}} T_{12}\nonumber \\
&= 2\sigma^0_{\alpha\dot{\alpha}}P_0 + \frac{2}{\ell}\sigma^i_{\alpha\dot{\alpha}}J^i + \frac{2}{\ell}\sigma^0_{\alpha\dot{\alpha}}T_{12} \\
\{\chi_{1\alpha}, \chi_{2\dot{\alpha}}^\dag\} &= (\sigma^\mu_{\alpha\dot{\alpha}} - \sigma^0_{\alpha\dot{\beta}}\bar{\sigma}^{\mu\dot{\beta}\beta}\sigma^0_{\beta\dot{\alpha}})P_\mu - \frac{i}{2\ell}(\sigma^0_{\alpha\dot{\beta}}\udind{(\bar{\sigma}^{\mu\nu})}{\dot{\beta}}{\dot{\alpha}} - \duind{(\sigma^{\mu\nu})}{\alpha}{\beta}\sigma^0_{\beta\dot{\alpha}})M_{\mu\nu} \nonumber \\
&= 2\lp P_i + \frac{i}{\ell}K^i\rp \sigma^i_{\alpha\dot{\alpha}} \\
\{\chi_{2\alpha}, \chi_{1\dot{\alpha}}^\dag\} &= 2\lp P_i - \frac{i}{\ell}K^i\rp \sigma^i_{\alpha\dot{\alpha}} \\
\{\chi_{2\alpha}, \chi_{2\dot{\alpha}}^\dag\} &= 2\sigma^0_{\alpha\dot{\alpha}}P_0 - \frac{2}{\ell}\sigma^i_{\alpha\dot{\alpha}}J^i + \frac{2}{\ell}\sigma^0_{\alpha\dot{\alpha}}T_{12} \ .
\end{align}
The Hamiltonian and angular momenta act diagonally on $\chi_1$ and $\chi_{2}$, but the momenta and boosts mix the two spinors
\begin{align}
[P^0, \chi_{1}] &= -\frac{1}{2\ell}\chi_{1}, \quad [P^0, \chi_{2\alpha}] = +\frac{1}{2\ell}\chi_{2\alpha} \\
[J^i, \chi_{1}] &= -\frac{1}{2}\sigma^i\chi_{1}, \quad [J^i, \chi_{2}] = -\frac{1}{2}\sigma^i\chi_{1} \\
[P^i, \chi_{1}] &= \frac{1}{2\ell}\sigma^i\chi_{2}, \quad [P^i, \chi_{2}] = -\frac{1}{2}\sigma^i\chi_{2} \\
[K^i, \chi_{1}] &= -\frac{i}{2}\sigma^i\chi_{2}, \quad [K^i, \chi_{2}] = -\frac{i}{2}\sigma^i\chi_{1} \ .
\end{align}
There are two possible supersymmetry subalgebras to choose from, $\{P^0,~J^i,~T_{12},~\chi_{1},~\chi_{1}^\dag\}$ and $\{P^0,~J^i,~T_{12},~\chi_{2},~\chi_{2}^\dag\}$.
Both sets contain all the generators needed to describe the residual symmetry of the worldline.

The set we identify with that of the massive quivers is the one containing the $\chi_1$ supersymmetries.
To complete the identification with eqs.~\eqref{eq:alg1}-\eqref{eq:alg6}, we should choose $\ell = 1/\Omega$, $\rho=2$, and 
\begin{equation}
-P_0 = P^0 \to H|_{\rho=2} = H_0 + \Omega R, \quad T_{12}\to R, \quad J^i\to J^i, \quad \chi_{1}\to Q, \quad \chi_{1}^\dag\to Q^\dag
\end{equation}
The superalgebras of the gauge-fixed worldline of the superparticle traveling in an AdS$_4$ supergravity background and the massive quiver thus match identically in the $R$-frame $\rho = 2$.


\section{Conventions and identities for the dimensional reduction}
\label{app:reductiondetails}

This appendix outlines the conventions used for the spherical harmonics on $S^3$ along with the coordinate systems used.
Many of these results are taken from \cite{Kim:2003rza}, mostly appendices B and C.


\subsection{Spherical harmonics on $S^3$}

The indices $a,~b,\dots$ will be `curved' indices while the hatted indices $\hat{a},~\hat{b},\dots$ will be `flat'.
The curved indices can be raised with the inverse metric $g^{ab}$.
Flat indices can be raised or lowered with the Kronecker delta $\delta^{\hat{a}\hat{b}}$, $\delta_{\hat{a}\hat{b}}$.

The metric on the sphere $S^3$ of radius $\rsp$ is given by
\beq
ds^2 = \rsp^2(d\theta^2 + \sin^2(\theta)d\psi^2 + \sin^2(\theta)\sin^2(\psi)d\chi^2) \ .
\eeq
The Ricci curvature tensor and scalar are given by
\beq
\Rcal_{ab} = \frac{2}{\rsp^2}g_{ab},\quad \Rcal = g^{ab}\Rcal_{ab} = \frac{6}{\rsp^2}\ .
\eeq
The curved differentials are given by
\beq
dx^1 = d\theta, \quad dx^2 = d\psi, \quad dx^3 = d\chi
\eeq
and the triads $e^{\hat{a}} = e^{\hat{a}}_a dx^a$ by
\beq
e^{\hat{1}} = d\theta, \quad e^{\hat{2}} = \sin(\theta)d\psi, \quad e^{\hat{3}} = \sin(\theta)\sin(\psi) d\chi\ .
\eeq
The matrices $\sigma_a$ are the Pauli matrices pulled back to the sphere
\beq
\sigma_a = \sigma_{\hat{a}}e^{\hat{a}}_a
\eeq
with $\sigma_{\hat{a}}$ the numerical Pauli matrices.

Scalar, spinor, and vector functions on $S^3$ can be expanded in a complete basis $Y_{(0)}^{\ell m}$, $Y_{(1/2)\alpha}^{\ell m \pm}$, $Y_{(1)a}^{\ell m\pm}$.
The representation of the harmonics under the full $\mathrm{SU}(2)_\ell\times \mathrm{SU}(2)_r$ are given in Table \ref{tab:harmonics}.
\begin{table}[t!]
\centering
\begin{tabular}{|c|c|c|c|}
\hline
Spin & Harmonic & Rep of $\mathrm{SU}(2)_\ell\times \mathrm{SU}(2)_r$ & Dimension \\
\hline
0 & $Y_{(0)}^{\ell m}$ & $(\ell+1, \ell+1)$ & $(\ell+1)^2$ \\
\hline
\multirow{2}{*}{$\displaystyle\frac{1}{2}$} & $Y_{(1/2)\alpha}^{\ell m +}$ & $(\ell+2, \ell+1)$ & \multirow{2}{*}{$(\ell+1)(\ell+2)$} \\
& $Y_{(1/2)\alpha}^{\ell m -}$ & $(\ell+1, \ell+2)$ & \\
\hline
\multirow{2}{*}{1} & $Y_{(1)a}^{\ell m +}$ & $(\ell+3, \ell+1)$ & \multirow{2}{*}{$(\ell+1)(\ell+3)$} \\
& $Y_{(1)a}^{\ell m -}$ & $(\ell+1, \ell+3)$ & \\
\hline
\end{tabular}
\caption{The scalar, spinor, and vector harmonics on $S^3$ classified by their representation under $\mathrm{SU}(2)_\ell\times \mathrm{SU}(2)_r$.}
\label{tab:harmonics}
\end{table}
They are normalized such that
\begin{align}
\frac{1}{2\pi^2\rsp^3}\int_{S^3} Y_{(0)}^{\ell m}Y_{(0)}^{\ell' m'} &= \delta^{\ell\ell'}\delta^{mm'} \\
\frac{1}{2\pi^2\rsp^3}\int_{S^3} \lp Y_{(1/2)\alpha}^{\ell m \pm}\rp^* Y_{(1/2)\alpha}^{\ell' m'\pm} &= \delta^{\ell\ell'}\delta^{mm'} \\
\frac{1}{2\pi^2\rsp^3}\int_{S^3} Y_{(1)a}^{\ell m \pm}Y_{(1)}^{\ell' m' \pm a} &= \delta^{\ell\ell'}\delta^{mm'}\ .
\end{align}
They satisfy the eigenvalue equations 
\begin{align}
\label{eq:lbscalar}
\nabla^2Y_{(0)}^{\ell m} &= -\frac{1}{\rsp^2}\ell(\ell+2)Y_{(0)}^{\ell m}, %
& 1\leq m\leq (\ell+1)^2 \\
\label{eq:lbspinor}
\sigma^a_{\alpha\beta}\nabla_a Y_{(1/2)\beta}^{\ell m \pm} &= \pm\frac{i}{\rsp}(\ell+\tfrac{3}{2})Y_{(1/2)\alpha}^{\ell m}, %
& 1\leq m\leq (\ell+1)(\ell+2) \\
\label{eq:lbvector}
\nabla^2Y_{(1)a}^{\ell m\pm} &= -\frac{1}{\rsp^2}[\ell(\ell+4)+2]Y_{(1)a}^{\ell m\pm}, %
& 1\leq m\leq (\ell+1)(\ell+3),
\end{align}
where $\nabla_a$ is the covariant derivative on $S^3$, and $\nabla^2 = \nabla^a\nabla_a$.
For the vector harmonics, we have the additional identities
\begin{align}
\label{eq:transversality}
\nabla^a Y_{(1)a}^{\ell m \pm} &= 0 \\
\label{eq:vectorricci}
\nabla_a\nabla_b Y_{(1)}^{\ell m \pm a} &= [\nabla_a, \nabla_b]Y_{(1)c}^{\ell m \pm a} = \Rcal^a_{\phantom{a}cab}Y_{(1)}^{\ell m \pm c} = \frac{2}{\rsp^2}Y_{(1)b}^{\ell m \pm}\ .
\end{align}

The truncation process of section \ref{sec:reduction} relies on the $\ell=0$ spherical harmonics and so the remainder of this section will mostly outline their properties and identities.
The lowest spherical scalar harmonic is just a constant,
\beq
Y_{(0)}^{00} = 1\ .
\eeq
For convenience we introduce new variables for the lowest spherical spinor and vector harmonics.
\begin{align}
S_\alpha^{\hat{\beta}\pm} &\equiv Y_{(1/2)\alpha}^{0\hat{\beta}\pm}, \quad \hat{\beta}=1,~2 \\
V_a^{\hat{a}\pm} &\equiv Y_{(1)a}^{0\hat{a}\pm}, \quad \hat{a}=1,~2,~3\ .
\end{align}


\subsection{Curved and flat spinor indices}

The curved spinor indices for the spinor harmonics as currently defined are of $\mathrm{SO}(3)$ type.
To make connection to a theory on $\rst$ and we promote them to $\mathrm{SO}(4)\sim \mathrm{SU}(2)_l\times \mathrm{SU}(2)_r$ type.
The conjugates now come with dotted indices and we can use the four dimensional Pauli matrices with greek indices ($\mu$, $\nu$, \dots) ranging from 0 to 3.
Our conventions for the Pauli matrices are those used by Wess and Bagger \cite{Wess:1992cp}
\beq
\sigma^\mu = (-I_2, \sigma^i), \quad \bar{\sigma}^\mu = (-I_2, -\sigma^i)\ .
\eeq
The hatted spinor indices become flat indices for a representation of $\mathrm{SU}(2)_l\times \mathrm{SU}(2)_r$.
We will not choose to raise or lower flat spinor indices with the Levi-Civita tensor.
Instead, we will lower them automatically upon conjugation.
That is
\beq
(S_{\hat{\alpha}}^{\pm})^\dag_{\dot{\alpha}} \equiv (S^{\hat{\alpha}\pm}_\alpha)^*\ .
\eeq
A conjugate spinor can be expanded, for example, as
\beq
\psi^{\dag}_{\dot{\alpha}} = (S_{\hat{\alpha}}^+)^{\dag}_{\dot{\alpha}}\psitt^{\dag\hat{\alpha}}\ .
\eeq
After the reduction, we will be left with the $\mathrm{SU}(2)_\ell$ symbols $\duind{\delta}{\hat{\alpha}}{\hat{\beta}}$ and $\duind{(\sigma^{\hat{a}})}{\hat{\alpha}}{\hat{\beta}}$.

Flat spinor indices are contracted according the NW-SE convention.
If $\psi$ and $\chi$ are flat spinors, we have
\begin{align}
\psi\chi &= \psi^{\hat{\beta}}\chi_{\hat{\beta}} = \epsilon^{\hat{\beta}\hat{\alpha}}\psi_{\hat{\alpha}}\chi_{\hat{\beta}} \\
\psi^\dag\chi^\dag &= \psi_{\hat{\beta}}^\dag\chi^{\dag\hat{\beta}} = \epsilon_{\hat{\beta}\hat{\alpha}}\psi^{\dag\hat{\alpha}}\chi^{\dag\hat{\beta}}
\end{align}
The quiver matrix models are written in terms of the flat spinor indices, but the hats have been removed.
Further, the Levi-Civita symbol is explicitly written between spinor variables and accounts for the appearance of various minus signs.


\subsection{Killing equations and explicit representations}

Using equation \eqref{eq:lbspinor}, the spinors $S$ satisfy the Killing spinor equation
\beq
\label{eq:killingspinor}
\nabla_a S^{\hat{\alpha}\pm} = \pm \frac{i}{2\rsp}\sigma_a\bar{\sigma}_0 S^{\hat{\alpha}\pm}\ .
\eeq
Explicitly the Killing spinors are given by
\begin{align}
S^{\hat{\alpha}\pm} &= \exp\lp \pm \frac{i}{2}\theta \sigma_{\hat{1}}\rp \exp\lp \frac{i}{2}\psi\sigma_{\hat{3}}\rp \exp\lp \frac{i}{2}\chi\sigma_{\hat{1}}\rp S_0^{\hat{\alpha}\pm} \\
S_0^{\hat{1}\pm} &= (1, 0)^T, \quad S_0^{\hat{2}\pm} = (0,1)^T\ .
\end{align}
The Killing vectors on $S^3$ are given by the lowest mode vector spherical harmonic and are related to the spinors $S^{\hat{\alpha}\pm}$ via the Pauli matrices
\beq
\label{eq:vpexp}
(S_{\hat{\alpha}}^\pm)^\dag\sigma_a S^{\hat{\beta}\pm} = \duind{(\sigma_{\hat{a}})}{\hat{\alpha}}{\hat{\beta}}V_a^{\hat{a}\pm}\ .
\eeq
Inverting this relation we have
\beq
\label{eq:vpexpinvert}
V_a^{\hat{a}\pm} = \frac{1}{2}\Tr(\sigma^{\hat{a}}S^{\pm\dag} \sigma_a S^\pm)\ .
\eeq
It is can be shown using \eqref{eq:killingspinor} that the $V_a^{\hat{a}\pm}$ satisfy the Killing vector equations
\beq
\nabla_a V_b^{\hat{a}\pm} + \nabla_b V_a^{\hat{a}\pm} = 0\ .
\eeq
They are given explicitly by
\begin{align}
V_a^{\hat{1}\pm}dx^a &= \cos(\psi)e^{\hat{1}} - \cos(\theta)\sin(\psi) e^{\hat{2}} \pm \sin(\theta)\sin(\psi) e^{\hat{3}} \\
V_a^{\hat{2}\pm}dx^a &= \sin(\psi)\cos(\chi)e^{\hat{1}} + (\cos(\theta)\cos(\psi)\cos(\chi) \mp \sin(\theta)\sin(\chi))e^{\hat{2}} \nonumber \\
&\qquad - (\cos(\theta)\sin(\chi) \pm \sin(\theta)\cos(\psi)\cos(\chi))e^{\hat{3}} \\
V_a^{\hat{3}\pm} dx^a &= \sin(\psi)\sin(\chi)e^{\hat{1}} + (\cos(\theta)\cos(\psi)\sin(\chi) \pm \sin(\theta)\cos(\chi))e^{\hat{2}} \nonumber \\
&\qquad + (\cos(\theta)\cos(\chi) \mp \sin(\theta)\cos(\psi)\sin(\chi))e^{\hat{3}}\ .
\end{align}


\subsection{Spinor and vector identities}

The solutions of \eqref{eq:killingspinor} are orthonormalized and form a complete basis
\begin{align}
\label{eq:spinors0}
(S_{\hat{\alpha}}^\pm)^\dag_{\dot{\alpha}}\bar{\sigma}_0^{\dot{\alpha}\alpha} S^{\hat{\beta}\pm}_{\alpha} &= %
(S_{\hat{\alpha}}^\pm)^\dag\bar{\sigma}_0 S^{\hat{\beta}\pm} = \duind{\delta}{\hat{\alpha}}{\hat{\beta}} \\
\sum_{\hat{\alpha}=1}^2 (S_\alpha^{\hat{\alpha}\pm})(S_{\hat{\alpha}}^\pm)^\dag_{\dot{\alpha}} &= \sigma_{0\alpha\dot{\alpha}}\ .
\end{align}
They also can be chosen to transform the epsilon symbol in curved space to that of flat space
\beq
\label{eq:spinorep}
\epsilon^{\alpha\beta}S_\alpha^{\hat{\alpha}\pm}S_\beta^{\hat{\beta}\pm} = \epsilon^{\hat{\alpha}\hat{\beta}}\ .
\eeq
Conjugating this equation we have
\beq
\epsilon^{\dot{\alpha}\dot{\beta}}(S_{\hat{\alpha}}^\pm)^{\dag}_{\dot{\alpha}}(S_{\hat{\beta}}^\pm)^{\dag}_{\dot{\beta}} = \epsilon^{\hat{\alpha}\hat{\beta}} = \epsilon_{\hat{\beta}\hat{\alpha}}\ .
\eeq
This property is very important because it guarantees that spinor bilinears do not produce higher mode spherical harmonics.
Contracting each side with a Killing spinor relates the two kinds of spinors.
\beq
\bar{\sigma}_{0}^{\dot{\alpha}\alpha}\epsilon_{\hat{\beta}\hat{\alpha}}S^{\hat{\alpha}\pm}_{\alpha} %
=\bar{\sigma}_{0}^{\dot{\alpha}\alpha} \epsilon^{\dot{\gamma}\dot{\beta}}(S_{\hat{\alpha}}^\pm)^{\dag}_{\dot{\gamma}}(S_{\hat{\beta}}^\pm)^{\dag}_{\dot{\beta}}S^{\hat{\alpha}\pm}_{\alpha} %
= \bar{\sigma}_{0}^{\dot{\alpha}\alpha}\sigma_{0\alpha\dot{\gamma}}\epsilon^{\dot{\gamma}\dot{\beta}}(S_{\hat{\beta}}^\pm)^{\dag}_{\dot{\beta}}
= \epsilon^{\dot{\alpha}\dot{\beta}}(S_{\hat{\beta}}^\pm)^{\dag}_{\dot{\beta}}\ .
\eeq
Contracting with another spinor we have
\beq
\epsilon_{\hat{\alpha}\hat{\beta}}S^{\hat{\alpha}\pm}_{\alpha}S^{\hat{\beta}\pm}_{\beta} %
= -\sigma_{0\alpha\dot{\alpha}}\epsilon^{\dot{\alpha}\dot{\beta}}(S_{\hat{\beta}}^\pm)^{\dag}_{\dot{\beta}}S^{\hat{\beta}\pm}_{\beta} %
= -\sigma_{0\alpha\dot{\alpha}}\sigma_{0\beta\dot{\beta}}\epsilon^{\dot{\alpha}\dot{\beta}} %
= -\epsilon_{\beta\alpha} %
= \epsilon_{\alpha\beta}\ .
\eeq
Conjugating we have
\beq
\epsilon^{\hat{\alpha}\hat{\beta}}(S_{\hat{\alpha}}^\pm)^{\dag}_{\dot{\alpha}}(S_{\hat{\beta}}^\pm)^{\dag}_{\dot{\beta}} %
= \epsilon_{\dot{\beta}\dot{\alpha}}\ .
\eeq

As a consequence of the orthonormality of the Killing spinors, the Killing vectors are also orthonormal
\beq
\label{eq:vhortho}
V_a^{\hat{a}\pm}V^{\hat{b}\pm a} = \delta^{\hat{a}\hat{b}}\ .
\eeq
Using \eqref{eq:vpexpinvert} we can prove
\begin{align}
V_a^{\hat{a}\pm}V_b^{\hat{a}\pm} &= \frac{1}{4}\duind{(\sigma^{\hat{a}})}{\hat{\alpha}}{\hat{\beta}}(S^{\pm\dag}_{\hat{\beta}} (-1)\bar{\sigma}_a S^{\hat{\alpha}\pm})\duind{(\sigma^{\hat{a}})}{\hat{\gamma}}{\hat{\delta}}(S^{\pm\dag}_{\hat{\delta}}(-1)\bar{\sigma}_b S^{\hat{\gamma}\pm}) \\
&= \frac{1}{4} (\duind{\delta}{\hat{\alpha}}{\hat{\beta}}\duind{\delta}{\hat{\gamma}}{\hat{\delta}} + 2\epsilon_{\hat{\alpha}\gamma}\epsilon^{\hat{\beta}\hat{\delta}})((S^{\pm}_{\hat{\beta}})^{\dag}_{\dot{\beta}} \bar{\sigma}_a^{\phantom{a}\dot{\beta}\alpha} S^{\hat{\alpha}\pm}_\alpha)((S^{\pm}_{\hat{\delta}})^{\dag}_{\dot{\delta}} \bar{\sigma}_b^{\phantom{b}\dot{\delta}\gamma} S^{\hat{\gamma}\pm}_\gamma) \\
&= \frac{1}{2}\bar{\sigma}_a^{\phantom{a}\dot{\beta}\alpha}\bar{\sigma}_b^{\phantom{b}\dot{\delta}\gamma}\epsilon_{\dot{\delta}\dot{\beta}}\epsilon_{\alpha\gamma} \\
&= \frac{1}{2}\bar{\sigma}_a^{\phantom{a}\dot{\beta}\alpha}\epsilon^{\dot{\delta}\dot{\alpha}}\sigma_{b\alpha\dot{\alpha}}\epsilon_{\dot{\delta}\dot{\beta}} \\
&= -\frac{1}{2}\Tr(\bar{\sigma}_a\sigma_b) \\
&= -\frac{1}{2}(-2g_{ab}) \\
&= g_{ab}\ .
\end{align}

As a consequence of \eqref{eq:vpexp} and \eqref{eq:vhortho} we have
\beq
\label{eq:ssv}
(S_{\hat{\alpha}}^\pm)^\dag \sigma_a S^{\hat{\beta}\pm}V^{\hat{a}\pm a} = \duind{(\sigma_{\hat{b}})}{\hat{\alpha}}{\hat{\beta}}V_a^{\hat{b}\pm}V^{\hat{a}\pm a} = \duind{(\sigma^{\hat{a}})}{\hat{\alpha}}{\hat{\beta}}\ .
\eeq
Contracting each side with a Killing spinor we have the relation
\beq
\sigma_a S^{\hat{\beta}\pm}V^{\hat{a}\pm a} = (\bar{\sigma}_0S^{\hat{\alpha}+})\duind{(\sigma^{\hat{a}})}{\hat{\alpha}}{\hat{\beta}}\ .
\eeq
As a result of equation \eqref{eq:transversality} the Killing vectors are divergenceless
\beq
\label{eq:divergenceless}
\nabla^a V_a^{\hat{a}\pm} = 0\ .
\eeq
As a result of equation \eqref{eq:lbvector} we have
\beq
\label{eq:lpvector}
\nabla^2 V_a^{\hat{a}\pm} = -\frac{2}{\rsp^2}V_a^{\hat{a}\pm}\ .
\eeq
Using \eqref{eq:vpexp} and the Killing spinor equation, one can show
\begin{align}
\nabla_a V_b^{\hat{a}\pm} &= \pm \frac{1}{\rsp}\epsilon_{abc}V^{\hat{a}\pm c}\ .
\end{align}
As a result of this we have
\beq
\label{eq:vectortriple}
V^{\hat{a}\pm a}V^{\hat{b}\pm b}\nabla_a V_b^{\hat{c}\pm} = \pm \frac{1}{\rsp} \epsilon_{abc}V^{\hat{a}\pm a}V^{\hat{b}\pm b}V^{\hat{c}\pm c} = \pm \frac{1}{\rsp}\det(V)\epsilon^{\hat{a}\hat{b}\hat{c}} = \pm \frac{1}{\rsp}\epsilon^{\hat{a}\hat{b}\hat{c}} \ .
\eeq
Contracting each side with a vector harmonic and using the Killing vector equation we have
\beq
V^{\hat{b}\pm b}\nabla_b V_a^{\hat{a}\pm} = \pm\frac{1}{\rsp}\epsilon^{\hat{a}\hat{b}\hat{c}}V_a^{\hat{c}\pm}\ .
\eeq



\begin{thebibliography}{10}

\bibitem{Douglas:1996sw}
M.~R. Douglas and G.~W. Moore, {\it {D-branes, quivers, and ALE instantons}},
  \href{http://arxiv.org/abs/hep-th/9603167}{{\tt hep-th/9603167}}.

\bibitem{Douglas:1996yp}
M.~R. Douglas, D.~N. Kabat, P.~Pouliot, and S.~H. Shenker, {\it {D-branes and
  short distances in string theory}},  {\em Nucl. Phys.} {\bf B485} (1997)
  85--127, [\href{http://arxiv.org/abs/hep-th/9608024}{{\tt hep-th/9608024}}].

\bibitem{Douglas:2000qw}
M.~R. Douglas, B.~Fiol, and C.~Romelsberger, {\it {The Spectrum of BPS branes
  on a noncompact Calabi-Yau}},  {\em JHEP} {\bf 09} (2005) 057,
  [\href{http://arxiv.org/abs/hep-th/0003263}{{\tt hep-th/0003263}}].

\bibitem{Denef:2002ru}
F.~Denef, {\it {Quantum quivers and Hall / hole halos}},  {\em JHEP} {\bf 0210}
  (2002) 023, [\href{http://arxiv.org/abs/hep-th/0206072}{{\tt
  hep-th/0206072}}].

\bibitem{Watamura:1983hj}
S.~Watamura, {\it {Spontaneous Compactification and CP$_N$: $\mathrm{SU}(3)
  \times \mathrm{SU}(2) \times \mathrm{U}(1), \sin^2 \theta_\text{W}, g_3 /
  g_2$ and $\mathrm{SU}(3)$-Triplet Chiral Fermions in Four-dimensions}},  {\em
  Phys. Lett.} {\bf B136} (1984) 245.

\bibitem{Nilsson:1984bj}
B.~E.~W. Nilsson and C.~N. Pope, {\it {Hopf Fibration of Eleven-dimensional
  Supergravity}},  {\em Class. Quant. Grav.} {\bf 1} (1984) 499.

\bibitem{Aharony:2008ug}
O.~Aharony, O.~Bergman, D.~L. Jafferis, and J.~Maldacena, {\it {N=6
  superconformal Chern-Simons-matter theories, M2-branes and their gravity
  duals}},  {\em JHEP} {\bf 0810} (2008) 091,
  [\href{http://arxiv.org/abs/0806.1218}{{\tt arXiv:0806.1218}}].

\bibitem{Berenstein:2002jq}
D.~E. Berenstein, J.~M. Maldacena, and H.~S. Nastase, {\it {Strings in flat
  space and pp waves from $\mathcal{N}=4$ super Yang-Mills}},  {\em JHEP} {\bf
  0204} (2002) 013, [\href{http://arxiv.org/abs/hep-th/0202021}{{\tt
  hep-th/0202021}}].

\bibitem{Banks:1996vh}
T.~Banks, W.~Fischler, S.~Shenker, and L.~Susskind, {\it {M theory as a matrix
  model: A Conjecture}},  {\em Phys.Rev.} {\bf D55} (1997) 5112--5128,
  [\href{http://arxiv.org/abs/hep-th/9610043}{{\tt hep-th/9610043}}].

\bibitem{Kim:2003rza}
N.~Kim, T.~Klose, and J.~Plefka, {\it {Plane wave matrix theory from
  $\mathcal{N}=4$ super Yang-Mills on $\mathbb{R} \times S^3$}},  {\em
  Nucl.Phys.} {\bf B671} (2003) 359--382,
  [\href{http://arxiv.org/abs/hep-th/0306054}{{\tt hep-th/0306054}}].

\bibitem{Duff:1986hr}
M.~Duff, B.~Nilsson, and C.~Pope, {\it {Kaluza-Klein Supergravity}},  {\em
  Phys.Rept.} {\bf 130} (1986) 1--142.

\bibitem{Grana:2005jc}
M.~Grana, {\it {Flux compactifications in string theory: A Comprehensive
  review}},  {\em Phys. Rept.} {\bf 423} (2006) 91--158,
  [\href{http://arxiv.org/abs/hep-th/0509003}{{\tt hep-th/0509003}}].

\bibitem{Douglas:2006es}
M.~R. Douglas and S.~Kachru, {\it {Flux compactification}},  {\em Rev. Mod.
  Phys.} {\bf 79} (2007) 733--796,
  [\href{http://arxiv.org/abs/hep-th/0610102}{{\tt hep-th/0610102}}].

\bibitem{Anninos:2013mfa}
D.~Anninos, T.~Anous, F.~Denef, and L.~Peeters, {\it {Holographic
  Vitrification}},  {\em JHEP} {\bf 1504} (2015) 027,
  [\href{http://arxiv.org/abs/1309.0146}{{\tt arXiv:1309.0146}}].

\bibitem{Denef:2000nb}
F.~Denef, {\it {Supergravity flows and D-brane stability}},  {\em JHEP} {\bf
  08} (2000) 050, [\href{http://arxiv.org/abs/hep-th/0005049}{{\tt
  hep-th/0005049}}].

\bibitem{Denef:2007vg}
F.~Denef and G.~W. Moore, {\it {Split states, entropy enigmas, holes and
  halos}},  {\em JHEP} {\bf 11} (2011) 129,
  [\href{http://arxiv.org/abs/hep-th/0702146}{{\tt hep-th/0702146}}].

\bibitem{Anninos:2011vn}
D.~Anninos, T.~Anous, J.~Barandes, F.~Denef, and B.~Gaasbeek, {\it {Hot Halos
  and Galactic Glasses}},  {\em JHEP} {\bf 01} (2012) 003,
  [\href{http://arxiv.org/abs/1108.5821}{{\tt arXiv:1108.5821}}].

\bibitem{Witten:1998xy}
E.~Witten, {\it {Baryons and branes in anti-de Sitter space}},  {\em JHEP} {\bf
  07} (1998) 006, [\href{http://arxiv.org/abs/hep-th/9805112}{{\tt
  hep-th/9805112}}].

\bibitem{Ivanov:2014ima}
E.~Ivanov and S.~Sidorov, {\it {New type of $\mathcal{N}=4$ supersymmetric
  quantum mechanics}},  {\em AIP Conf.Proc.} {\bf 1606} (2014) 374--385.

\bibitem{Ivanov:2013ova}
E.~Ivanov and S.~Sidorov, {\it {Deformed Supersymmetric Mechanics}},  {\em
  Class.Quant.Grav.} {\bf 31} (2014) 075013,
  [\href{http://arxiv.org/abs/1307.7690}{{\tt arXiv:1307.7690}}].

\bibitem{Festuccia:2011ws}
G.~Festuccia and N.~Seiberg, {\it {Rigid Supersymmetric Theories in Curved
  Superspace}},  {\em JHEP} {\bf 1106} (2011) 114,
  [\href{http://arxiv.org/abs/1105.0689}{{\tt arXiv:1105.0689}}].

\bibitem{Assel:2015nca}
B.~Assel, D.~Cassani, L.~Di~Pietro, Z.~Komargodski, J.~Lorenzen, and
  D.~Martelli, {\it {The Casimir Energy in Curved Space and its Supersymmetric
  Counterpart}},  {\em JHEP} {\bf 07} (2015) 043,
  [\href{http://arxiv.org/abs/1503.05537}{{\tt arXiv:1503.05537}}].

\bibitem{Merlatti:2000ed}
P.~Merlatti, {\it {M theory on $AdS_4 \times Q^{111}$: The Complete
  $\mathrm{Osp}(2|4) \times \mathrm{SU}(2) \times \mathrm{SU}(2) \times
  \mathrm{SU}(2)$ spectrum from harmonic analysis}},  {\em Class.Quant.Grav.}
  {\bf 18} (2001) 2797--2826, [\href{http://arxiv.org/abs/hep-th/0012159}{{\tt
  hep-th/0012159}}].

\bibitem{Myers:1999ps}
R.~C. Myers, {\it {Dielectric branes}},  {\em JHEP} {\bf 9912} (1999) 022,
  [\href{http://arxiv.org/abs/hep-th/9910053}{{\tt hep-th/9910053}}].

\bibitem{Nishioka:2008ib}
T.~Nishioka and T.~Takayanagi, {\it {Fuzzy Ring from M2-brane Giant Torus}},
  {\em JHEP} {\bf 10} (2008) 082, [\href{http://arxiv.org/abs/0808.2691}{{\tt
  arXiv:0808.2691}}].

\bibitem{Kachru:1999vj}
S.~Kachru and J.~McGreevy, {\it {Supersymmetric three cycles and supersymmetry
  breaking}},  {\em Phys. Rev.} {\bf D61} (2000) 026001,
  [\href{http://arxiv.org/abs/hep-th/9908135}{{\tt hep-th/9908135}}].

\bibitem{Gutierrez:2010bb}
N.~Gutierrez, Y.~Lozano, and D.~Rodriguez-Gomez, {\it {Charged particle-like
  branes in ABJM}},  {\em JHEP} {\bf 1009} (2010) 101,
  [\href{http://arxiv.org/abs/1004.2826}{{\tt arXiv:1004.2826}}].

\bibitem{Denef:2008wq}
F.~Denef, {\it {Les Houches Lectures on Constructing String Vacua}},  in {\em
  {String theory and the real world: From particle physics to astrophysics.
  Proceedings, Summer School in Theoretical Physics, 87th Session, Les Houches,
  France, July 2-27, 2007}}, pp.~483--610, 2008.
\newblock \href{http://arxiv.org/abs/0803.1194}{{\tt arXiv:0803.1194}}.

\bibitem{Freedman:2012zz}
D.~Z. Freedman and A.~Van~Proeyen, {\em {Supergravity}}.
\newblock Cambridge U. Press, Cambridge, 2012.

\bibitem{Moore:1998pn}
G.~W. Moore, {\it {Arithmetic and attractors}},
  \href{http://arxiv.org/abs/hep-th/9807087}{{\tt hep-th/9807087}}.

\bibitem{Billo:1999ip}
M.~Billo, S.~Cacciatori, F.~Denef, P.~Fre, A.~Van~Proeyen, et~al., {\it {The
  0-brane action in a general D = 4 supergravity background}},  {\em
  Class.Quant.Grav.} {\bf 16} (1999) 2335--2358,
  [\href{http://arxiv.org/abs/hep-th/9902100}{{\tt hep-th/9902100}}].

\bibitem{Wess:1992cp}
J.~Wess and J.~Bagger, {\em {Supersymmetry and supergravity}}.
\newblock Princeton U. Press, Princeton, 1992.

\bibitem{Brunner:1999jq}
I.~Brunner, M.~R. Douglas, A.~E. Lawrence, and C.~Romelsberger, {\it {D-branes
  on the quintic}},  {\em JHEP} {\bf 08} (2000) 015,
  [\href{http://arxiv.org/abs/hep-th/9906200}{{\tt hep-th/9906200}}].

\bibitem{Douglas:2000ah}
M.~R. Douglas, B.~Fiol, and C.~Romelsberger, {\it {Stability and BPS branes}},
  {\em JHEP} {\bf 09} (2005) 006,
  [\href{http://arxiv.org/abs/hep-th/0002037}{{\tt hep-th/0002037}}].

\bibitem{Sen:1985ph}
D.~Sen, {\it {Supersymmetry in the Space-time $\mathbb{R} \times S^3$}},  {\em
  Nucl.Phys.} {\bf B284} (1987) 201.

\bibitem{Sohnius:1982fw}
M.~Sohnius and P.~C. West, {\it {The Tensor Calculus and Matter Coupling of the
  Alternative Minimal Auxiliary Field Formulation of $N=1$ Supergravity}},
  {\em Nucl.Phys.} {\bf B198} (1982) 493.

\bibitem{Nishioka:2014zpa}
T.~Nishioka and I.~Yaakov, {\it {Generalized indices for $ \mathcal{N} = 1$
  theories in four-dimensions}},  {\em JHEP} {\bf 12} (2014) 150,
  [\href{http://arxiv.org/abs/1407.8520}{{\tt arXiv:1407.8520}}].

\bibitem{Delduc:2006yp}
F.~Delduc and E.~Ivanov, {\it {Gauging N=4 Supersymmetric Mechanics}},  {\em
  Nucl. Phys.} {\bf B753} (2006) 211--241,
  [\href{http://arxiv.org/abs/hep-th/0605211}{{\tt hep-th/0605211}}].

\bibitem{Gaiotto:2004ij}
D.~Gaiotto, A.~Strominger, and X.~Yin, {\it {Superconformal black hole quantum
  mechanics}},  {\em JHEP} {\bf 0511} (2005) 017,
  [\href{http://arxiv.org/abs/hep-th/0412322}{{\tt hep-th/0412322}}].

\bibitem{Denef:2007yt}
F.~Denef, D.~Gaiotto, A.~Strominger, D.~Van~den Bleeken, and X.~Yin, {\it
  {Black Hole Deconstruction}},  {\em JHEP} {\bf 1203} (2012) 071,
  [\href{http://arxiv.org/abs/hep-th/0703252}{{\tt hep-th/0703252}}].

\bibitem{Bena:2013dka}
I.~Bena and N.~P. Warner, {\it {Resolving the Structure of Black Holes:
  Philosophizing with a Hammer}},  \href{http://arxiv.org/abs/1311.4538}{{\tt
  arXiv:1311.4538}}.

\bibitem{Raeymaekers:2015sba}
J.~Raeymaekers and D.~V.~d. Bleeken, {\it {Microstate solutions from black hole
  deconstruction}},  \href{http://arxiv.org/abs/1510.00583}{{\tt
  arXiv:1510.00583}}.

\end{thebibliography}

\providecommand{\href}[2]{#2}\begingroup\raggedright\endgroup

\end{document}